\pdfoutput=1
\documentclass[11pt,aps,a4paper,eqsecnum,amsmath,amssymb
              ,superscriptaddress,notitlepage,nofootinbib
              ]{revtex4-1}

\usepackage{graphicx}

\usepackage{verbatim}
\usepackage[dvipsnames]{xcolor}

\usepackage{tikz}
\usepackage[rm]{subfigure}

\begin{document}
\preprint{}
\title{
  On the critical behaviour  of the integrable $q$-deformed $OSp(3|2)$ superspin
  chain
}  

\author{Holger Frahm}
\author{Konstantin Hobu\ss{}}
\affiliation{%
Institut f\"ur Theoretische Physik, Leibniz Universit\"at Hannover,
Appelstra\ss{}e 2, 30167 Hannover, Germany}

\author{M\'arcio J. Martins}
\affiliation{%
Departamento de F\'isica, Universidade Federal de S\~ao Carlos,
C.P. 676, 13565-905 S\~ao Carlos (SP), Brazil}

\date{\today}

\begin{abstract}
This paper is concerned with the investigation of the 
massless regime 
of an integrable spin chain based on the 
quantum group deformation 
of the 
$OSp(3|2)$ superalgebra. The finite-size properties of the 
eigenspectra    
are computed by solving the respective Bethe ansatz 
equations for
large system sizes allowing us to uncover the 
low-lying critical
exponents. 
We present evidences that critical exponents appear to be
built in terms of composites of anomalous dimensions of 
two Coulomb gases with distinct radii and 
the exponents associated to $Z(2)$ degrees of freedom. This view is
supported by the fact that 
the $S=1$ XXZ integrable chain spectrum is present
in some of the sectors of our superspin chain at a particular
value of the deformation parameter. We find that the
fine structure of finite-size effects 
is very rich for a typical anisotropic spin chain.
In fact, we argue on the existence
of a family of states with the same conformal dimension 
whose lattice degeneracies are apparently lifted 
by logarithmic corrections. On the other hand we also
report on states of the spectrum whose 
finite-size corrections seem to be governed by
a power law behaviour. We finally observe that under toroidal 
boundary conditions 
the ground state dependence on the twist angle has
two distinct
analytical structures. 

\end{abstract}


\maketitle

\section{Introduction}
The interest in the study of one-dimensional 
spin chains goes back
at least
to the exact solution found by Bethe 
for the eigenspectrum of the spin-1/2 
isotropic Heisenberg model \cite{Bethe31}. Over
the years it becomes clear that the spin chain 
Hamiltonians
can be generated by means of the path integral 
formulation
of the partition function of classical two dimensional
lattice statistical models \cite{Baxter:book}. An important 
class of
such systems are the vertex models whose 
statistical configurations are specified by assigning 
spin variables to each
bond of the lattice. It turns out that the theory
of integrable models makes it possible to associate
to each Lie algebra an exactly solved vertex model
whose weights are parametrized by trigonometric 
functions \cite{Bazhanov85,Jimb86}. In general, the spin chains
derivable from such vertex models have a massless
region and their critical properties are expected
to be described by (1+1)-dimensional conformal field 
theories. For instance, in the case of simple 
laced Lie algebras
it is believed that the respective spin chains 
are lattice
realizations of field theories in the category of
the Wess-Zumino-Witten models \cite{AfHa87}. By way
of contrast the understanding of the critical 
properties of spin chains based on super groups
is still matter of investigation since the 
identification of the underlying field theory
appears to be more involved. For example,
recently it was found that the isotropic 
spin chains 
invariant by the
$OSp(n|2m)$ superalgebra have
a number of states in the eigenspectrum
leading to the same conformal 
dimension \cite{FrMa15,FrMa18}. This
degeneracy in the critical dimensions grows with 
the lattice 
size and it has been seen as the 
signature of the presence of non-compact degrees of freedom in the continuum limit \cite{JaRS03}. Recall
here that this scenario has been advocated before 
to be present in
the context of spin chains 
built out of staggered vertex 
models \cite{EsFS05,SaSc07,IkJS08,FrMa11,FrMa12,IkJS12,CaIK13,FrSe14}.

In this paper we initiate an investigation 
of the 
critical behaviour of a spin chain derived from
a vertex model based on the quantum superalgebra
$U_q[OSp(3|2)]$. This superspin chain has the 
peculiar
feature of having one state whose energy per 
site has no
finite-size correction for arbitrary values
of the deformation parameter $q$ in the critical
region. This property is probably 
related to the fact
that the $R$-matrix of this vertex model can be
expressed in terms of the generators 
of a braid-monoid
algebra introduced by
Birman, Wenzel and Murakami \cite{BiWe89,Mura87}. 
For this vertex model
the corresponding product 
of two Temperley-Lieb operators 
gives rise to a loop with a trivial factor  
independent of the deformation 
parameter $q$ \cite{GaMa06}. Therefore, the 
situation is
similar to that of the spin-1/2
Heisenberg with 
anisotropy
$\Delta=1/2$ \cite{AlBB88,Stroganov01} except that 
the vanishing of finite-size correction 
is valid for both even and odd lattice size 
with periodic boundary conditions. 

We shall take this particular state as our point of reference to study the
finite-size corrections in the low energy spectrum of the $U_q[OSp(3|2)]$
superspin chain on a ring of size $L$.  Assuming periodic conditions one
expects that a given eigenenergy $E_n(L)$ of the superspin chain Hamiltonian
in the gapless regime, for large $L$, behaves as \cite{Card86a,Card86b}
\begin{equation}
  E_n(L) = L \varepsilon_{\infty} +  \frac{2\pi v_F}{L} X_n
  + o\left(\frac{1}{L}\right)\,,
\end{equation}
where $\varepsilon_{\infty}$ is the energy per site of the reference state,
$v_F$ is the Fermi velocity of the massless low-lying excitations and $X_n$
denotes the anomalous dimensions of the corresponding operator in the field
theory.

Here we are interested to uncover the structure of the conformal dimensions
$X_n$ with the deformation parameter $q$ for the compact part of the
eigenspectrum. We will see that the dependence of such critical exponents with
$q$ can be expressed in terms of the conformal content of two Coulomb gases
with distinct compactification radii.  It turns out that in the limit
$q \rightarrow 1$ one of the Coulomb gas does not contribute to the critical
behaviour since the respective vortex exponent diverges. This helps to clarify
the reason for the abundance of states with divergent critical exponents found
in our previous work on the isotropic $OSp(3|2)$ superspin chain
\cite{FrMa15}.  Interestingly enough, we find that the feature of having a
family of distinct states with the same leading finite-size correction appears
to persist for generic values of $q \neq 1$ studied in this paper.

This paper is organized as follows. In the next section we describe the Bethe
ansatz solution for the spectrum of the $U_q[OSp(3|2)]$ superspin chain for
two suitable distinct grading basis with generic toroidal boundary conditions.
We use this solution to compute certain thermodynamic limit properties such as
the low momenta dispersion relation for low lying excitations. In
Section~\ref{secFINITESIZE} we elaborate on the leading finite-size
corrections in certain spectral sectors of the superspin chain using the
so-called root density method. This provides us some insights to propose a
generic ansatz for the the behaviour of the anomalous dimensions with the
deformation parameter.
In addition we point out a correspondence between the Bethe ansatz description
and the energies of certain eigenstates of the superspin chain with those of
the integrable spin $S=1$ Heisenberg model for a particular value of the
deformation parameter.
This relationship turns out to be useful to verify our working proposal for
the conformal dimensions. In Section~\ref{secNUMERICS} we use the Bethe ansatz
solution to compute the eigenenergies of the superspin chain for many states
in nine sectors corresponding to different eigenvalues of the generators of
the Cartan subalgebra of $OSp(3|2)$.
Altogether we have investigate the finite-size corrections of around seventy
distinct states. This has helped us to fix some free parameter and stablish
some constraints among the quantum numbers entering our proposal for the
conformal dimensions. The analysis of the subleading finite-size corrections
suggest that for some states we have a combination of power law and
logarithmic behaviour.

\section{The Model and the Bethe Ansatz}
It is known that one-dimensional quantum spin chains can be obtained by taking
certain limits of the Boltzmann weights of classical two-dimensional vertex
models. In this paper we investigate a spin chain which can be derived from a
vertex model based on the quantum deformation of the vector representation of
the Lie superalgebra $OSp(3|2)$. The Boltzmann weights of this vertex model
can be encoded on the so called $R$-matrix which acts on the tensor product of
two five-dimensional spaces, $V_1\otimes V_2$. We shall denote the $R$-matrix
by $R_{12}(\lambda)$ where $\lambda$ represents the spectral parameter
parameterizing the weights. The expression of such $R$-matrix was explicitly
exhibited in Ref.~\cite{GaMa04}.
It turns out that the $R$-matrix commutes with two distinct $U(1)$ symmetries. An immediate consequence of this fact is the following property,
\begin{equation}
    \left[ R_{12}(\lambda), \mathcal{G} \otimes \mathcal{G} \right] = 0\,,
\end{equation}
where $\mathcal{G}$ is a diagonal matrix represented here as
\begin{equation}
  \label{boundMA}
  {\mathcal{G}}=\left(\begin{array}{ccccc}
      e^{i\phi_1} & 0 &0 &0 &0\\
      0 & 1 &0 &0 &0  \\
      0 & 0& \mathrm{e}^{i\phi_2}&0 &0\\
      0 & 0& 0 & \mathrm{e}^{2i\phi_2}&0\\
      0 & 0& 0 & 0& e^{i(2\phi_2-\phi_1)}
    \end{array}\right),\quad
  0 \leq \phi_1,\phi_2 \leq \pi\,.
\end{equation}
Using the matrix ${\mathcal{G}}$ one can study general toroidal boundary
conditions for the corresponding superspin chain preserving the integrability
of the model.  Although the focus of this work in on the case of genuine
periodic chains, i.e.\ $(\phi_1,\phi_2)=(0,0)$ or ${\mathcal{G}}= \mathbf{1}$,
it will turn out to be convenient to formulate the problem for the generic
case. Specifically, we shall investigate the spectrum of the the antiferromagnetic superspin chain with $L$ sites subject to these boundary conditions.  The Hamiltonian with nearest neighbour interactions is expressed in terms of the $R$-matrix and the twist matrix $\mathcal{G}$ and reads
\begin{equation}
\label{HAM}
{\mathcal{H}} = i \sum_{j=1}^{L-1}\frac{\partial}{\partial \lambda} 
R_{jj+1}(\lambda)|_{\lambda=0} 
+i{\mathcal{G}}_{L}^{-1}\frac{\partial}{\partial \lambda} 
R_{L1}(\lambda)|_{\lambda=0} {\mathcal{G}}_L\,.
\end{equation}

The diagonalization of this Hamiltonian can be performed within the algebraic
Bethe ansatz leading to a set of algebraic equations for the rapidities
parameterizing the eigenspectrum.  The basic tools have been already discussed
in \cite{GaMa04} and here we will present only the main results.  The presence of
two $U(1)$ symmetries allows to
decompose the Hilbert space of the model into sectors according to the
eigenvalues of the corresponding charges.  Labeling these sectors by a pair of
integers $(n_1,n_2)$ there exist two convenient reference states which can be
used for the Bethe ansatz, namely the fully polarized states in the sectors
$(0,L)$ and $(L,0)$.  Selecting one of these also fixes the ordering of
bosonic and fermionic basis states to $bfbfb$ and $fbbbf$, respectively. It is
well known that for models based on superalgebras the explicit form of the
Bethe equations depends on the choice of this grading.  Below we will use
whichever turns out to be more convenient for the particular state, therefore
both formulations are presented here.

\subsection{Bethe ansatz in $bfbfb$ grading}
The eigenstates of the Hamiltonian in the grading $bfbfb$ are parameterized
by two sets of complex numbers $\lambda_j^{(1)}$, $j=1,\ldots,L-n_2$ and
$\lambda_j^{(2)}$, $j=1,\ldots,L-n_1-n_2$, satisfying the Bethe equations
\begin{equation}
\label{betheBFBFB}
\begin{aligned}
 \left[\frac{\sinh(\lambda_{j}^{(1)}+i\gamma/2)}
 {\sinh(\lambda_{j}^{(1)}-i\gamma/2)}\right]^{{L}} = &\,
 \mathrm{e}^{i\phi_1} \prod_{k=1}^{{L-n_1-n_2}}
  \frac{\sinh(\lambda_{j}^{(1)}-\lambda_{k}^{(2)}+i\gamma/2)}
 {\sinh(\lambda_{j}^{(1)}-\lambda_{k}^{(2)}-i\gamma/2)},\quad j=1,\cdots,{L-n_2}\,,
 \\
 \prod_{k=1}^{{L-n_2}}
  \frac{\sinh(\lambda_{j}^{(2)}-\lambda_{k}^{(1)}+i\gamma/2)}
 {\sinh(\lambda_{j}^{(2)}-\lambda_{k}^{(1)}-i\gamma/2)} = &\,
 \mathrm{e}^{i(\phi_1-\phi_2)} \prod_{\stackrel{k=1}{k \ne j }}^{L-n_1-n_2}
  \frac{\sinh(\lambda_{j}^{(2)}-\lambda_{k}^{(2)}+i\gamma)}
 {\sinh(\lambda_{j}^{(2)}-\lambda_{k}^{(2)}-i\gamma)}\,
  \frac{\sinh(\lambda_{j}^{(2)}-\lambda_{k}^{(2)}-i\gamma/2)}
 {\sinh(\lambda_{j}^{(2)}-\lambda_{k}^{(2)}+i\gamma/2)}\,,
 \\ 
 &\qquad j=1,\cdots,{L-n_1-n_2} \,,
\end{aligned}
\end{equation}
where the parameter $0 \leq \gamma < \pi $ is related to the quantum
deformation.  Note that in this grading the algebra inclusion
$OSp(3|2) \supset OSp(1|2)$ becomes explicit: the second set of equations
coincides with those for an inhomogeneous $OSp(1|2)$ vertex model
\cite{FrMa15,FrMa18,GaMa04}.  In terms of these parameters the corresponding
energy is
\begin{equation}
\label{eneBFBFB}
  E\left(\{\lambda_j^{(a)}\},L\right)
  = -\sum^{L-n_2}_{j=1} \frac{2 \sin\gamma}
  {\cos\gamma-\cosh(2\lambda_j^{(1)})} -2L \cot\gamma\,.
\end{equation}

This choice of grading turns out to be particularly convenient to study the
thermodynamic limit of the superspin chain in the subsector with charges
$(n_1=0,n_2)$. In this subsector the numbers of Bethe roots are the same for
both levels which helps to simplify the analysis. Comparing the spectrum
obtained by exact diagonalization of the Hamiltonian with our numerical
solution of the Bethe equations (\ref{betheBFBFB}) for small lattice sizes we
find that the Bethe root configurations for levels in these sectors are
dominated by pairs of complex conjugate rapidities with
$\mathrm{Im}\left(\lambda_j^{(1)}\right) \simeq \pm 3\gamma/4$ and
$\mathrm{Im}\left(\lambda_j^{(2)}\right) \simeq \pm \gamma/4$.
To compute bulk properties of the superspin chain we concentrate our analysis
of the Bethe equations to the case of genuine periodic boundary conditions,
$\phi_1=\phi_2=0$. In this case we find that the differences between the real
centers of these two-strings on level-$1$ and $2$ become exponentially small
for large $L$.  This motivates the following string hypothesis involving two
$bfbfb$-Bethe roots from each level for the analysis of the thermodynamic
limit, $L\to\infty$:
\begin{equation}
\label{stringBFBFB}
  \lambda^{(1)}_{j,\pm} \simeq \xi_j \pm i\frac{3\gamma}{4}\,,\quad
  \lambda^{(2)}_{j,\pm} \simeq \xi_j \pm i\frac{\gamma}{4}\,,\quad
  \xi_j\in\mathbb{R}\,.
\end{equation}
Rewriting the Bethe equations (\ref{betheBFBFB}) in terms of the real $\xi_j$
we find that that the second set of Bethe equations is satisfied.  Therefore
we can restrict our study to the analysis of the remaining first level
equations which, after taking the logarithm, read
\begin{equation}
\label{betheSTR2}
\begin{aligned}
 L &\left[\Phi\left(\xi_j,\frac{5\gamma}{4}\right)
        -\Phi\left(\xi_j,\frac{\gamma}{4}\right) \right] 
   = -2\pi Q_j +\\
   &+ \sum_{k=1}^{(L-n_2)/2}
      \left[ \Phi\left(\xi_j-\xi_k,\frac{3\gamma}{2}\right)
      +\Phi\left(\xi_j-\xi_k,\gamma\right) - \Phi\left(\xi_j-\xi_k,\frac{\gamma}{2}\right) 
      \right]\, ,\quad
j=1,\ldots,\frac{L-n_2}{2}\,,
\end{aligned}
\end{equation}
where $ \Phi(x,\gamma) = 2\arctan\left[ \tanh x \, \cot\gamma\right]$.  The
numbers $Q_j$ define the possible branches of the logarithm.  They are
integers or half-integers, depending on the parity of the number of strings
$(L-n_2)/2$ (which is an integer for the states considered).

In this formulation the thermodynamic limit can be studied within the root
density approach \cite{YaYa69} in which the roots $\xi_j$ are expected
to fill the entire real axis. Their density $\sigma(\xi)$ can be defined
as,
\begin{equation}
\sigma(x)= \frac{d Z(x)}{d x}
\end{equation}
with the counting function $\left.Z(x)\right|_{x=\xi_j}=Q_j/L$.
In the limit $L \rightarrow \infty$ the Bethe equations (\ref{betheSTR2}) turn into an integral equation for the density $\sigma(x)$ given by,
\begin{equation}
\label{intBFBFB}
\begin{aligned}
    2\pi \sigma(x) =\, &\Phi'\left(x,\frac{\gamma}{4}\right)
    -\Phi'\left(x,\frac{5\gamma}{4}\right) +\\
    &+\int_{-\infty}^{\infty} \mathrm{d}y\,
    \left [
      \Phi'\left(x-y,\frac{3\gamma}{2}\right)
      + \Phi'\left(x-y,\gamma\right) 
      - 
        \Phi'\left(x-y,\frac{\gamma}{2}\right) \right] \sigma(y)
\end{aligned}
\end{equation}
where $\Phi'(x,\gamma)= \frac{2 \sin(2\gamma)}{\cosh(2x)-\cos(2\gamma)}$. The integral equation (\ref{intBFBFB}) can be solved by the standard Fourier transform method leading to
\begin{equation}
\label{rho}
\sigma(x)= \frac{1}{\gamma \cosh(2\pi x/\gamma)}\,.
\end{equation}
Using this expression we can calculate the energy per site
$\varepsilon_\infty$ in the infinite volume limit to obtain,
\begin{equation}
\label{e0dens}
\varepsilon_\infty = -\int_{-\infty}^{\infty} \mathrm{d}x\,\frac{\sinh\left(\frac{\gamma x}{2}\right) 
\cosh\left(\frac{(\pi-3\gamma/2)x}{2}\right)}
{\sinh\left(\frac{\pi x}{2}\right) \cosh\left(\frac{\gamma x}{4}\right)}
-2\cot(\gamma) = -2\cot{\frac{\gamma}{2}}\,.
\end{equation}
Note that the density of roots and the energy density coincide with the
corresponding expressions (\ref{S1dens}) and (\ref{S1einf}) of the spin $S=1$
XXZ model. Furthermore, since the energy and momentum of elementary excitations
above the ground state are given in terms of the root density as
\begin{equation}
  \varepsilon(x)= 2 \pi \sigma(x),\quad p(x)=\int_{x}^{\infty}
  \mathrm{d}y\,\varepsilon(y)\,. 
\end{equation}
we find that the superspin chain, too, has gapless excitations with linear
dispersion
\begin{equation}
\label{DIS}
    \varepsilon(p) \sim v_F |p|
\end{equation}
with Fermi velocity $v_F=2\pi/\gamma$.

\subsection{Bethe ansatz in $fbbbf$ grading}
Alternatively, the algebraic Bethe ansatz can be done in the grading $fbbbf$.
In this case the eigenstates are parameterized by roots
of a different set of Bethe equations, 
\begin{equation}
\begin{aligned}
\label{betheFBBBF}
  \left[\frac{\sinh(\lambda_{j}^{(1)}+i\gamma/2)}
             {\sinh(\lambda_{j}^{(1)}-i\gamma/2)}\right]^{{L}} = &\,
 e^{i\phi_1} \prod_{k=1}^{{L-n_1-n_2}}
  \frac{\sinh(\lambda_{j}^{(1)}-\lambda_{k}^{(2)}+i\gamma/2)}
       {\sinh(\lambda_{j}^{(1)}-\lambda_{k}^{(2)}-i\gamma/2)},\quad j=1,\cdots,{L-n_1}\,,  
 \\
 \prod_{k=1}^{{L-n_1}}
  \frac{\sinh(\lambda_{j}^{(2)}-\lambda_{k}^{(1)}+i\gamma/2)}
 {\sinh(\lambda_{j}^{(2)}-\lambda_{k}^{(1)}-i\gamma/2)}= &\,
 e^{i\phi_2} \prod_{\stackrel{k=1}{k \ne j }}^{L-n_1-n_2}
  \frac{\sinh(\lambda_{j}^{(2)}-\lambda_{k}^{(2)}+i\gamma/2)}
 {\sinh(\lambda_{j}^{(2)}-\lambda_{k}^{(2)}-i\gamma/2)}\,,
 \quad j=1,\cdots,{L-n_1-n_2} \,.
\end{aligned}
\end{equation}
These equations are related to (\ref{betheBFBFB}) by a particle-hole transformation in rapidity space \cite{BarX92}.  This transformation implies that the second level roots $\lambda^{(2)}$ coincide in the two formulations while the first level ones, $\lambda^{(1)}$, depend on the choice of grading, $bfbfb$ and $fbbbf$. In the following we shall use the same notation but specify the underlying grading, whenever specific root configurations are discussed.

In $fbbbf$ grading the energy of the corresponding eigenstates is given in terms
of the Bethe roots by
\begin{equation}
\label{eneFBBBF}
  E(\{\lambda_j^{(a)}\},L)= \sum^{L-n_1}_{j=1}
    \frac{2 \sin\gamma}{\cos\gamma-\cosh(2\lambda_j^{(1)})}\,.
\end{equation}

Again, the analysis of the thermodynamic limit is simplified when we consider
charge sectors where the number of rapidities $\lambda_j^{(a)}$ on level
$a=1,2$, i.e.\ $(n_1,n_2=0)$ for $fbbbf$ grading.  Here we find that the Bethe
root configurations for low energy states of (\ref{HAM}) are dominated by
pairs of complex conjugate rapidities with imaginary parts $\simeq\pm\gamma/4$
on both levels. As the system size $L$ grows, the root configurations on the
two levels become exponentially close.  Therefore, we can proceed as above and
to study these levels in the thermodynamic limit, $L \to \infty$, we rewrite
the Bethe equations in terms of the real centers $\xi_j$ of the root complexes
containing four $fbbbf$ Bethe roots
\begin{equation}
  \label{stringFBBBF}
  \lambda^{(1)}_{j,\pm}\simeq \xi_j \pm i\frac{\gamma}{4}\,,\quad
  \lambda^{(2)}_{j,\pm}\simeq \xi_j \pm i\frac{\gamma}{4}\,,\quad
  \xi_j \in \mathbb{R}\,. 
\end{equation}
As a result the second level of the Bethe equations are satisfied\footnote{%
  Strictly speaking this occurs for the twist $\phi_2=\pi$ because the
  emergence of a minus sign on left hand side of the second level Bethe
  equations (\ref{betheFBBBF}) when $\lambda^{(1)}_j=\lambda^{(2)}_j$. Here,
  however, we are interested in properties of the thermodynamic limit which is
  assumed to be independent of specific toroidal boundary condition.}
and the spectrum in these sectors is parameterized in terms of the string
equations 
\begin{equation}
\label{betheSTR1}
\begin{aligned}
   L \left[\Phi\left(\xi_j,\frac{3\gamma}{4}\right) +\Phi\left(\xi_j,\frac{\gamma}{4}\right) \right] 
   &= 2\pi Q_j + \sum_{k=1}^{(L-n_1)/2} \left[
      2\Phi\left(\xi_j-\xi_k,\frac{\gamma}{2}\right)
      +\Phi\left(\xi_j-\xi_k,\gamma\right) \right], \\
  & j=1,\ldots,\frac{L-n_1}{2}\,.
\end{aligned}
\end{equation}
Again, the numbers $Q_j$ defining the possible branches of the logarithm are
integers or half-integers, depending on the parity of the integer $(L-n_1)/2$.

Note that the same set of equations is obtained when the Bethe equations
(\ref{S1bethe}) of the integrable spin $S=1$ XXZ model are rewritten using the
string hypothesis (\ref{S1string}), see Appendix~\ref{appS1}.  This
identification extends to the expressions for the corresponding
eigenenergies. Therefore, we can rely on existing results for the spin $S=1$
model to obtain certain properties of the superspin chain in the thermodynamic
limit $L\to\infty$: in complete agreement with what we found for the $(0,n_2)$
charge sectors based on the Bethe ansatz in grading $bfbfb$ the energy density
and dispersion of gapless excitations are given by (\ref{e0dens}) and
(\ref{DIS}), respectively.

\section{Finite-size spectrum}
\label{secFINITESIZE}
To initiate our investigation of the operator content of the $q$-deformed
$OSp(3|2)$ superspin chain we have studied its spectrum for small system sizes
by exact diagonalization of (\ref{HAM}) with toroidal boundary conditions
$(\phi_1=0, \phi_2)$.  Based on the numerical results we observe that the
lowest energy in the charge sector $(n_1,n_2)=(0,0)$ is
\begin{equation}
  \label{e00exact}
  E_{(0,0)}(L) \equiv -2L\cot\frac{\gamma}{2}\,
\end{equation}
without any finite-size corrections independent of $\phi_2$.  Note that this
is exactly the energy $L\varepsilon_\infty$ obtained within the root density
approach for $L\to\infty$, see Eq.~(\ref{e0dens}). As we will see below,
however, this is not the ground state of the superspin chain for finite
$\gamma$.  Inspite of this we shall take this level as a point of reference
and compute the effective scaling dimensions, $X_\text{eff}$, for the states
considered below from
\begin{equation}
  X_{\text{eff}}(L) = \frac{L}{2\pi v_F} \left(E(L)-L\varepsilon_\infty\right)\,.
\label{definition_Xeff}
\end{equation}

\subsection{Root density approach}
Before we present our numerical results for the spectrum of scaling dimensions
based on solutions of the Bethe equations \eqref{betheBFBFB} and
\eqref{betheFBBBF}, respectively, we extend our analytic treatment of the
thermodynamic limit in the previous section to get first insights into the
finite-size scaling of the lowest states in the charge sectors $(n_1=0,n_2)$
and $(n_1,n_2=0)$.  Following de Vega and Woynarovich \cite{VeWo85} we compute
the corresponding finite-size energy gaps at large but finite $L$ within the
root density approach based on the respective string hypotheses.

Starting from the $bfbfb$ string equations (\ref{betheSTR2}) we find for the
lowest state in the sector $(0,n_2)$ we find
\begin{equation}
  E_{(0,n_2)}(L) - L\varepsilon_\infty \simeq \frac{2\pi v_F}{L}\left(n_2^2
    \frac{\gamma}{4\pi}\right) + o\left(\frac{1}{L}\right)\,, 
  \quad n_2=1,2,3,\ldots
  \label{conjecture_0_n2}
\end{equation}
As discussed in Appendix~\ref{appS1} the root density approach neglects
possible contributions to the scaling dimensions due to deviations of the
Bethe roots from the string hypothesis (\ref{stringBFBFB}).  Here, however, it
appears that we are in a fortunate situation as far as the correct $1/L$
behaviour is concerned: our prediction (\ref{conjecture_0_n2}) reproduces the
exact energy (\ref{e00exact}) for $n_2=0$ and is confirmed by our numerical
analysis based on the full Bethe equations in the sectors $(0,n_2>0)$, see
Section~\ref{secNUMERICS} below.

Similarly, we can study the finite-size scaling of the lowest states in the
sectors $(n_1,0)$ based on the $fbbbf$ string hypothesis
(\ref{stringFBBBF}).  In fact, the large $L$ corrections resulting from
(\ref{betheSTR1}) have already been studied for the XXZ chain
 \cite{AlMa89}. Adapting this approach to the present model we find that, for $n_1$ odd, the finite-size scaling is given by 
\begin{equation}
\label{conjecture_n1_0}
  E_{(n_1,0)}(L) - L\varepsilon_\infty \simeq \frac{2\pi v_f}{L}\left(n_1^2
    \frac{(\pi-\gamma)}{4\pi}  -{\frac14}
  \right)  + o\left(\frac{1}{L}\right)\,,
  \quad n_1=1,3,5,\ldots
\end{equation}
Once again, there arise subtleties due to the approximations entering the string hypothesis: deviations from (\ref{stringFBBBF}), either in the imaginary parts of the roots forming the strings or between the string centers on level one and two, can modify the scaling dimensions substantially.  
In spite of that we shall see that the numerical analysis performed in Section~\ref{secNUMERICS} will in fact confirm the proposal (\ref{conjecture_n1_0}) for the lowest state when $n_1$ is odd.

\subsection{Relation to the spin $S=1$ XXZ model}
\label{sec:rel2XXZ}
Here we provide additional support for the proposed finite-size spectrum by
uncovering relations between the $OSp(3|2)$ superspin chain and the $S=1$ XXZ
model: at the particular choice of $\gamma=\pi/2$ for the deformation
parameter the charge sectors $(0,n_2)$ and $(n_1,2)$ of the superspin chain
can be shown to contain the eigenenergies of the spin $S=1$ XXZ model at the
same anisotropy.

We begin by considering the $bfbfb$ Bethe equations in the subsector
$(0,n_2)$.  It is straightforward to see that for $\gamma=\pi/2$ the root
configuration
\begin{equation}
 \lambda^{(1)}_j = \lambda^{(2)}_j +i\frac{\pi}{2}
 = \mu_j+i\frac{\pi}{2}\,,\qquad j=1,\cdots,L-n_2\,,
\end{equation}
automatically satisfies the second set of Bethe equations
(\ref{betheBFBFB}). The remaining first level Bethe equations constrain the
rapidities $\mu_j$ by the relations
\begin{equation}
\label{betheXXZSOP1}
\left[\frac{\sinh(\mu_{j}+i\pi/4)}{\sinh(\mu_{j}-i\pi/4)}\right]^{{L}}=
 (-1)^{n_2+1} \prod_{\stackrel{k=1}{k \neq j}}^{{L-n_2}}
  \frac{\sinh(\mu_{j}-\mu_{k}+i\pi/4)}
 {\sinh(\mu_{j}-\mu_{k}-i\pi/4)}\,,
\quad j=1,\cdots,{L-n_2}\,.
\end{equation}
These are exactly the Bethe equations (\ref{S1bethe}) of the spin $S=1$ XXZ
model with twisted boundary conditions $\varphi=0$ ($\varphi=\pi$) in the
sector with odd (even) magnetization $n_2$.  Furthermore, the corresponding
energy $(\ref{eneBFBFB})$ of the superspin chain coincides with the expression
(\ref{S1energy}) for the XXZ model. Therefore there exists a direct
correspondence between some energy levels of the superspin chain in the
subsectors $(0,n_2)$ and the spectrum of the $S=1$ XXZ model with suitably
twisted boundary conditions for $\gamma=\pi/2$.  Note that this observation
supports our proposal (\ref{conjecture_0_n2}) based on the root density
method. In fact, from the conformal content (\ref{S1fse}) of the periodic,
i.e.\ $\varphi=0$, Heisenberg XXZ $S=1$ model with anisotropy $\gamma=\pi/2$
in the sector with odd $n=n_2$ and vorticity $m=0$ we find for $\gamma=\pi/2$:
\begin{equation}
  E_{n_2,0}^{\mathrm{XXZ}}(L) - L\varepsilon_0
  \,\,{\stackrel{\gamma=\frac{\pi}{2}}{=}}\,\, \frac{2\pi v_f}{L^2}
   \left(\frac{n_2^2}{8}+\frac{1}{8}-\frac{c}{12}\right)= 
  \frac{2\pi v_f}{L^2}\,\left(\frac{n_2^2}{8}\right)\,,
\end{equation}
in perfect agreement with our proposal. The same holds for $n_2$ even using the operator content of the XXZ $S=1$ model but now with \emph{anti}periodic boundary conditions, $\varphi=\pi$. 

Similarly, we now consider the $fbbbf$ Bethe equations in the sector
$(n_1,2)$. Substituting a root configuration where the second level roots
$\lambda_j^{(2)}$ coincide with $L-n_1-2$ of the first level ones, i.e.\
\begin{equation}
    \lambda_j^{(1)} =\lambda_j^{(2)} \equiv \mu_j\,,
  \quad \mathrm{for~}j=1,\ldots, L-n_1-2\,,
\end{equation}
into the second set of Bethe equations (\ref{betheFBBBF}) we see that they are
fulfilled for arbitrary values of the $\mu_j$ provided that the two remaining
first level roots are
$\lambda^{(1)}_{L-n_1-1}=\Lambda$ and $\lambda^{(1)}_{L-n_1} = \Lambda + i\pi/2$.
For even $L-n_1$ the first level Bethe equations for such a configuration can be
satisfied provided that $\Lambda$ is chosen to be one of the roots
of\footnote{%
  Note that $\Lambda=0$ is a solution of this equation for any set
  $\{\mu_j\}_{j=1}^{L-n_1-2}$ which is invariant under $\mu\leftrightarrow
  -\mu$.}
\begin{equation}
    \left[ \frac{\sinh(\Lambda + i\pi/4)}{\sinh(\Lambda-i\pi/4)} \right]^L
    = \prod_{{k=1}}^{{L-n_1-2}}
  \frac{\sinh(\Lambda-\mu_{k}+i\pi/4)}
 {\sinh(\Lambda-\mu_{k}-i\pi/4)}\,,
\end{equation}
and the set of rapidities $\{\mu_j\}_{j=1}^{L-n_1-2}$ satisfies the Bethe
equations (\ref{S1bethe}) of the integrable $S=1$ XXZ model with
\emph{anti}periodic boundary conditions, $\varphi=\pi$, in the sector with
magnetization $n=n_1+2$
\begin{equation}
\label{betheXXZSOP}
\left[\frac{\sinh(\mu_{j}+i\pi/4)}
 {\sinh(\mu_{j}-i\pi/4)}\right]^{{L}}=
 -\prod_{\stackrel{k=1}{k \neq j}}^{{L-n_1-2}}
  \frac{\sinh(\mu_{j}-\mu_{k}+i\pi/4)}
 {\sinh(\mu_{j}-\mu_{k}-i\pi/4)},\quad j=1,\cdots,{L-n_1-2} 
\end{equation}
We further note that the contributions of the roots $\lambda^{(1)}=\Lambda$,
$\Lambda+i\pi/2$ to the corresponding energy (\ref{eneFBBBF}) of the superspin
chain cancel each other for $\gamma=\pi/2$. Thus we have established another
one-to-one correspondence between certain eigenenergies of the superspin
chain, now in the charge sectors $(n_1,2)$, and the spectrum of the
\emph{anti}periodic spin $S=1$ Heisenberg model with even magnetization $n_1$
at $\gamma=\pi/2$. We have checked numerically that the complete spectrum of
the latter appears in that of the superspin chain for lengths up to $L=8$.

The spectral inclusions appearing for deformation parameter $\gamma=\pi/2$ are
summarized as
\begin{equation}
\begin{aligned}
 \mathrm{Spec}[\mathrm{XXZ}(\varphi=0)]_n &\subset \mathrm{Spec}[OSp(3|2)]_{(0,n)}\,, 
\quad\mathrm{for~}n\mathrm{~odd}\,, \\
 \mathrm{Spec}[\mathrm{XXZ}(\varphi=\pi)]_n &\subset \mathrm{Spec}[OSp(3|2)]_{(0,n)}\,, 
\quad\mathrm{for~}n\mathrm{~even}\,, \\
 \mathrm{Spec}[\mathrm{XXZ}(\varphi=\pi)]_{n+2} &\subset \mathrm{Spec}[OSp(3|2)]_{(n,2)}\,,  
\quad\mathrm{for~}n\mathrm{~even}\,, 
\end{aligned}
\end{equation}
where $XXZ(\varphi)$ refers to the Heisenberg XXZ $S=1$ model with toroidal
boundary conditions $\varphi$. 
In Table~\ref{tabL7} we exhibit these inclusions explicitly for $L=7$.
\begin{table}[t]
\begin{center}
\begin{tabular}{|c||c|}
  \hline
$OSp(3|2)_{(n_1,n_2)}$ & XXZ$(\varphi)_n$ \\ \hline \hline
$(0,0)$ & $n=0$, $\varphi=\pi$    \\ \hline
$(0,1)$ & $n=1$, $\varphi=0$      \\ \hline 
$(0,2)$ & $n=2$, $\varphi=\pi$      \\ \hline
$(0,3)$ & $n=3$, $\varphi=0$     \\ \hline
$(0,4)$, $(2,2)$ & $n=4$, $\varphi=\pi$    \\ \hline
$(0,5)$ & $n=5$, $\varphi=0$      \\ \hline
$(0,6)$, $(4,2)$ & $n=6$, $\varphi=\pi$   \\ \hline
$(0,7)$ & $n=7$, $\varphi=0$   \\ \hline
\end{tabular}
\caption{Spectral inclusion for $L=7$ between sectors of the $OSp(3|2)$ and the Heisenberg XXZ spin-$S=1$ chains for $\gamma=\pi/2$. }
\label{tabL7}
\end{center}
\end{table}

We emphasize again that these inclusions are particular for the model with the
specially chosen deformation parameter $\gamma=\pi/2$ where we observe
additional degeneracies in the finite-size spectrum of the superspin chain.
Away from this special point we have no evidence for such a relation with the
XXZ model and the eigenspectrum does not appear to be invariant under
$\gamma\leftrightarrow\pi-\gamma$.

\section{Numerical results}
\label{secNUMERICS}
In this section we analyze the finite-size scaling of the low-lying
excitations of the superspin chain with an even number of lattice sites in a given charge sector $(n_1,n_2)$. Specifically we shall consider the nine distinct sectors with $n_1$ and $n_2$ taking values from the set $\{0,1,2\}$. 

As mentioned earlier, the root configurations of the low lying levels are dominated by the string complexes (\ref{stringBFBFB}) and (\ref{stringFBBBF}) depending on the grading used.  Apart from these a typical solution to the Bethe equations contains a finite number of roots which do not belong to one of these complexes.  A complete classification of the patterns formed by these additional roots for the $U_q[OSp(3|2)]$ superspin chain is not known.  In our numerical work we have observed the following configurations:
\begin{subequations}
\label{def:stringnotation}
\begin{enumerate}
  \item[$1^v_a$:] $1$-strings with parity $v=\pm1$ on level $a=1,2$:
  \begin{equation}
  \label{def:string1}
      \lambda^{(a)} =\xi+i\frac{\pi}{4}(1-v)\,,
  \end{equation}
  \item[$2^+_a$:] $2$-strings with parity $v=1$ on level $a=1,2$:
  \begin{equation}
  \label{def:string2}
    \lambda^{(a)}_\pm \simeq \xi \pm i\frac{\gamma}{4}\,,
  \end{equation}
  \item[$\bar{2}^v_a$:] wide $2$-strings with parity $v=\pm1$ on level $a=1,2$:
  \begin{equation}
  \label{def:string2w}
    \lambda^{(a)}_\pm \simeq \xi +i\frac{\pi}{4}(1-v) \pm i\frac{3\gamma}{4},
  \end{equation}
  
  \item[${3}^v_{12}$:] mixed $3$-strings with parity $v=\pm1$, combining two level-$1$ and one level-$2$ roots as:
  \begin{equation}
  \label{def:string3a}
    \lambda^{(1)}_\pm\simeq\xi + i\frac{\pi}{4}(1-v) \pm i\frac{\gamma}{2}\,,\quad
      \lambda^{(2)} = \xi+i\frac{\pi}{4}(1-v)\,,
  \end{equation}
  \item[${3}^v_{21}$:] mixed $3$-strings with parity $v=\pm1$, combining one level-$1$ and two level-$2$ roots:
  \begin{equation}
  \label{def:string3b}
    \lambda^{(1)}=\xi + i\frac{\pi}{4}(1-v)\,,\quad
      \lambda^{(2)}_\pm\simeq\xi +i\frac{\pi}{4}(1-v) \pm i\frac{\gamma}{2}\,,
  \end{equation}

\end{enumerate}
where  $\xi \in \mathbb{R}$.

We note that both the root complexes (\ref{stringBFBFB}) and (\ref{stringFBBBF}) and the string configurations (\ref{def:string2})--(\ref{def:string3b}) of length $>1$ may appear strongly deformed in Bethe root configurations for finite $L$ or when their real center $\xi$ becomes large.  In such cases it may not be possible to discriminate between $2$-strings, wide $2$-strings, and the components of mixed three strings with $\mathrm{Im}(\lambda)\notin\{0,\pi/2\}$.  In these cases the corresponding roots appear as 
\begin{enumerate}
    \item[$z_a$:] pairs of complex conjugate Bethe roots on level $a=1,2$:
    \begin{equation}
    \label{def:stringx}
        \lambda_\pm^{(a)} = \xi \pm i\eta\,,\quad
        \xi\in\mathbb{R}\,,\quad 
        0 < |\eta| < \frac{\pi}{2}\,,
    \end{equation}
\end{enumerate}
where we have used that solutions to the Bethe equations are defined modulo $i\pi$ only.
\end{subequations}

To describe the root configurations parameterizing a particular eigenstate
below we introduce the following notation: using the patterns
(\ref{def:stringnotation}) we will indicate only the roots which are not part
of the dominant string complexes, i.e.\ Eqs.~(\ref{stringBFBFB}) and
(\ref{stringFBBBF}) depending on the grading.
As an example, consider an excitation which is described by a certain number of $fbbbf$ complexes and, in addition, $k$ real roots and one $1^-$-string on the first level as well as a wide ${2}^-$-string on the second level.  Such a configuration will be indicated by the short notation $f:[(1^+_1)^k,1^-_1,\bar{2}^-_2]$. 
Note that the number of additional $fbbbf$ string complexes in this root configuration is fixed by the charges $(n_1,n_2)$ which determine the total number of Bethe roots for a given grading and system size $L$.
In addition to such root patterns with finite $\xi$ we have also observed solutions containing strings which are located at $\pm\infty$.  For these we do not have to distinguish different parities and extend our notation as, e.g., $[1_1]_{+\infty}$ for a root $\lambda^{(1)}=+\infty$.
As a complement to this qualitative description the full set of data used for our finite-size analysis is available online \cite{FrHM19.data}.

Considering our observations in Section~\ref{secFINITESIZE} we expect that
the critical regime of the superspin chain is in the interval
$\gamma\in[0,\pi]$.  Let us emphasize, however, that for much of the numerical
analysis in this work we have concentrated on the region $\gamma\le\pi/2$
where we find that most of the states considered keep their basic root
structure, independent of the deformation parameter. 
As a working hypothesis for the respective lowest scaling dimensions 
we propose that they can be expressed in terms of the sum of
several distinct parts.
The underlying $U(1)$ symmetries of the 
superspin chain 
give rise to two Gaussian fields
with distinct compactification ratios depending on the deformation parameter $\gamma$.
Motivated by our preliminary finite-size analysis within the root density approach we propose that these Coulomb gas
contributions to the anomalous dimensions are given by
\begin{equation}
  \label{GAUSconjecture_full}
  \Xi_{n_1,m_1}^{n_2,m_2} =
  n_1^2\,\frac{\pi-\gamma}{4\pi} + m_1^2\, \frac{\pi}{\pi-\gamma} 
  + n_2^2\,\frac{\gamma}{4\pi}   + m_2^2\, \frac{\pi}{4\gamma}
\end{equation}
where $m_1$ and $m_2$ take into account the vortex companions of the spin
excitations $n_1$ and $n_2$.  
By the same token we propose that the contribution of the Gaussian fields 
to the conformal spin is
\begin{equation}
  \label{GAUSconjecture_spin}
    \sigma_{n_1,m_1}^{n_2,m_2} = n_1 m_1 + \frac12 n_2 m_2\,.
\end{equation}

In addition we anticipate that there are contributions to the anomalous dimensions and to the conformal spin coming from fields associated with discrete symmetries. 
Support for this expectation comes from the inclusion of levels of the $S=1$ integrable XXZ chain in the spectrum of the superspin chain discussed in Section~\ref{sec:rel2XXZ}. Recall here that the critical properties of the XXZ chain is known to be described in terms of composites of Gaussian 
and $Z(2)$ fields, see Appendix~\ref{appS1}. Putting these
informations together
we propose that the conformal data for the $U_q[OSp(3|2)]$ 
superspin chain can be described by
\begin{equation}
  \label{conjecture_full}
  \begin{aligned}
  X_{n_1,m_1}^{n_2,m_2} & = \Xi_{n_1,m_1}^{n_2,m_2} +x_0\,,\\
  s_{n_1,m_1}^{n_2,m_2} & = \sigma_{n_1,m_1}^{n_2,m_2} \pm s_0\,.
  \end{aligned}
\end{equation}
where $x_0$ and $s_0$ account for the contributions 
of the potential
discrete degrees of freedom to the conformal properties.

Here $x_0$ is assumed to be independent of
$\gamma$ and will be determined from the
numerical solution of the Bethe equations (\ref{betheBFBFB}) and
(\ref{betheFBBBF}) for system sizes up to $L=8192$.
Before considering this
task we would like to make the following remark.
We expect that the quantum
numbers in (\ref{conjecture_full}) will be subjected to selection rules
which will play a role in selecting one out of the possible values allowed
for $x_0$. This can been seen for instance by comparing
the proposal (\ref{conjecture_full}) with the expected conformal 
dimensions 
of the isotropic $OSp(3|2)$ superspin chain. We see that in the limit 
$\gamma \rightarrow 0$ 
the conformal dimensions (\ref{conjecture_full}) do not depend on 
the quantum number $n_2$ while
the modes $m_2 \ne 0$ decouple from the 
low energy spectrum. 
This comparison on the subspace
of states $n_2=m_2=0$ reveals us that for $L$ even 
the quantum numbers $n_1$ and $m_1$ satisfy the following rule,
\begin{eqnarray}
&& \bullet~~ \mathrm{for}~~n_1~~ even \rightarrow m_1= \pm \frac{1}{2}, 
\pm \frac{3}{2}, \dots, \nonumber \\
&& \bullet~~ \mathrm{for}~~n_1~~ odd \rightarrow m_1= 0,\pm 1,\pm 2,\dots,
\end{eqnarray}
where in both cases we have $x_0=-1/4$.

In next section we shall see that other values for $x_0$ are possible when 
the space of states are enlarged to include states with $n_2$ and $m_2 \neq 0$. This is an
indication of the presence of additional degrees of freedom besides 
the Gaussian fields in the operator content of the $U_q[OSp(3|2)]$ superspin chain.
In addition to that we find that the vortex
quantum number $m_1$ appears to take values on $\mathbb{Z}/2$ while the
quantum number $m_2 \in \mathbb{Z}$.
%

\subsection{Sector $(0,0)$}
\label{sec:sub00}
We have already mentioned above that the ground state in this sector has
energy $L\epsilon_\infty$ without finite-size corrections, Eq.~(\ref{e00exact}).  In terms of our
proposal (\ref{conjecture_full}) this corresponds to a primary operator with
scaling dimension $X_{0,0}^{0,0}=\Xi_{0,0}^{0,0}=0$.  This state has
zero momentum, consistent with a conformal spin $s=s_0=0$ of the corresponding operator.

\begin{figure}[ht]
   \includegraphics[width=0.95\textwidth]{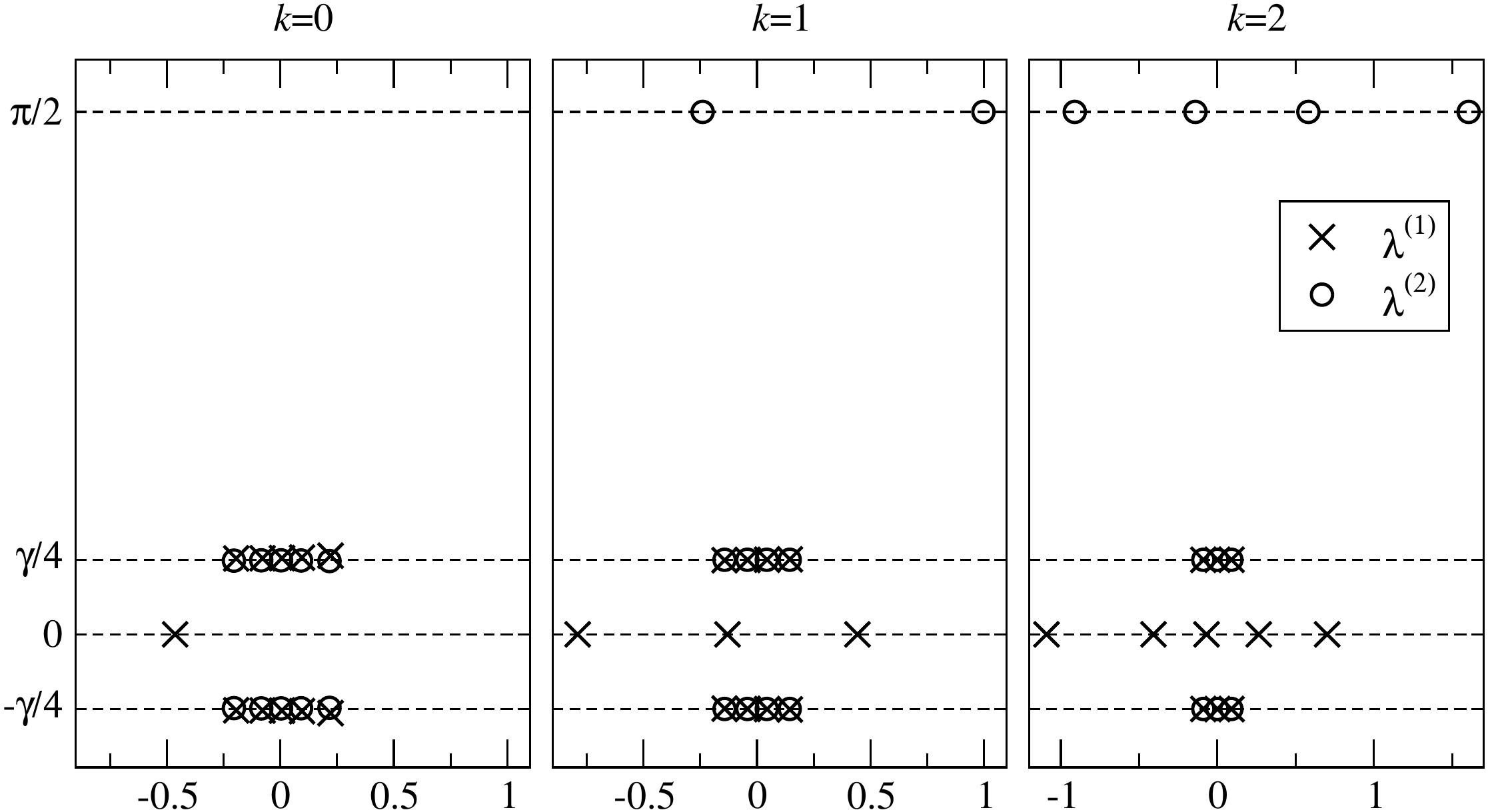}
    \caption{Finite part of the $fbbbf$ root configurations (\ref{tower00}) with $k=0,1,2$ for $L = 12$ and $\gamma = 2\pi/7$. 
    }
  \label{fig:00_tower-roots}
\end{figure}

For the first excitation in this charge sector we find that its $fbbbf$ root
configuration contains a single root $\lambda^{(1)}=-\infty$ on the first
level and a two-string $\lambda_\pm^{(2)}=-\infty$ on the second level.  As a consequence, the remaining finite roots have to satisfy the Bethe equations (\ref{betheFBBBF}) for $n_1=1$, $n_2=1$ in the presence of twists $(\phi_1,\phi_2)=(2\gamma,\gamma)$.  They are arranged in $(L-2)/2$ $fbbbf$ two-string complexes (\ref{stringFBBBF}) and an additional first level root $\lambda^{(1)}\in\mathbb{R}$, i.e.\ $f:[1^+_1]\oplus[1_1,2_2]_{-\infty}$ in the notation introduced in Eqs.~(\ref{def:stringnotation}) above.  In fact, this is the first of a family of excitations with zero momentum in this sector in which $k$ of the string complexes are replaced by $2k$ real roots $\lambda^{(1)}$ and $2k$ second level roots with $\mathrm{Im}(\lambda^{(2)})=\pi/2$, i.e.\
\begin{equation}
  \label{tower00}
  \begin{aligned}
  \left\{\lambda^{(1)} \right\}
  &= \left\{\xi_j^{(1)}\pm i\frac{\gamma}{4},\,
    \xi_j^{(1)}\in\mathbb{R}\right\}_{j=1}^{(L-2)/2-k}
  \cup \left\{\lambda_j \in\mathbb{R}\right\}_{j=1}^{2k+1}
  \cup\left\{-\infty\right\}\,\\
  \left\{\lambda^{(2)} \right\}
  &= \left\{\xi_j^{(2)}\pm i\frac{\gamma}{4},\,
    \xi_j^{(2)}\in\mathbb{R}\right\}_{j=1}^{(L-2)/2-k}
  \cup \left\{\lambda_j \in\mathbb{R}+i\frac{\pi}{2}\right\}_{j=1}^{2k}
  \cup \left\{-\infty,-\infty\right\}\,,
\end{aligned}
\end{equation}
or $f:[(1_1^+)^{(2k+1)},(1_2^-)^{2k}] \oplus[1_1,2_2]_{-\infty}$ for short, see Figure~\ref{fig:00_tower-roots}.

\begin{figure}[ht]
  \includegraphics[width=0.7\textwidth]{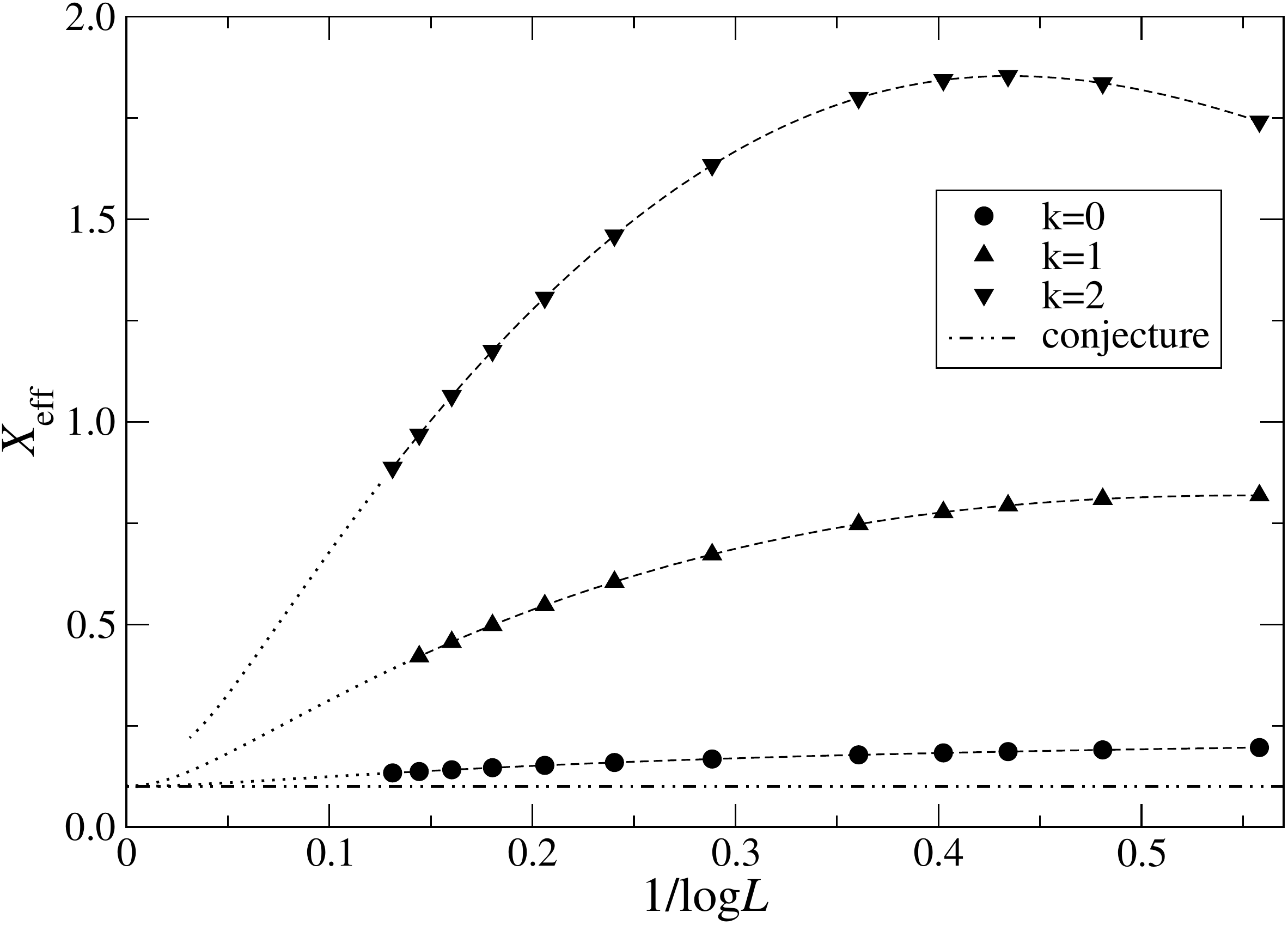}
  \caption{Effective scaling dimensions extracted from the finite-size
    behaviour of the eigenenergies of the superspin chain in sector $(0,0)$
    described by $fbbbf$ root configurations (\ref{tower00}) with $k=0,1,2$
    for $\gamma=2\pi/7$. Dotted lines show the extrapolation of the
    finite-size data assuming a rational dependence on $1/\log L$, the
    dashed-dotted line is our conjecture (\ref{conj00-tower}) for this
    anisotropy.}
  \label{fig:00_tower-scaling}
\end{figure}

We have solved the Bethe equations (\ref{betheFBBBF}) for these configurations
with $k=0,1,2$ for chains of up to $L=2048$ sites.  The effective scaling
dimensions (\ref{definition_Xeff}) computed from the resulting finite-size
energies show strong subleading corrections to scaling, see
Figure~\ref{fig:00_tower-scaling}.
Assuming a rational dependence on $1/\log L$ we have extrapolated the
finite-size data and find that these levels form a 'tower' starting at
\begin{equation}
  \label{conj00-tower}
  X_{0,\frac12}^{0,0} = \Xi_{0,\frac12}^{0,0} -\frac14
  =\frac{\pi}{4(\pi-\gamma)} - \frac14\,
\end{equation}
with the dominant subleading corrections vanishing as a power of $1/\log L$ in the thermodynamic limit $L\to\infty$, similar as in the isotropic model \cite{FrMa15}.

\begin{figure}[ht]
  \includegraphics[width=0.7\textwidth]{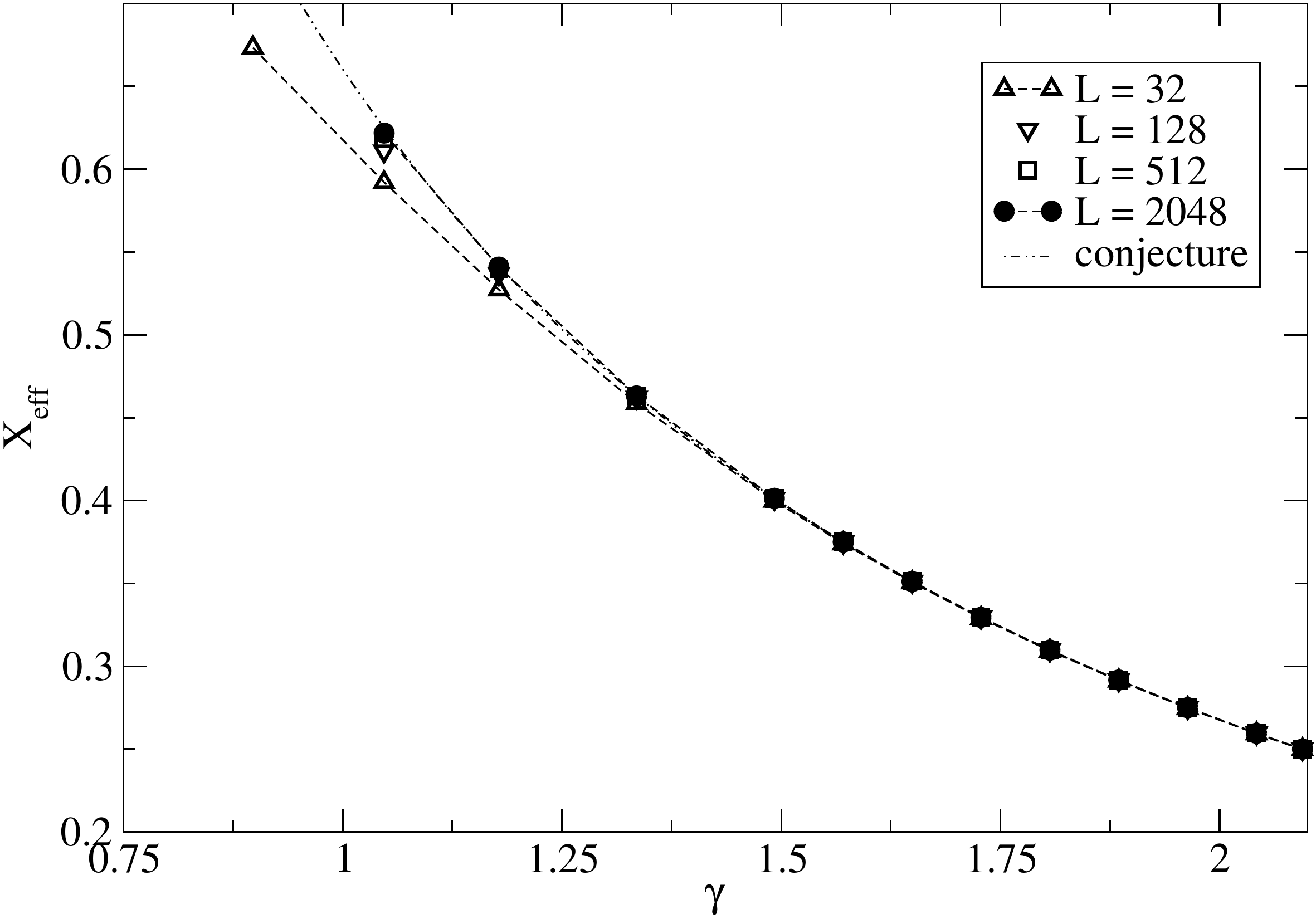}
  \caption{Effective scaling dimensions of the second excitation in the sector $(0,0)$ as a function of $\gamma$ for various system sizes. The dashed-dotted
    line is our conjecture (\ref{conj00-s2}) for this state.}
  \label{fig:00-s2-scaling}
\end{figure}

The next excitation for small system sizes has momentum $p=0$. 
This state is described by root configurations $b:[1_1^+]\oplus[1_1,2_2]_{-\infty}$ or $f:[1_1]_\infty\oplus[1_1,2_2]_{-\infty}$, depending on the grading used. It is studied most conveniently in the $bfbfb$ formulation where the
finite roots satisfy the corresponding Bethe equations (\ref{betheBFBFB}) with $n_1=n_2=1$ and twist angles $(\phi_1,\phi_2) = (2\gamma, \gamma)$.  The effective scaling dimensions of this state as computed from the finite-size energies in chains with up to $L=2048$ sites extrapolate to
\begin{equation}
  \label{conj00-s2}
  X_{0,0}^{0,1} = \Xi_{0,0}^{0,1}-\frac18 = \frac{\pi}{4\gamma} -\frac18\,,
\end{equation}
see Figure~\ref{fig:00-s2-scaling}.
This observation can be underpinned by a relation to the $S=1$ XXZ chain similar to the ones discussed in Section~\ref{sec:rel2XXZ}: the finite roots in the $fbbbf$ configuration satisfy the Bethe equations (\ref{betheFBBBF}) in sector $(n_1,n_2)=(2,0)$ with twist angles $(\phi_1,\phi_2)=(2\gamma,2\gamma)$.  Furthermore, at $\gamma=\pi/2$, we find that $\lambda_j^{(1)}=\lambda_j^{(2)}\equiv \mu_j$ for these roots, which satisfy the Bethe equations (\ref{S1bethe}) of the XXZ model with periodic boundary conditions for this value of $\gamma$. As a result the proposal (\ref{conj00-s2}) agrees with the finite size scaling of the lowest zero-momentum excitation in the sector with magnetization $n=2$ of the XXZ model with periodic boundary conditions, Eq.~(\ref{S1fse}),
\begin{equation}
  E_{2,0}^{\mathrm{XXZ}}(L) - L\varepsilon_0
  \,\,{\stackrel{\gamma=\frac{\pi}{2}}{=}}\,\, \frac{2\pi v_F}{L}
   \left(\frac{2^2}{8}-\frac{c}{12}\right)= 
  \frac{2\pi v_F}{L}\,\left(\frac{3}{8}\right)\,.
\end{equation}

\begin{figure}[ht]
  \includegraphics[width=0.7\textwidth]{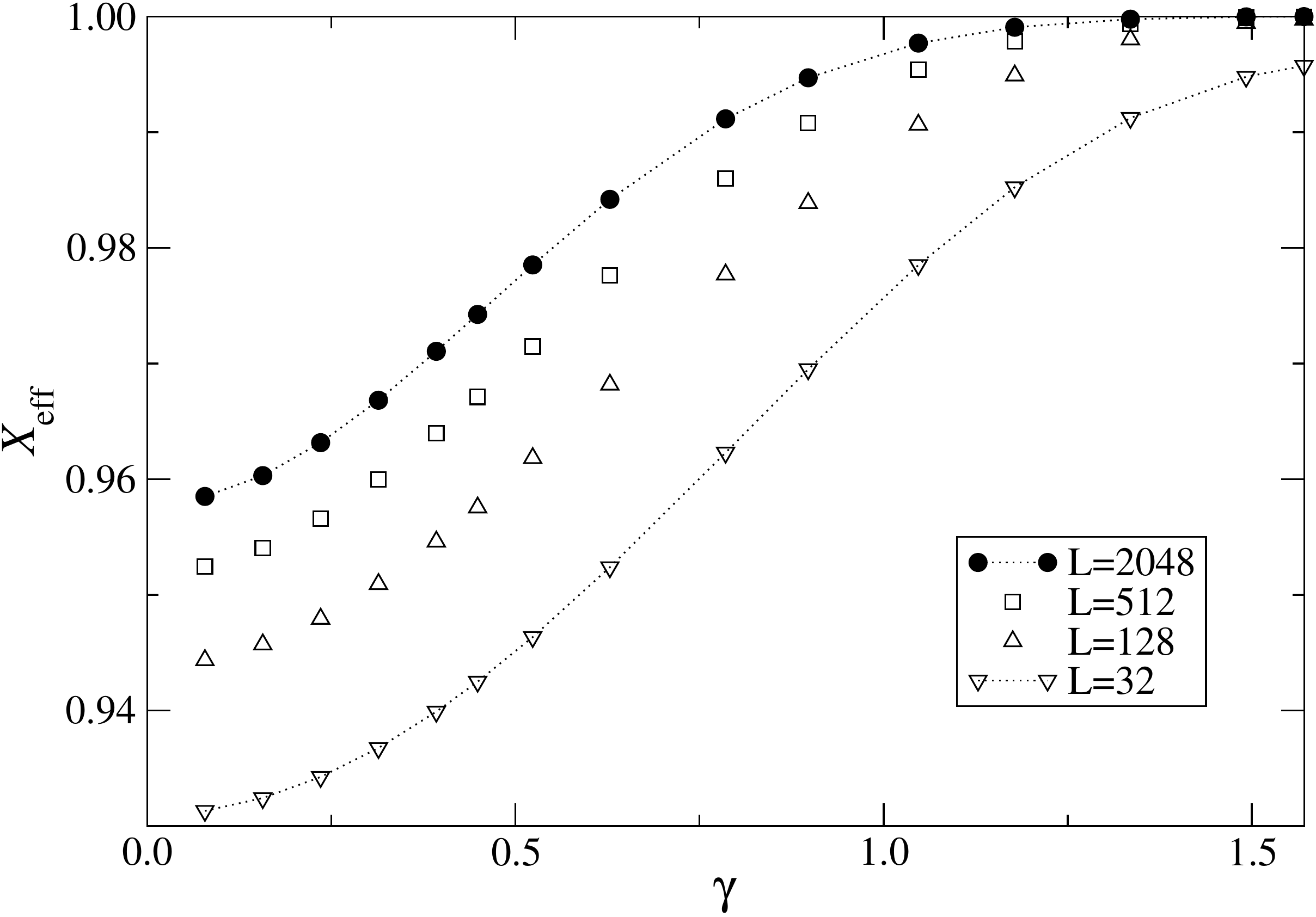}
  \caption{Effective scaling dimensions of the descendent of the ground state of charge sector $(0,0)$ as a function of $\gamma$ for various system sizes. Our conjecture for this state is (\ref{conj00-s4}), $X_{\textrm{eff}}\equiv1$, independent of $\gamma$.}
  \label{fig:00-s4-scaling}
\end{figure}

Continuing our finite-size analysis of the low energy states in the charge
sector $(0,0)$ we have identified an excitation with conformal spin $s=1$ 
described by a root configuration $f:[1_1]_{+\infty}\oplus[1_1,2_2]_{-\infty}$. In this case the finite roots satisfy the Bethe equations (\ref{betheFBBBF}) for $(n_1,n_2)=(2,0)$ in the presence of twists $(\phi_1,\phi_2)=(2\gamma,2\gamma)$.
From our numerical data we conclude that this level is a descendent of the lowest state in this sector with scaling dimension
\begin{equation}
  \label{conj00-s4}
  X = \Xi_{0,0}^{0,0} + 1 = 1\,,
\end{equation}
independent of $\gamma$, see Figure~\ref{fig:00-s4-scaling}.
Again we can relate this proposal to the finite-size scaling of an excitation appearing in the periodic $S=1$ XXZ model for $\gamma=\pi/2$. More precisely, this state corresponds to $E_{2,1}^{\mathrm{XXZ}}$, see Eq.~(\ref{S1fse}).

\begin{figure}[ht]
  \includegraphics[width=0.7\textwidth]{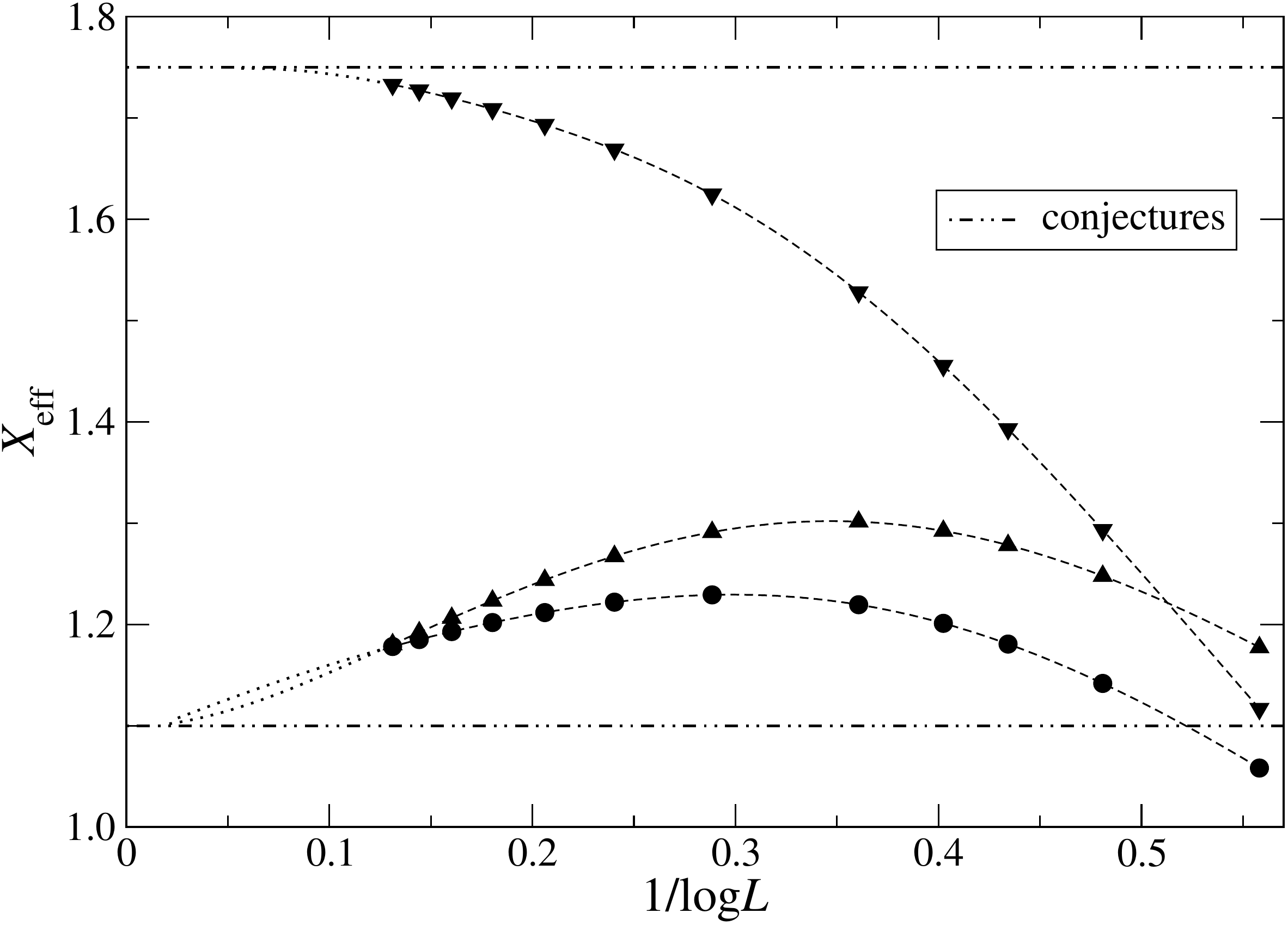}
  \caption{Extrapolation of the effective scaling dimensions of several spin $s=1$ states in charge sector $(0,0)$ for $\gamma=2\pi/7$. The conjectures are given in Eqs.~(\ref{conj00-s5.8}) and (\ref{conj00-s7}).}
  \label{fig:00-spin1}
\end{figure}

Among the remaining low energy states there are two levels with conformal spin $s=1$ and scaling dimension extrapolating to
\begin{equation}
  \label{conj00-s5.8}
  X = \Xi_{0,\frac12}^{0,0} -\frac14 + 1 = \frac{\pi}{4(\pi-\gamma)} -\frac14 + 1\,,
\end{equation}
see Figure~\ref{fig:00-spin1}.
The root structure of one of these states is
$f:[(1_1^+)^3,z_2]\oplus[1_1,2_2]_{-\infty}$.
We note that this
configuration is obtained by breaking one of $fbbbf$ string complexes in the first excitation, described by (\ref{tower00}) with $k=0$, into two real roots on level $1$ and 
complex pair on level $2$.

The roots for the other level are arranged as $f:[(1_1^+)^2]\oplus[1_1,2_2]_{-\infty}\oplus[1_1,2_2]_{+\infty}$. 
In this case the finite roots satisfy the Bethe equations (\ref{betheFBBBF}) for sector $(n_1,n_2)=(2,2)$ with periodic boundary conditions $(\phi_1,\phi_2)=(0,0)$.  This is one example of a more general situation, observed in exact diagonalization of systems with sizes up to $L=8$: many of the low lying eigenenergies in the sector $(2,2)$ are also present in the sector $(0,0)$. We defer the discussion of these common levels to Section~\ref{sec:sub22} where the low lying states of sector $(2,2)$ are studied.

Another spin $s=1$ level appearing in this sector is parameterized by a $bfbfb$ root configuration with the same string content as the second excitation discussed above, i.e.\ $b:[1_1^+]\oplus[1_1,2_2]_{-\infty}$.  Extrapolating its effective scaling dimension we find that it is a descendent of this excitation with
\begin{equation}
  \label{conj00-s7}
  X = \Xi_{0,0}^{0,1} -\frac18 + 1 = \frac{\pi}{4\gamma} -\frac18 + 1\,,
\end{equation}
which, for $\gamma=\pi/2$, can again be related to an eigenenergy in the magnetization $n=2$ sector of the spin $S=1$ XXZ model.
The finite-size scaling of this state is shown in Fig.~\ref{fig:00-spin1}, too.

Also among the low energy states is a spin $s=1$ level which, depending on the grading chosen, is described by root configurations $f:[(1_1^+)^3,z_2)]\oplus[1_1,2_2]_{-\infty}$ or $b:[1_1^+,1_1^-,z_2]\oplus[1_1,2_2]_{-\infty}\oplus[1_1]_{+\infty}$ for small $L$. In both gradings one of the real first-level roots, $[1_1^+]$, and the real position of the pair of complex conjugate roots, $[z_2]$, increases considerably as the system size grows.
We describe this behaviour in more detail in Appendix~\ref{app:roots}.
It is an indication for a change of the root configuration to a new pattern at some finite-size $L_*$ which we have not been able to identify, unfortunately.  The value of $L_*$ where this degeneration takes place depends on the anisotropy, e.g.\ $L_*\approx26$ for $\gamma=2\pi/7$. As a consequence of this scenario we do not have sufficient data for a reliable finite-size analysis of this level.

Our findings for the charge sector $(0,0)$ are summarized in Table~\ref{tab:sec00}.
\begin{table}[t]
\begin{ruledtabular}
\begin{tabular}{l|ccc|ccc|l}
     &  \multicolumn{3}{c|}{$X$} & \multicolumn{3}{c|}{$s$} & \\
 Eq. & $m_1$  & $m_2$ & $x_0$ & total spin & $ \sigma_{n_1,m_1}^{n_2,m_2}$ & $s_0$ & remark\\\hline
(\ref{e00exact}) & $0$ & $0$ & $0$ & $0$ & $0$ & $0$\\
(\ref{conj00-tower}) & $\frac12$ & $0$ & $-\frac14$ & $0$ & $0$ & $0$ & tower\\
(\ref{conj00-s2}) & $0$ & $1$ & $-\frac18$ & $0$ & $0$ & $0$ & \\
\end{tabular}
\end{ruledtabular}
\caption{\label{tab:sec00}Summary of the conformal data of primary fields identified in sector $(n_1,n_2)=(0,0)$. The parameterization is according to our proposal (\ref{conjecture_full}). 
In addition we have observed descendants of these primaries, namely (\ref{conj00-s4}) of (\ref{e00exact}), (\ref{conj00-s5.8}) of (\ref{conj00-tower}), and (\ref{conj00-s7}) of (\ref{conj00-s2}).}
\end{table}

\subsection{Sector $(0,1)$}
In this sector we have analyzed the eight lowest levels present in the spectrum of the superspin chain with $L=6$ sites and, in addition, some higher excitations which extrapolate to small effective scaling dimensions.  Based on our numerical finite-size analysis we find that the effective scaling dimensions of these states extrapolate to four different values fitting into our proposal (\ref{conjecture_full}).

\begin{figure}[ht]
  \includegraphics[width=0.7\textwidth]{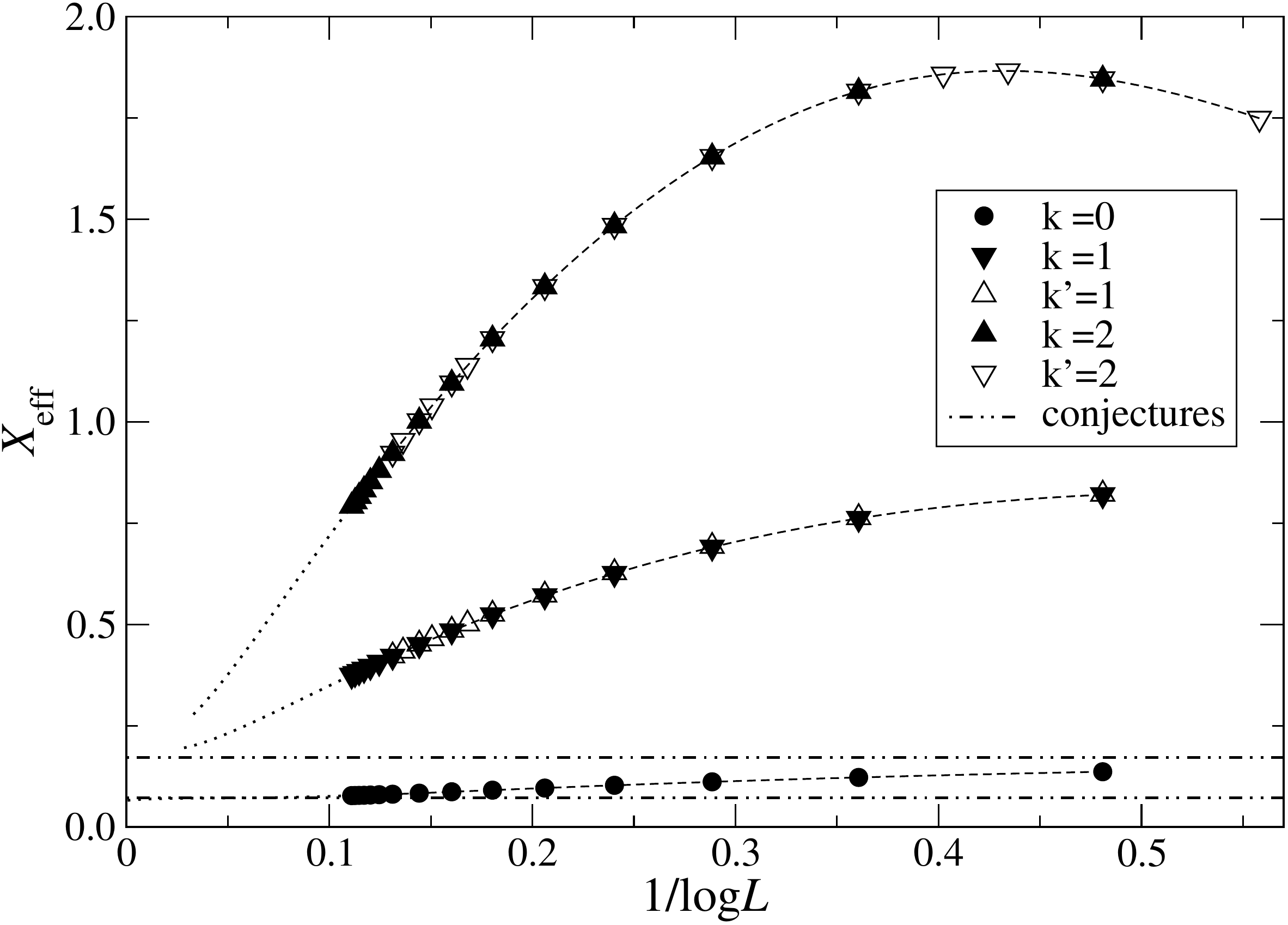}
  \caption{Similar as Fig.~\ref{fig:00_tower-scaling} but for the
    eigenenergies of the superspin chain in sector $(0,1)$ parameterized by the $fbbbf$ root configurations (\ref{tower01a}) with $k=0,1,2$ and (\ref{tower01b}) with $k'=1,2$ for $\gamma=2\pi/7$. The dashed-dotted lines are our conjectures (\ref{conj01-gs}) and (\ref{conj01-tower}) for this anisotropy.}
  \label{fig:01_tower-scaling}
\end{figure}

The lowest state is one of a family of levels with zero momentum, similar as in the $(0,0)$ sector, described by $fbbbf$ root configurations
\begin{equation}
  \label{tower01a}
  \begin{aligned}
  \left\{\lambda^{(1)} \right\}
  &= \left\{\xi_j^{(1)}\pm i\frac{\gamma}{4},\,
    \xi_j^{(1)}\in\mathbb{R}\right\}_{j=1}^{(L-2)/2-k}
  \cup \left\{\lambda_j \in\mathbb{R}\right\}_{j=1}^{2k}
  \cup\left\{\pm i\frac{\gamma}{2}\right\}\,\\
  \left\{\lambda^{(2)} \right\}
  &= \left\{\xi_j^{(2)}\pm i\frac{\gamma}{4},\,
    \xi_j^{(2)}\in\mathbb{R}\right\}_{j=1}^{(L-2)/2-k}
  \cup \left\{\lambda_j \in\mathbb{R}+i\frac{\pi}{2}\right\}_{j=1}^{2k}
  \cup \left\{0\right\}\,,
\end{aligned}
\end{equation}
or $f:[(1_1^+)^{2k},(1_2^-)^{2k},3_{12}^+]$ with integer $k\geq0$.  Here we find another set of levels parameterized by $fbbbf$ Bethe roots arranged as
\begin{equation}
  \label{tower01b}
  \begin{aligned}
  \left\{\lambda^{(1)} \right\}
  &= \left\{\xi_j^{(1)}\pm i\frac{\gamma}{4},\,
    \xi_j^{(1)}\in\mathbb{R}\right\}_{j=1}^{(L-2)/2-k'}
  \cup \left\{\lambda_j \in\mathbb{R}\right\}_{j=1}^{2k'+1}
  \cup\left\{i\frac{\pi}{2}\right\}\,\\
  \left\{\lambda^{(2)} \right\}
  &= \left\{\xi_j^{(2)}\pm i\frac{\gamma}{4},\,
    \xi_j^{(2)}\in\mathbb{R}\right\}_{j=1}^{(L-2)/2-k'}
  \cup \left\{\lambda_j \in\mathbb{R}+i\frac{\pi}{2}\right\}_{j=1}^{2k'-2}
  \cup \left\{i\frac{\pi}{2}, i\frac{\pi}{2} \pm i\frac{\gamma}{2}\right\}\,,
\end{aligned}
\end{equation}
or $f:[(1_1^+)^{(2k'+1)},(1_2^-)^{(2k'-1)}, 3_{21}^-]$ for positive integers $k'$.\footnote{We note that the mixed $3$-strings in these configurations are exact, i.e.\ the constituent rapidities are separated by $\gamma/2$ without deviations.}
We found that the levels (\ref{tower01a}) and (\ref{tower01b}) with $k=k'$ are almost degenerate, already for $L=6$ the relative difference of the energies at $\gamma=2\pi/7$ is $<10^{-3}$.  We have solved the the Bethe equations (\ref{betheFBBBF}) for these configurations with $k=0,1,2$, $k'=1,2$ and system sizes up to $L=8192$.  The scaling dimension for the lowest level ($k=0$) level extrapolates to
\begin{equation}
  \label{conj01-gs}
  X_{0,0}^{1,0} = \Xi_{0,0}^{1,0} = \frac{\gamma}{4\pi}\,.
\end{equation}
The remaining four levels ($k=1,2$ and $k'=1,2$) give rise to the same anomalous dimension, namely
\begin{equation}
  \label{conj01-tower}
  X_{0,\frac12}^{1,0} = \Xi_{0,\frac12}^{1,0} - \frac{1}{4} = \frac{\pi}{4(\pi-\gamma)} + \frac{\gamma}{4\pi} - \frac14\,.
\end{equation}
Again, this degeneracy is lifted for finite $L$ giving rise to a fine-structure due to strong subleading corrections to scaling.  The scaling of the five states giving (\ref{conj01-gs}) and (\ref{conj01-tower}) for $\gamma=2\pi/7$ is shown in Fig.~\ref{fig:01_tower-scaling}.

\begin{figure}[ht]
    \includegraphics[width=0.7\textwidth]{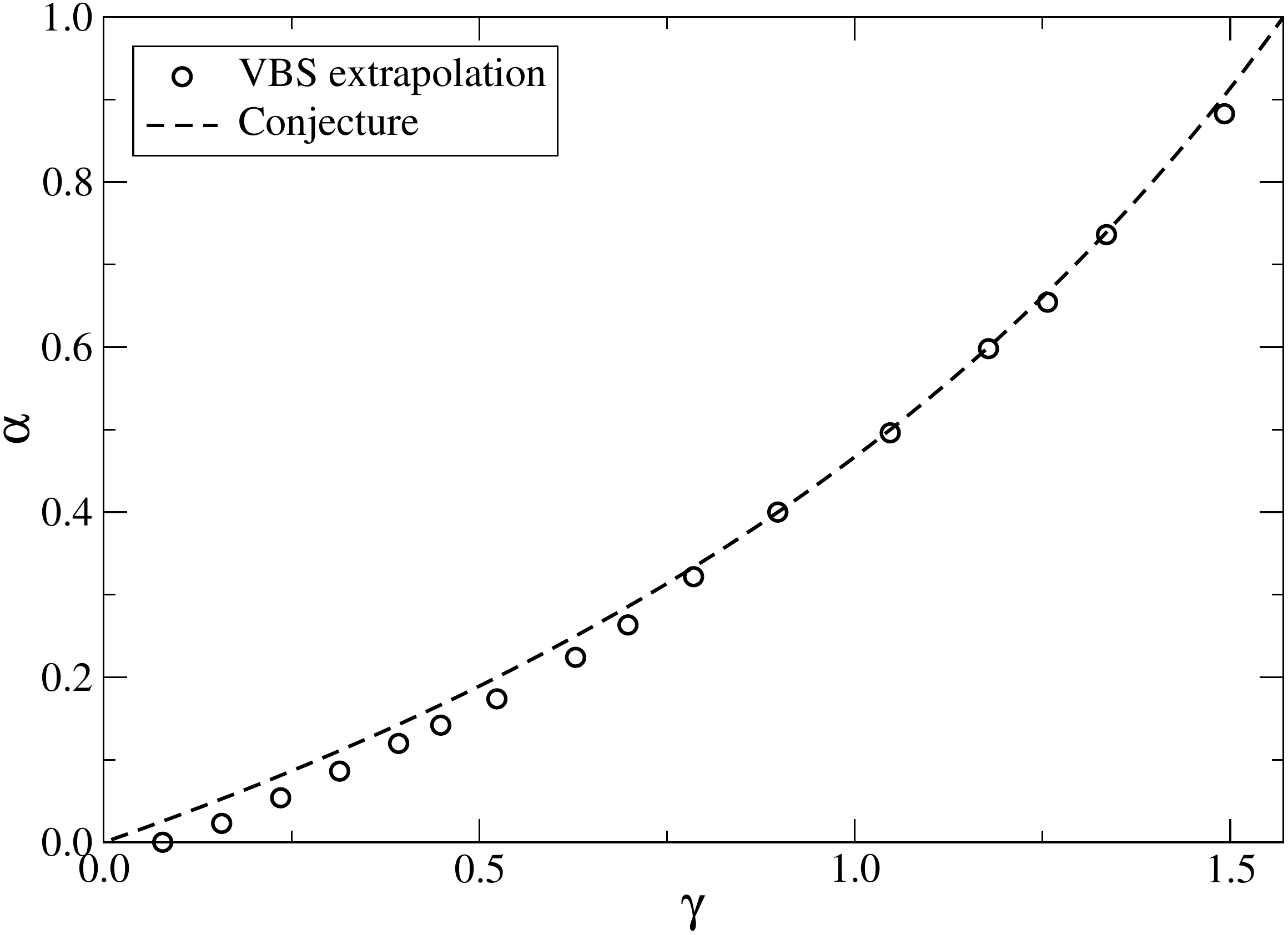}
    \caption{Exponent of the subleading corrections to scaling
      (\ref{eq:01-tower_subl}) for the ground state of the $(0,1)$-sector.}
    \label{fig:01_exponent_powerlawcorr}
\end{figure}

We have analyzed the scaling corrections for the $k=0$ state extrapolating to (\ref{conj01-gs}) in more detail: for $\gamma=\pi/2$ this state belongs to the class of levels discussed in Section~\ref{sec:rel2XXZ}: its energy coincides with the lowest eigenvalue of the $S=1$ XXZ model for magnetization $n=1$. This motivates to assume a power law dependence of the subleading terms on $1/L$. Extrapolating our numerical data using the VBS method \cite{HaBa81} we find
\begin{equation}
  \label{eq:01-tower_subl}
  X_{\textrm{eff}}(L) - X_{0,0}^{1,0} \propto L^{-\alpha}\,,
  \quad \alpha = \frac{\gamma}{\pi-\gamma}\,,
\end{equation}
see Fig.~\ref{fig:01_exponent_powerlawcorr}. 
Note that this behaviour coincides with that of the lowest state of the $S=1$ XXZ model for magnetization $n=1$ \cite{AlMa89} only for the anisotropy where we have established the correspondence to the superspin chain, i.e.\ at $\gamma=\pi/2$.  As $\gamma\to0$ the exponent $\alpha$ vanishes indicating the appearance of logarithmic corrections to scaling due to an operator in the theory becoming marginal.  This is in accordance with our previous study of the finite-size spectrum of the isotropic $OSp(3|2)$ chain \cite{FrMa15}.

For the other members of this tower, i.e.\ $k,k'>0$, we expect that the dominant subleading corrections are logarithmic.  A detailed analysis of the corrections to scaling in the combined presence of logarithms and power laws, however, would require data for significantly larger systems which are not accessible by the methods used here. 

\begin{figure}[ht]
  \includegraphics[width=0.7\textwidth]{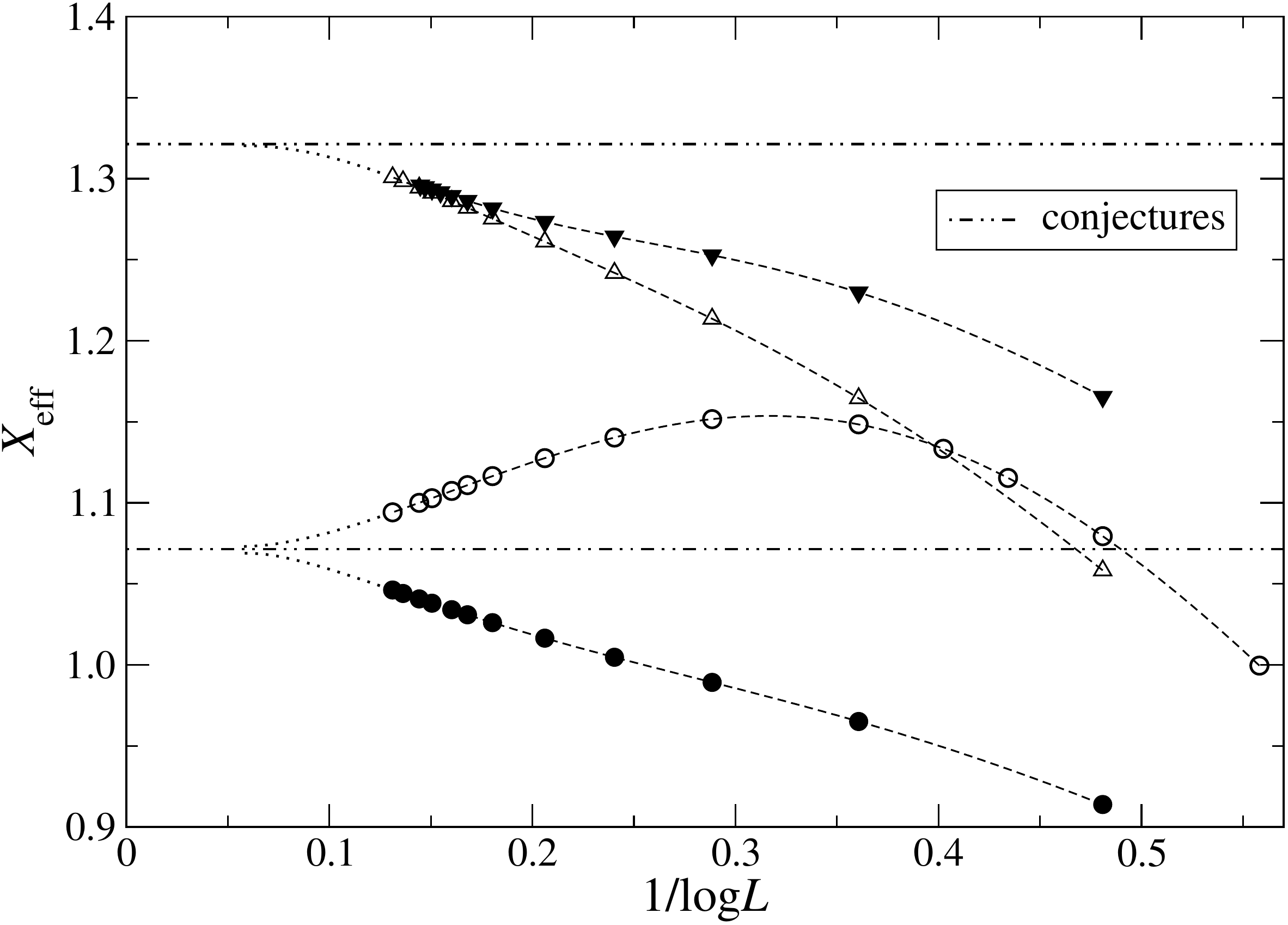}
  \caption{Similar as Fig.~\ref{fig:00_tower-scaling} but for the levels in sector $(0,1)$ extrapolating to (\ref{conj01-s4.6}) (circles) and (\ref{conj01-s1.5.7}) (triangles) in sector $(0,1)$ for $\gamma=2\pi/7$.}
  \label{fig:01_other-scaling}
\end{figure}

Two of the other low lying excitations are described by root configurations $f:[(1_1^-)^2,(1_2^-)]$ and $b:[1_1^-,1_2^+, \bar{2}_1^+,2_2^+]$, respectively.  Both carry conformal spin $s=1$ and they become degenerate in the thermodynamic limit where their effective scaling dimensions extrapolate to 
\begin{equation}
  \label{conj01-s4.6}
  X = \Xi_{0,0}^{1,0} + 1 = \frac{\gamma}{4\pi}  + 1\,,
\end{equation}
see Figure~\ref{fig:01_other-scaling},
indicating that these levels are descendents of (\ref{conj01-gs}).

\begin{figure}[ht]
  \includegraphics[width=0.7\textwidth]{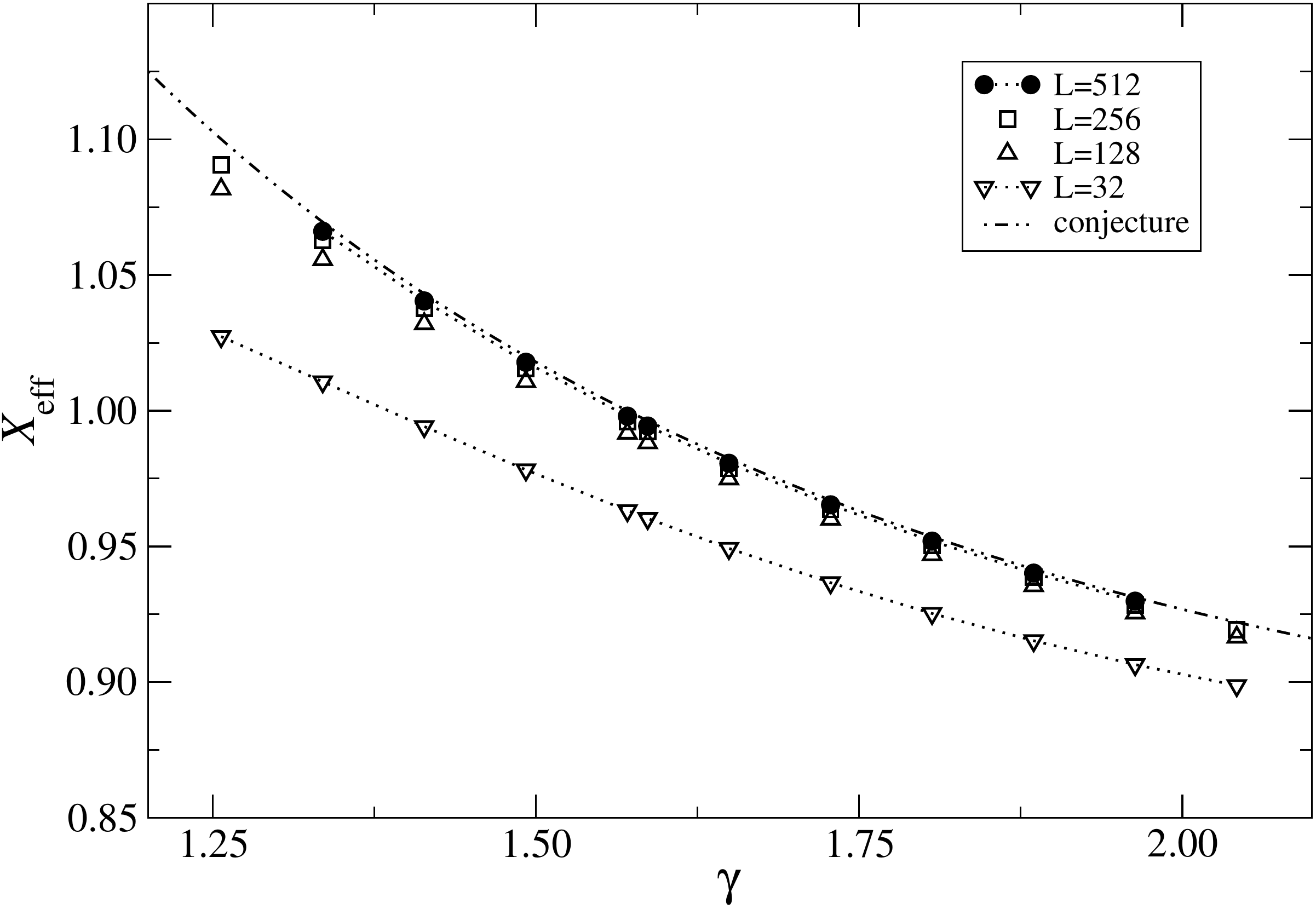}
  \caption{Effective scaling dimensions of the lowest energy singlet in sector $(0,1)$ with conjectured effective scaling dimension (\ref{conj01-s1.5.7}) as a function of $\gamma$ for various system sizes.}
  \label{fig:01-s1-scaling}
\end{figure}

In addition we have identified three low lying levels described by root configurations $b:[1_1^+,1_2^-]$, $f:[(1_1^+)^2,1_2^+,2_1^+,2_2^+]$, and $f:[(1_1^+)^2,(1_1^-)^2,1_2^+,\bar{2}_2^-]$, respectively.  Their effective scaling dimensions extrapolate to
\begin{equation}
  \label{conj01-s1.5.7}
  X_{0,0}^{1,1} = \Xi_{0,0}^{1,1} - \frac{1}{8} + \frac{1}{2} = \frac{\gamma}{4\pi} + \frac{\pi}{4\gamma} +\frac38\,.
\end{equation}
The first and third of these have conformal spin $s=0$, the second comes as a doublet of states with conformal spin $s=1$.
Again this is consistent with primaries being composites of fields with dimensions (\ref{GAUSconjecture_full}) and, according to (\ref{GAUSconjecture_spin}), conformal spin $n_2 m_2/2=1/2$ and an Ising energy operator with conformal weight $1/2$.
These factors can be combined to give a scaling dimension (\ref{conj01-s1.5.7}) and conformal spin  $s=0$ and $1$, respectively.
The finite-size scaling for one of the singlets and the doublet is shown in Figure~\ref{fig:01_other-scaling} for anisotropy $\gamma=2\pi/7$. For the lower energy singlet the Bethe equations have been solved only around $\gamma=\pi/2$, see Figure~\ref{fig:01-s1-scaling}.

In Table~\ref{tab:sec01} the results presented within this section are summed up.
\begin{table}[t]
\begin{ruledtabular}
\begin{tabular}{l|ccc|ccc|l}
     &  \multicolumn{3}{c|}{$X$} & \multicolumn{3}{c|}{$s$} & \\
 Eq. & $m_1$ & $m_2$ & $x_0$ & total spin & $\sigma_{n_1,m_1}^{n_2,m_2}$ & $s_0$ & remark\\\hline
(\ref{conj01-gs}) &$0$ & $0$ & $0$ & $0$ & $0$ & $0$ &  \\
(\ref{conj01-tower}) & $\frac12$  & $0$ & $-\frac14$ & $0$ &$0$ & $0$ & tower\\
(\ref{conj01-s1.5.7}) & $0$ & $1$ & $-\frac{1}{8}+\frac{1}{2}$ & $1,0$ & $\frac{1}{2}$ & $\pm\frac{1}{2}$ & Ising $(\frac{1}{2},0)$, $(0,\frac{1}{2})$\\ 
\end{tabular}
\end{ruledtabular}
\caption{\label{tab:sec01}Conformal data for the primaries identified in charge sector $(n_1,n_2)=(0,1)$, see also Table~\ref{tab:sec00}. 
We have also observed descendents of (\ref{conj01-gs}), see (\ref{conj01-s4.6}).}
\end{table}  
  
\subsection{Sector $(0,2)$}
\begin{figure}[p]
\centering
  \begin{minipage}{0.48\textwidth}
    \subfigure[]{\includegraphics[width=\textwidth]{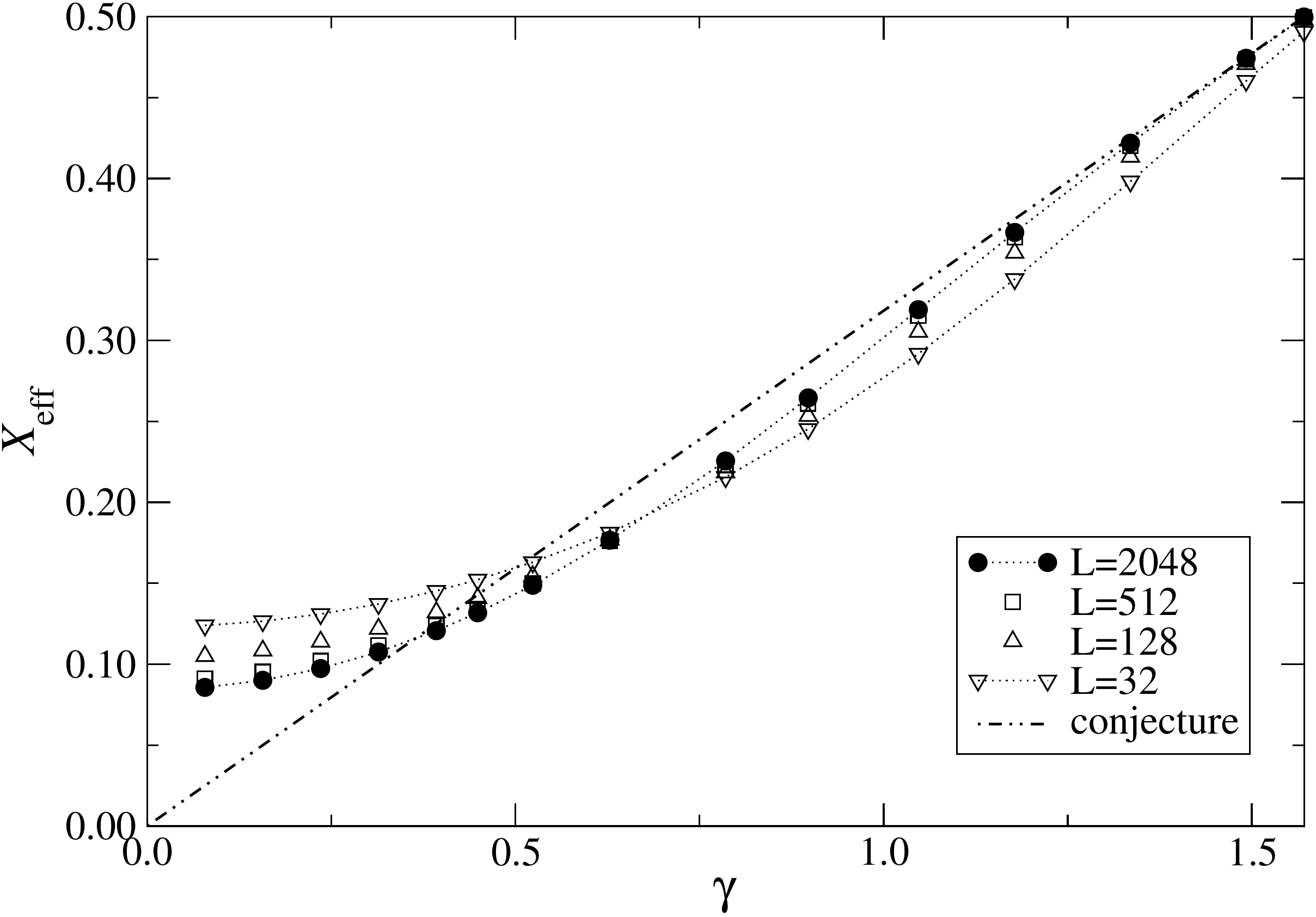}}
  \end{minipage}
  \addtocounter{subfigure}{2}
  \begin{minipage}{0.48\textwidth}
    \subfigure[]{\includegraphics[width=\textwidth]{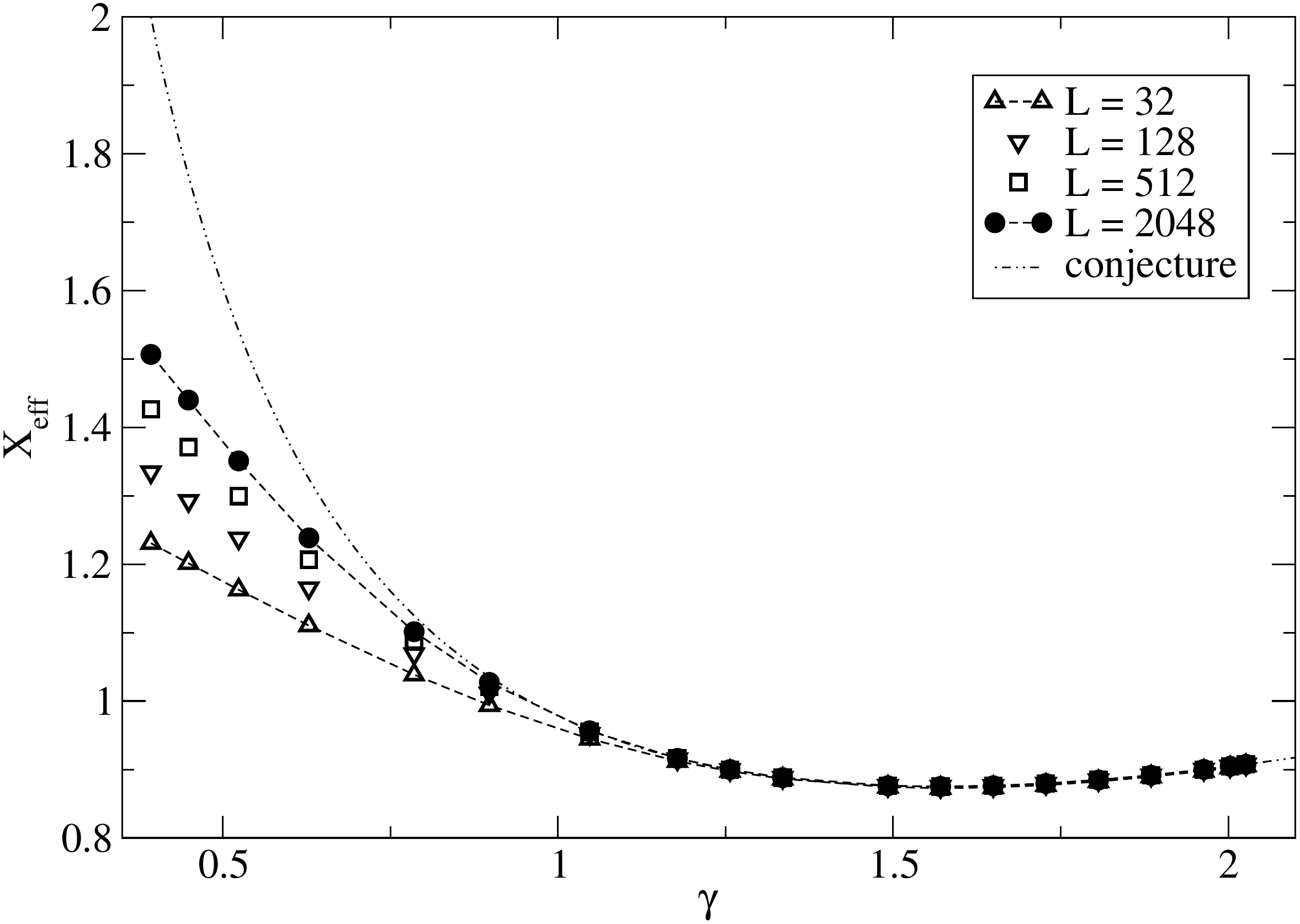}}
  \end{minipage}
  \addtocounter{subfigure}{-3}
  \begin{minipage}{0.48\textwidth}
    \subfigure[]{\includegraphics[width=\textwidth]{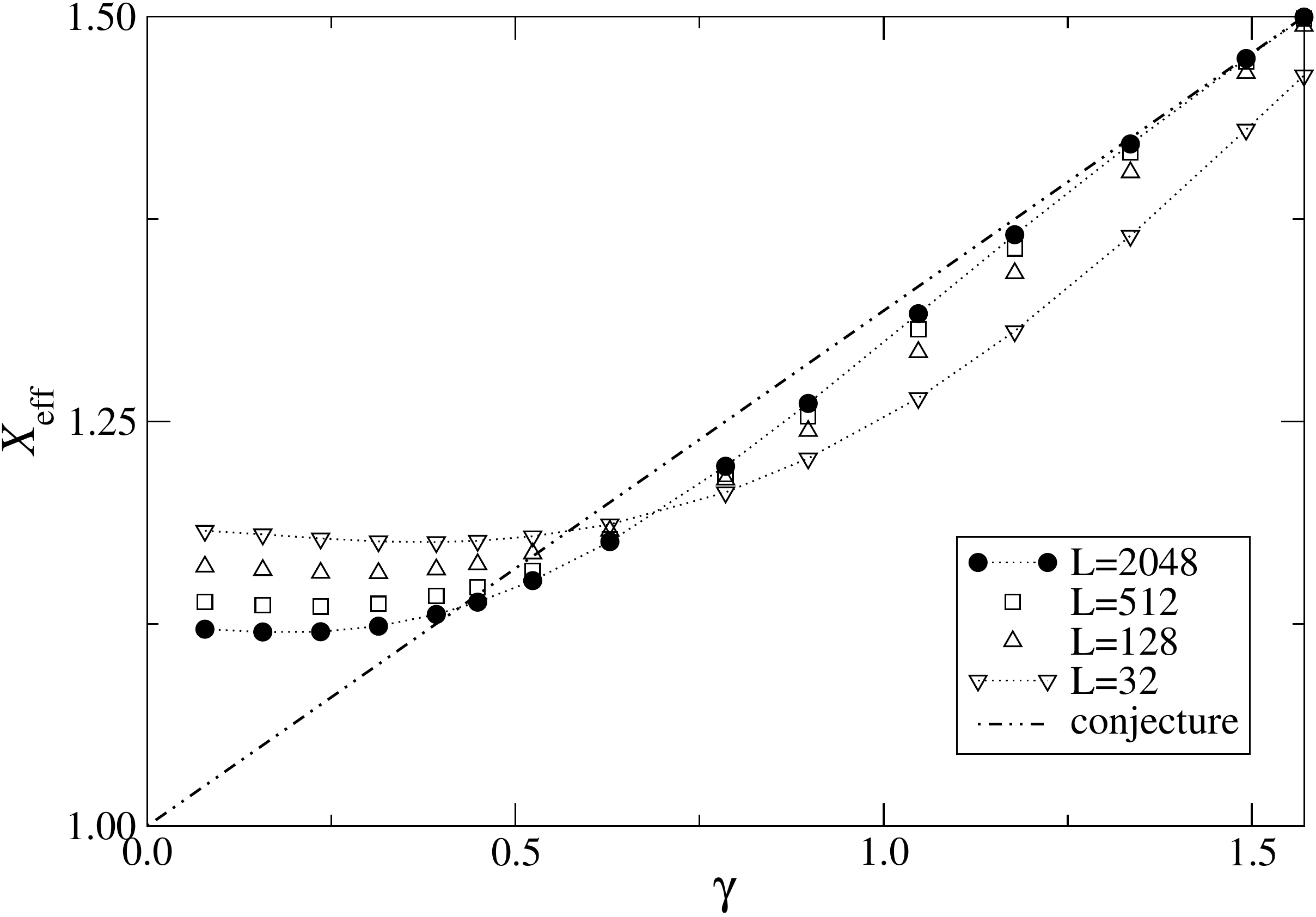}}
  \end{minipage}
  \addtocounter{subfigure}{2}
  \begin{minipage}{0.48\textwidth}
    \subfigure[]{\includegraphics[width=\textwidth]{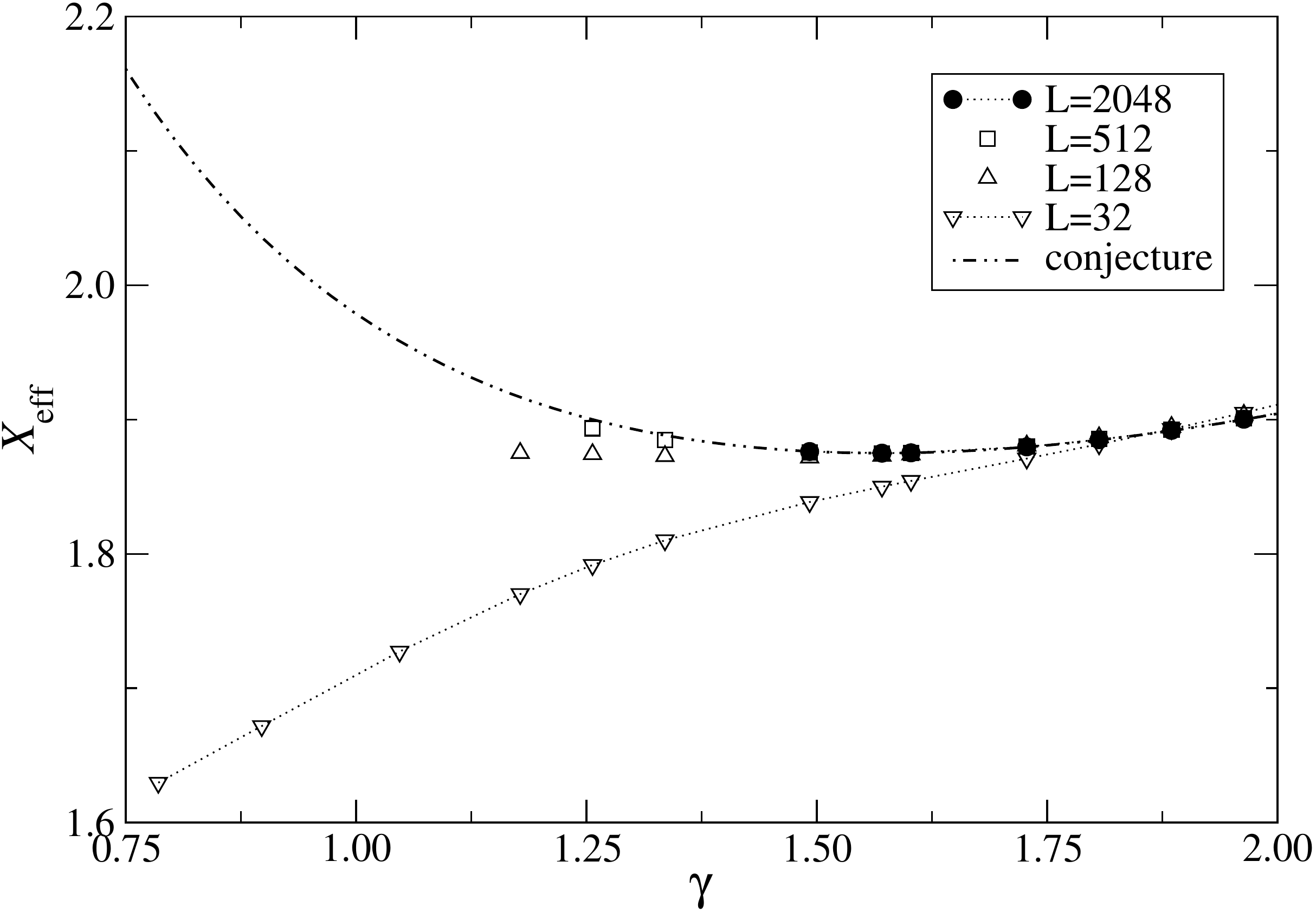}}
  \end{minipage}
  \addtocounter{subfigure}{-3}
  \begin{minipage}{0.48\textwidth}
    \subfigure[]{\includegraphics[width=\textwidth]{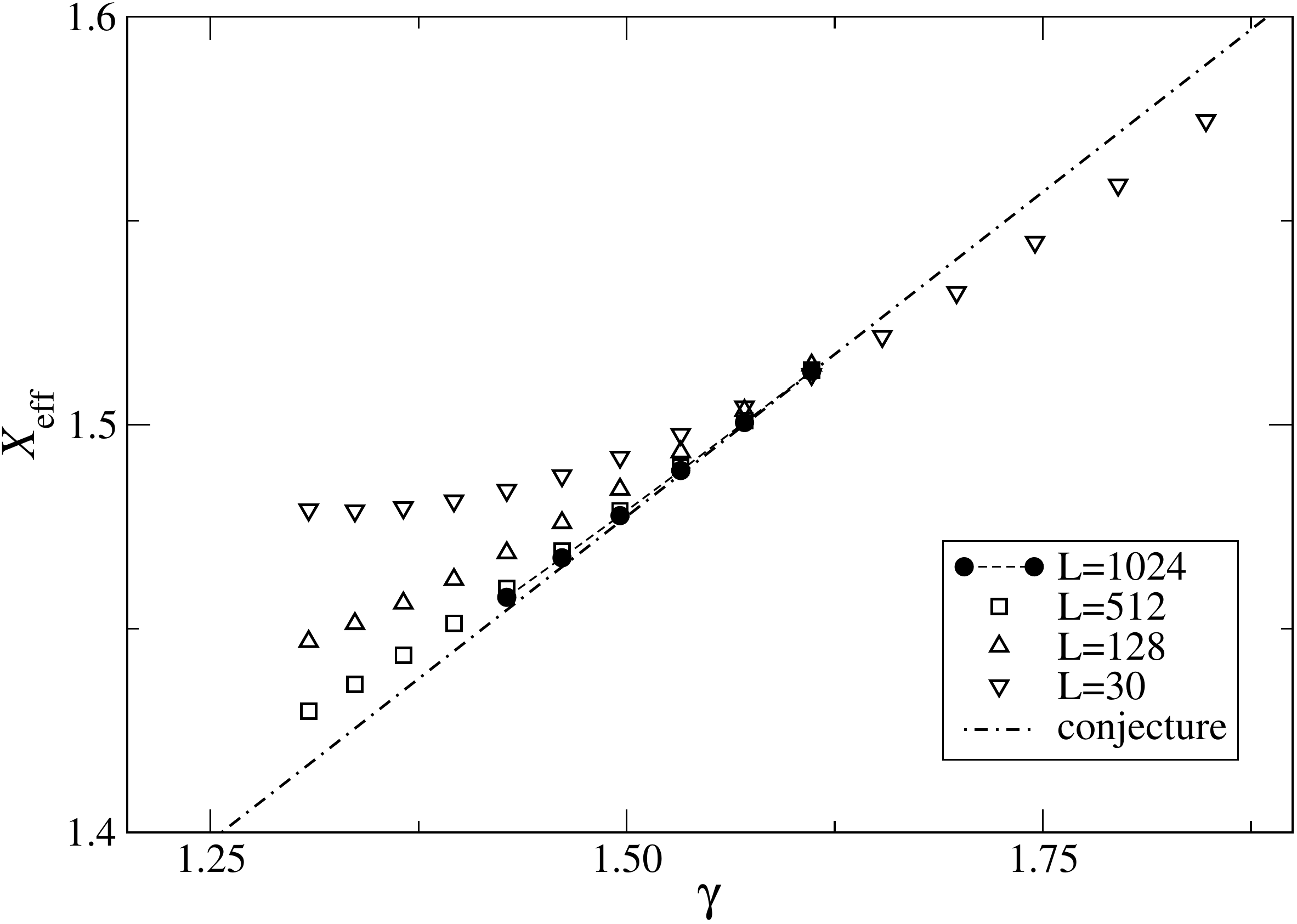}}
  \end{minipage}
  \addtocounter{subfigure}{2}
  \begin{minipage}{0.48\textwidth}
    \subfigure[]{\includegraphics[width=\textwidth]{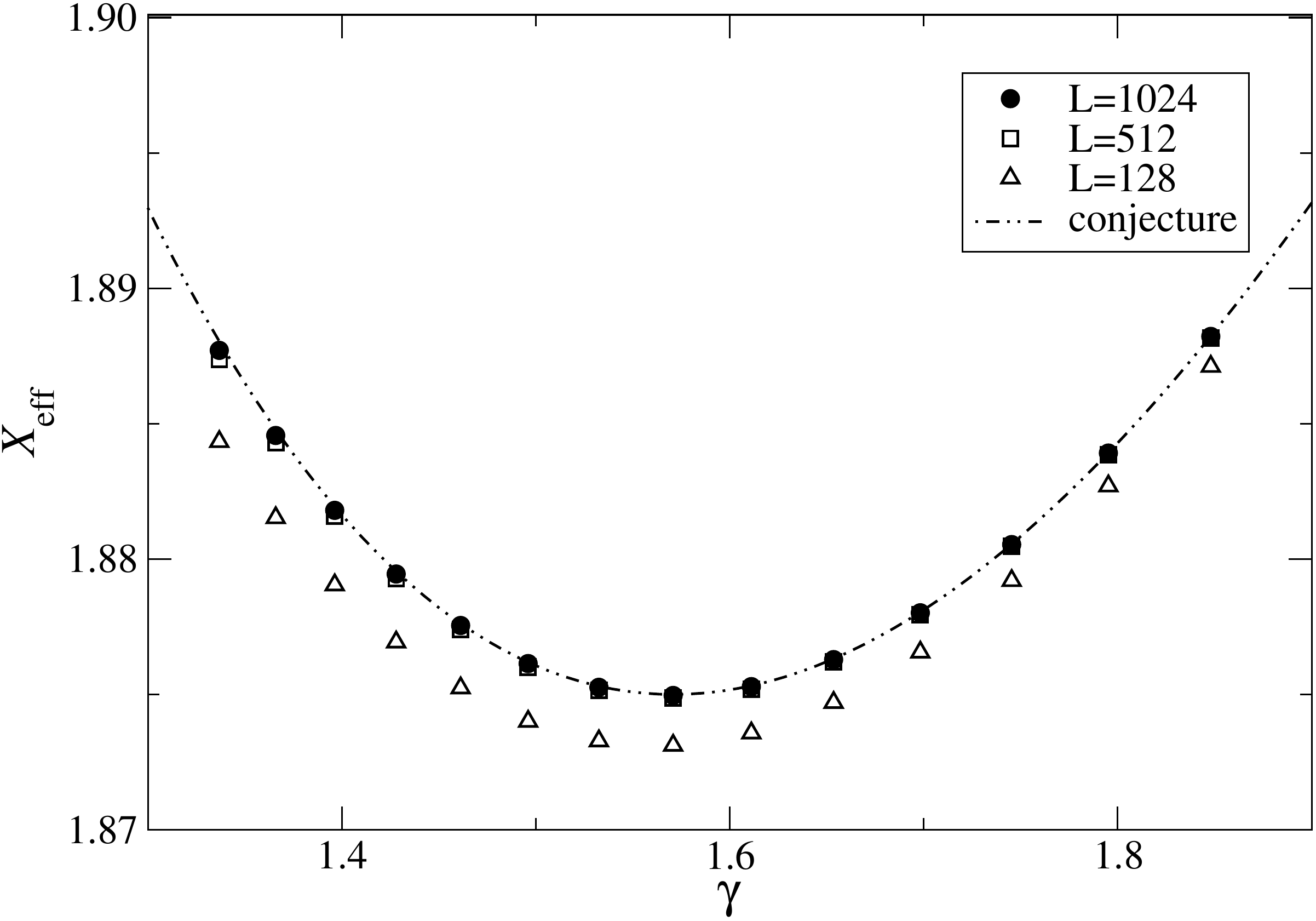}}
  \end{minipage}
  \caption{Effective scaling dimensions of several low energy states in sector $(0,2)$ as a function of $\gamma$ for various system sizes: displayed in the left panel are (a) the spin $s=0$ ground state with effective scaling dimension extrapolating to $X_{0,0}^{2,0}$, Eq.~(\ref{conj02-gs}), and in (b) and (c) two spin $s=1$ levels extrapolating to $X_{0,0}^{2,0}+1$, Eq.~(\ref{conj02-s4.6}. In the right panel the effective scaling dimension of (d) the spin $s=1$ level extrapolating to $X_{0,0}^{2,1}$, Eq.~(\ref{conj02-s3}), and the spin $s=0$ and $s=1$ excitations (e) and (f) extrapolating to $X_{0,0}^{2,1}+1$, Eq.~(\ref{conj02-s5.8}), are shown. Dotted lines connecting symbols are guides to the eye, dashed-dotted lines show the conjectured $\gamma$-dependence.}
  \label{fig:02-xvsg}
\end{figure}

Here the lowest state is most conveniently described in the $bfbfb$ grading where all Bethe roots are arranged in $(L-2)/2$ string complexes (\ref{stringBFBFB}).  The root configurations of the first and second excitation in this sector are obtained by breaking one of the string complexes into either the configurations $b:[(1_1^-)^2,1_2^+,1_2^-]$ or $b:[(1_1^-)^2,\bar{2}_2^-]$, i.e.\
\begin{subequations}\label{tower02}
\begin{align}
\label{tower02a}
  &\lambda^{(1)}_\pm = \pm \xi +i\frac{\pi}{2}\,,\quad
  \lambda^{(2)}_\pm = 0\,,\,i\frac{\pi}{2}\,, \\
\label{tower02b}
  &\lambda^{(1)}_\pm = \pm \xi +i\frac{\pi}{2}\,,\quad
  \lambda^{(2)}_\pm \simeq i\frac{\pi}{2} \pm i\frac {3\gamma}{4}\,,
\end{align}
\end{subequations}
with real $\xi$. These three states have zero momentum. For large $L$ the effective scaling dimension for the lowest level approaches
\begin{equation}
  \label{conj02-gs}
  X_{0,0}^{2,0} = \Xi_{0,0}^{2,0} = \frac{\gamma}{\pi}\,,
\end{equation}
see Figure~\ref{fig:02-xvsg}\,(a).
The subleading corrections are described by a power law in $1/L$.  At least for $\gamma\gtrsim\pi/3$ we find that they are described by same exponent as in (\ref{eq:01-tower_subl}) for the ground state in sector $(0,1)$.

The excitations described in (\ref{tower02}) belong to a family of excitations obtained by breaking more of the $bfbfb$ string complexes giving root configurations $b:[(1_1^-)^{2k},1_2^+,(1_2^-)^{2k-1}]$ or $b:[(1_1^-)^{2k},(1_2^-)^{2k-2},\bar{2}_2^-]$. We have analyzed the finite-size scaling of these levels for $k=1$ and $2$, indicating that the members of this family of excitations form another tower of scaling dimensions starting at
\begin{equation}
  \label{conj02-tower}
  X_{0,\frac12}^{2,0} = \Xi_{0,\frac12}^{2,0} - \frac{1}{4} = \frac{\pi}{4(\pi-\gamma)} + \frac{\gamma}{\pi} - \frac14\,.
\end{equation}
\begin{figure}[ht]
  \includegraphics[width=0.7\textwidth]{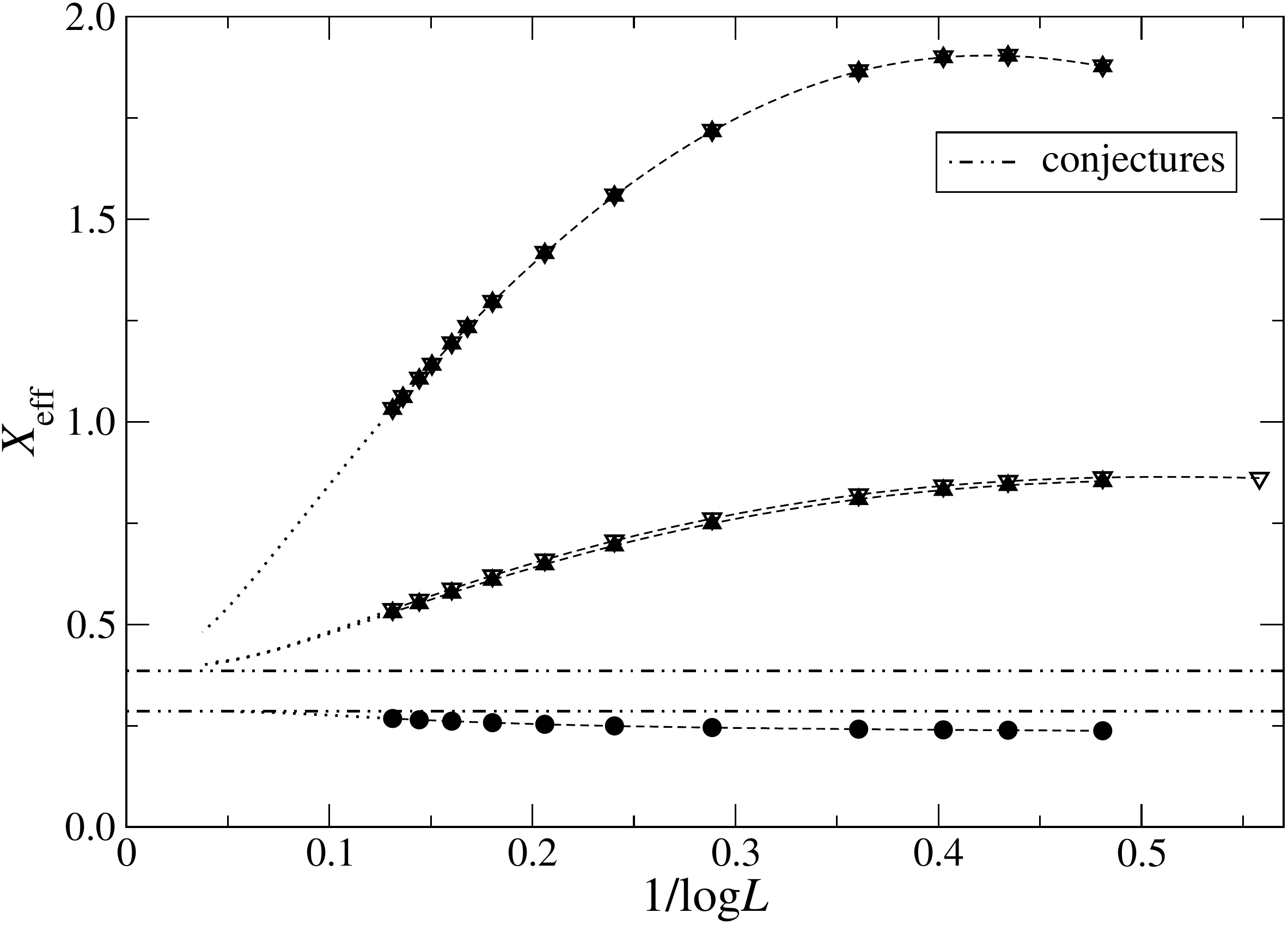}
  \caption{Similar as Fig.~\ref{fig:00_tower-scaling} but for the lowest eigenenergy and the first four  excitations forming a tower of scaling dimensions starting at (\ref{conj02-tower}) in sector $(0,2)$ for $\gamma=2\pi/7$. The dashed-dotted lines are our conjectures (\ref{conj02-gs}) and (\ref{conj02-tower}) for this anisotropy.}
  \label{fig:02_tower-scaling}
\end{figure}
For finite $L$ the degeneracy of these levels is lifted and for $\gamma>0$ the excitations are separated from the lowest state in this sector by a gap of order $1/L$ with strong subleading corrections, see Figure~\ref{fig:02_tower-scaling}.

There are two more levels in this sector whose root configurations are described in terms of $(L-2)/2$ $bfbfb$ string complexes. One of them leads to the scaling dimension 
\begin{equation}
  \label{conj02-s3}
  X_{0,0}^{2,1} = \Xi_{0,0}^{2,1} - \frac{1}{8} = \frac{\gamma}{\pi} + \frac{\pi}{4\gamma} - \frac18\,,
\end{equation}
and conformal spin $s=1$ in agreement with (\ref{GAUSconjecture_spin}).  The finite-size data for this state are shown in Figure~\ref{fig:02-xvsg}\,(d).

The other one has an effective scaling dimension extrapolating to
\begin{equation}
  \label{conj02-s4.6}
   X = \Xi_{0,0}^{2,0} + 1 = \frac{\gamma}{\pi} + 1\,,
\end{equation}
see Figure~\ref{fig:02-xvsg}\,(b). This level has spin $s=1$, indicating that this is a descendent of (\ref{conj02-gs}). Breaking one of the $bfbfb$ strings complexes we find another level, also with conformal spin $s=1$, described by a root configuration $b: [1_1^+, 1_1^-, 1_2^+, 1_2^-]$. The numerical solution of the Bethe equations for this state for sufficiently large systems is limited to anisotropies near $\gamma=\pi/2$ where the extrapolation of the finite-size gives again (\ref{conj02-s4.6}), as shown in Fig.~\ref{fig:02-xvsg}\,(c).

Next we report on two other low energy levels in this sector, a spin $s=0$ state with root configuration $b:[3_{12}^+,1_2^-]$ and a spin $s=1$ state with root configuration $b:[1_1^+,1_1^-,1_2^+,1_2^-]$. Both of them are found to extrapolate to the scaling dimension
\begin{equation}
  \label{conj02-s5.8}
  X = \Xi_{0,0}^{2,1}- \frac{1}{8} + 1  = \frac{\gamma}{\pi} + \frac{\pi}{4\gamma} - \frac18 +1 \,,
\end{equation}
see Figures~\ref{fig:02-xvsg}\,(e) and (f).
From these data we conclude that the zero spin level is a descendent of (\ref{conj02-s3}).  The $s=1$ level appears to be a primary, again the scaling dimension indicates the presence of an Ising field in the effective low energy description of the superspin chain.

\begin{figure}[ht]
  \includegraphics[width=0.7\textwidth]{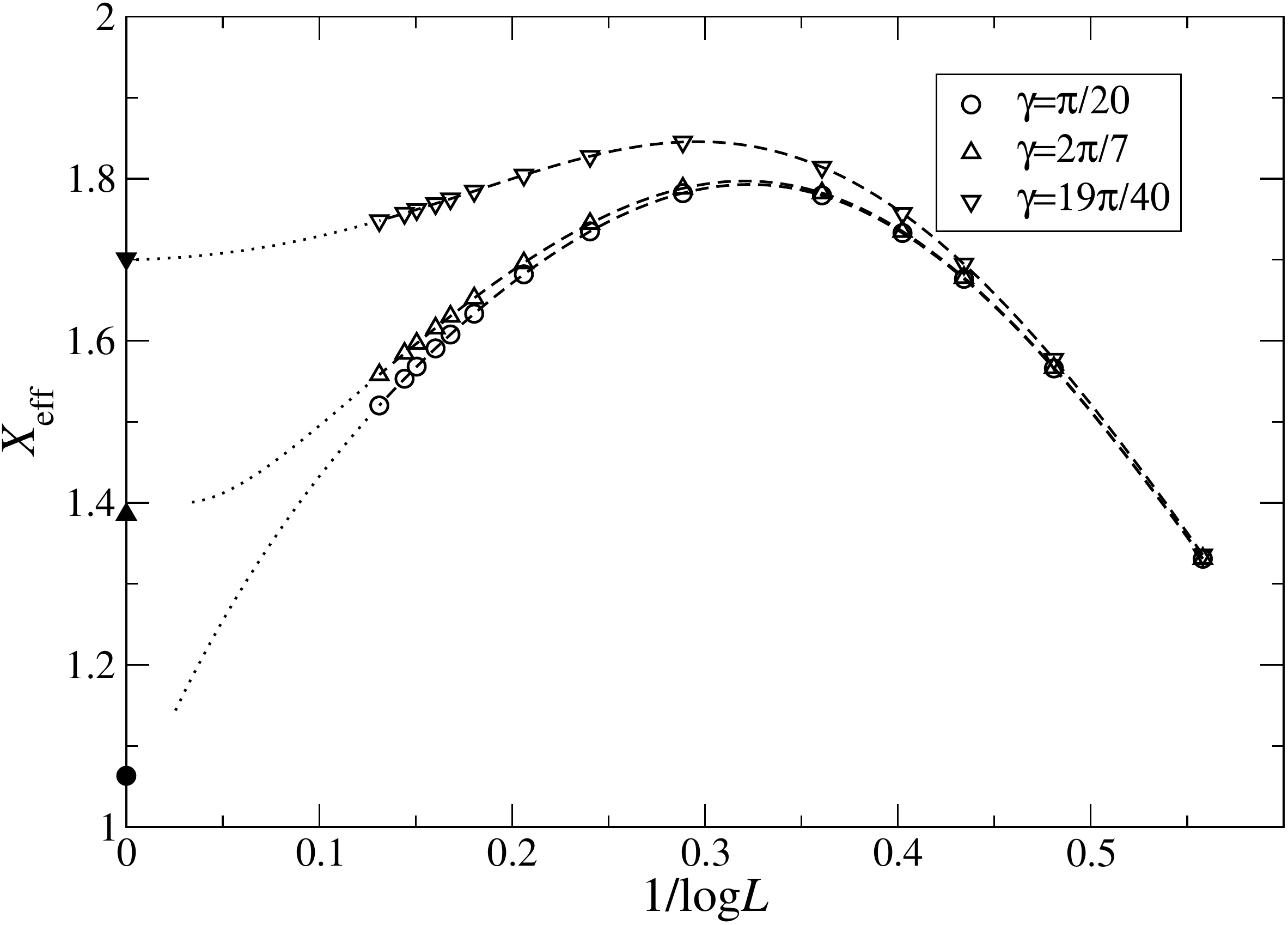}
  \caption{Effective scaling dimensions of the level extrapolating to (\ref{conj02-s7}) in sector $(0,2)$ as a function of $x=1/\log(L)$ for various values of the anisotropy $\gamma$.  Open (filled) symbols are the numerical data (the conjectured value in the thermodynamic limit $L\to\infty$). The dotted lines are the extrapolations assuming a rational dependence on $x$.}
  \label{fig:02_s7-scaling}
\end{figure}

Finally, we have identified a level with spin $s=1$ and effective scaling dimension extrapolating to
\begin{equation}
  \label{conj02-s7}
  X = \Xi_{0,\frac12}^{2,0} - \frac{1}{4} + 1 = \frac{\pi}{4(\pi-\gamma)} + \frac{\gamma}{\pi} - \frac14  +1\,.
\end{equation}
In $bfbfb$ grading its root configuration is $b:[(1_1^-)^2,1_2^+,1_2^-]$.
Based on this configuration we propose that this is a descendent of the state with roots (\ref{tower02a}) described above. The scaling of this level is displayed in Figure~\ref{fig:02_s7-scaling}.

To finish the investigation of the sector $(n_1,n_2) = (0,2)$ we present our findings for this sector in table~\ref{tab:sec02}.
\begin{table}[t]
\begin{ruledtabular}
\begin{tabular}{l|ccc|ccc|l}
     &  \multicolumn{3}{c|}{$X$} & \multicolumn{3}{c|}{$s$} & \\
 Eq. & $m_1$ &  $m_2$ & $x_0$ & total spin & $\sigma_{n_1,m_1}^{n_2,m_2}$ & $s_0$ & remark\\\hline
(\ref{conj02-gs}) &  $0$ & $0$ & $0$ & $0$ & $0$ & $0$ & \\
(\ref{conj02-tower}) &  $\frac12$ & $0$ & $-\frac14$ & $0$ & $0$ & $0$ & tower\\
(\ref{conj02-s3}) &  $0$ & $1$ & $-\frac{1}{8}$ & $1$ & $1$ & $0$ &\\
(\ref{conj02-s5.8}) & $0$ & $1$ & $-\frac{1}{8}+1$ & $1$ & $1$ & $0$ & Ising $(\frac{1}{2},\frac{1}{2})$ \\
\end{tabular}
\end{ruledtabular}
\caption{\label{tab:sec02}Conformal data for the levels studied in charge sector $(n_1,n_2)=(0,2)$ (see also Table~\ref{tab:sec00}).  We have also observed descendents of (\ref{conj02-gs}), see (\ref{conj02-s4.6}), one descendent of (\ref{conj02-tower}), see (\ref{conj02-s7}) and one descendent of (\ref{conj02-s3}), see (\ref{conj02-s5.8}).}
\end{table} 

\subsection{Sector $(1,0)$}
\begin{figure}[ht]
  \includegraphics[width=0.7\textwidth]{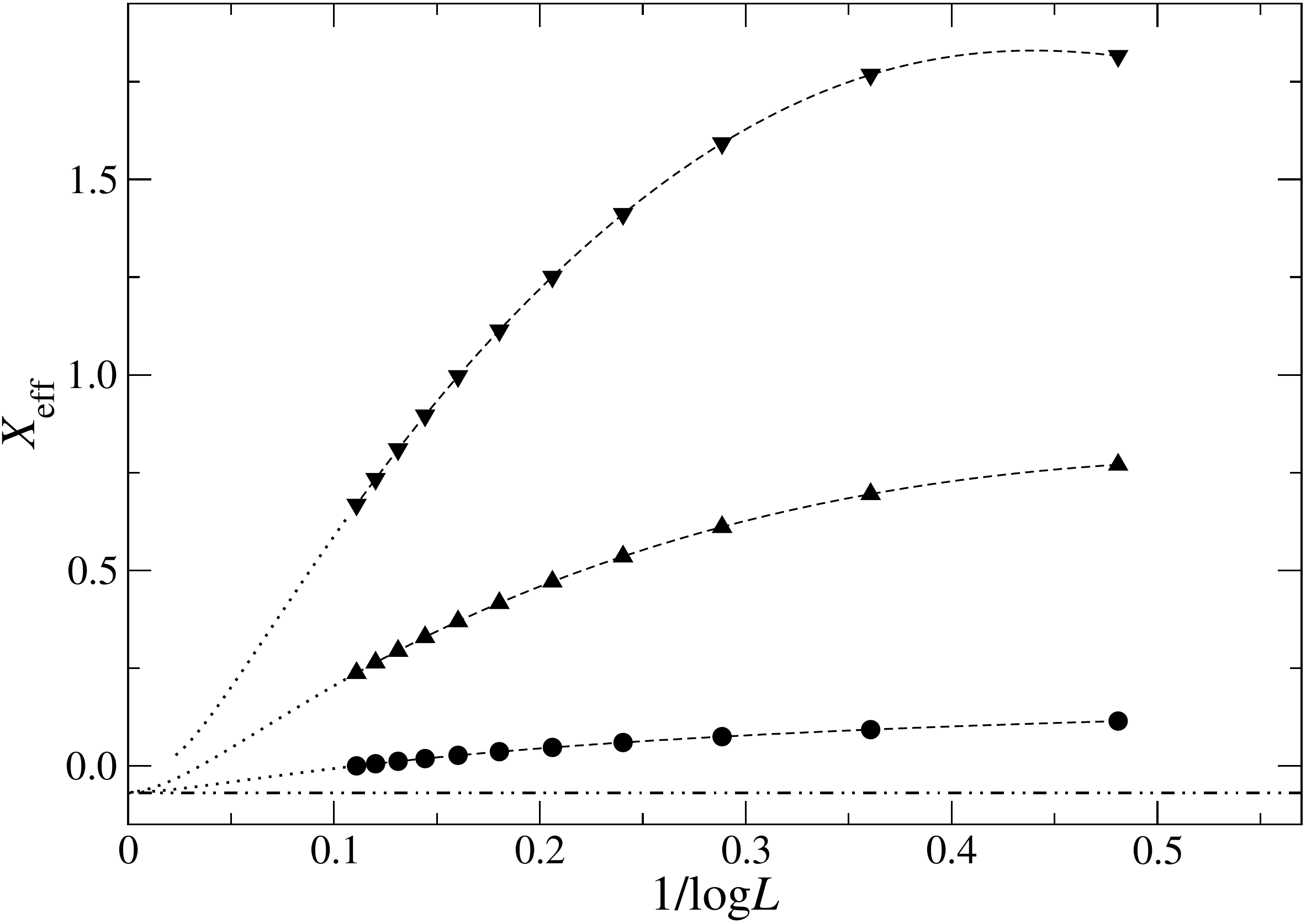}
  \caption{Similar as Fig.~\ref{fig:00_tower-scaling} but for the lowest eigenenergy and the related tower of levels in the spectrum of the superspin chain in sector $(1,0)$ for $\gamma=11\pi/40$. The dashed-dotted line is our conjecture (\ref{conj10-tower}) for this anisotropy.}
  \label{fig:10_tower-scaling}
\end{figure}

The lowest energy state in this sector is the overall ground state of the $OSp(3|2)$ superspin chain for $\gamma>0$.  It is described by 
a symmetric root configuration $f:[1_1^+,1_2^-]$ and has conformal spin $s=0$.
We have solved the Bethe equations (\ref{betheFBBBF}) for this state in systems with up to $L=8192$ sites.  The numerical finite-size data extrapolate to an effective scaling dimension
\begin{equation}
  \label{conj10-tower}
  X_{1,0}^{0,0} = \Xi_{1,0}^{0,0} - \frac{1}{4} = -\frac{\gamma}{4\pi}\,.
\end{equation}
Again, this root configuration can be used as a starting point to find related states forming a tower of scaling dimensions on top of (\ref{conj10-tower}).  Breaking one of the $fbbbf$ string complexes into
\begin{equation}
\label{tower10}
  \lambda^{(1)}_\pm = \pm \xi\,,\quad
  \lambda^{(2)}_- = \pm \eta+i\frac{\pi}{2}\,,
  \quad \xi,\eta\in\mathbb{R}\,,
\end{equation}
we obtain the first excitation in this sector.  Repeating this procedure leads to excitations $f:[(1_1^+)^{2k+1},(1_2^-)^{2k+1}]$.  For $k=1,2$ these excitations have conformal spin $s=0$ and their effective scaling dimensions extrapolate to (\ref{conj10-tower}) as $L\to\infty$.  Strong subleading corrections lift the degeneracy for finite $L$, see Figure~\ref{fig:10_tower-scaling}.

\begin{figure}[ht]
  \includegraphics[width=0.7\textwidth]{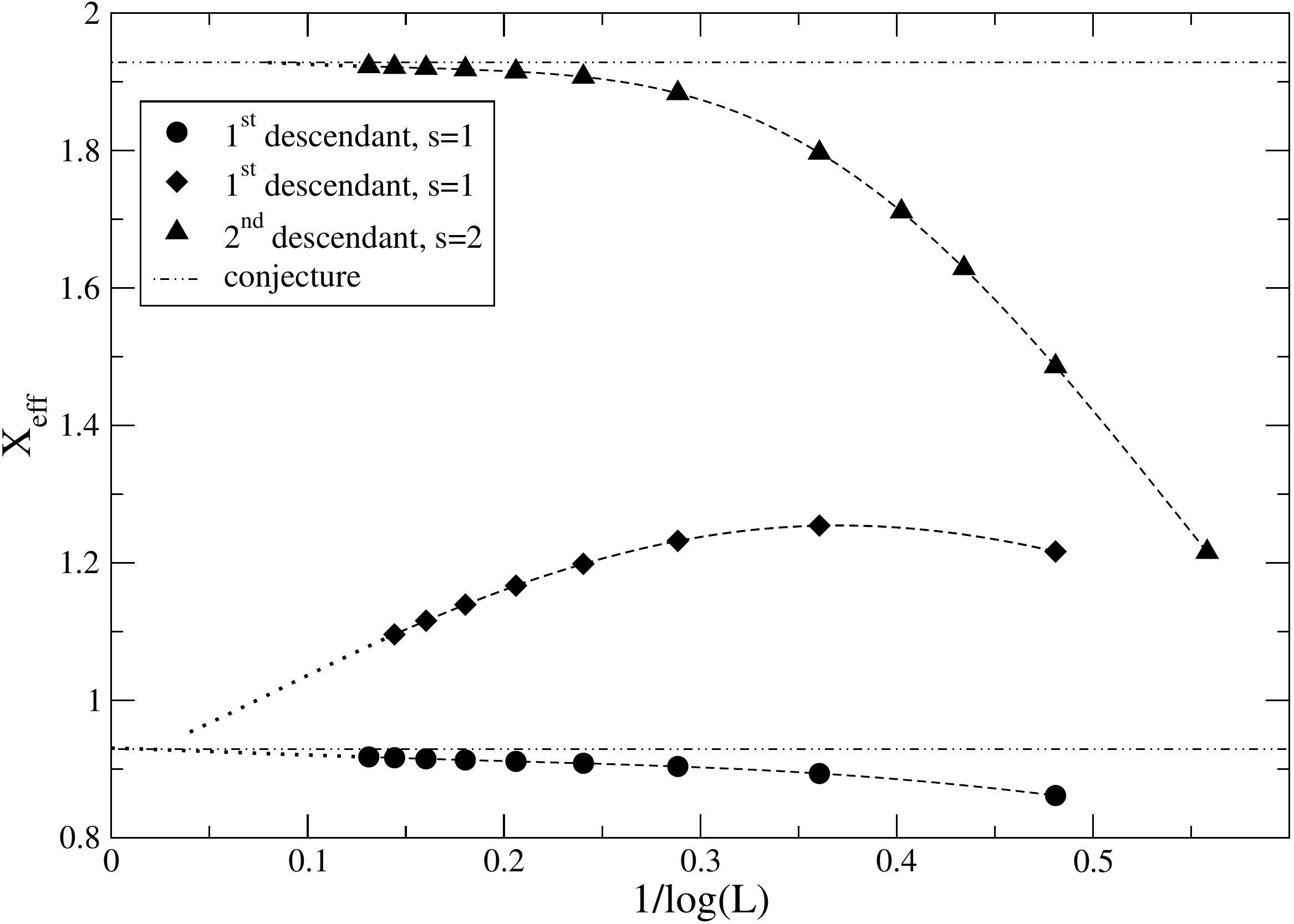}
  \caption{Similar as Fig.~\ref{fig:00_tower-scaling} but for the three lowest states with nonzero conformal spin in the spectrum of the superspin chain in sector $(1,0)$ for $\gamma=2\pi/7$. These states are descendants of the lowest two tower states. The dashed-dotted lines are our conjectures (\ref{conj10-s2.4}) and (\ref{conj10-s5}), respectively, for this anisotropy.}
  \label{fig:10_2nd_4th_5th-scaling}
\end{figure}

The extrapolation of our finite-size data for two spin $s=1$ levels in this sector described by $f:[1_1^-,1_2^+]$ and $f:[(1_1^+)^2,1_1^-,1_2^-,\bar{2}_2^+]$ root configurations, respectively, gives effective scaling dimensions
\begin{equation}
  \label{conj10-s2.4}
  X = \Xi_{1,0}^{0,0} - \frac{1}{4} + 1  = -\frac{\gamma}{4\pi} + 1\,.
\end{equation}
These levels are descendents of the lowest two states in the tower starting at (\ref{conj10-tower}). A potential second descendent of the lowest tower state is parameterized by Bethe roots arranged as $f:[1_1^-,1_2^+]$.  It has conformal spin $s=2$ and its scaling dimension extrapolates to
\begin{equation}
  \label{conj10-s5}
  X = \Xi_{1,0}^{0,0}- \frac{1}{4} + 2  = -\frac{\gamma}{4\pi} + 2\,,
\end{equation}
The effective scaling dimensions for these three states display strong subleading corrections, see Fig.~\ref{fig:10_2nd_4th_5th-scaling}.

\begin{figure}[ht]
  \includegraphics[width=0.7\textwidth]{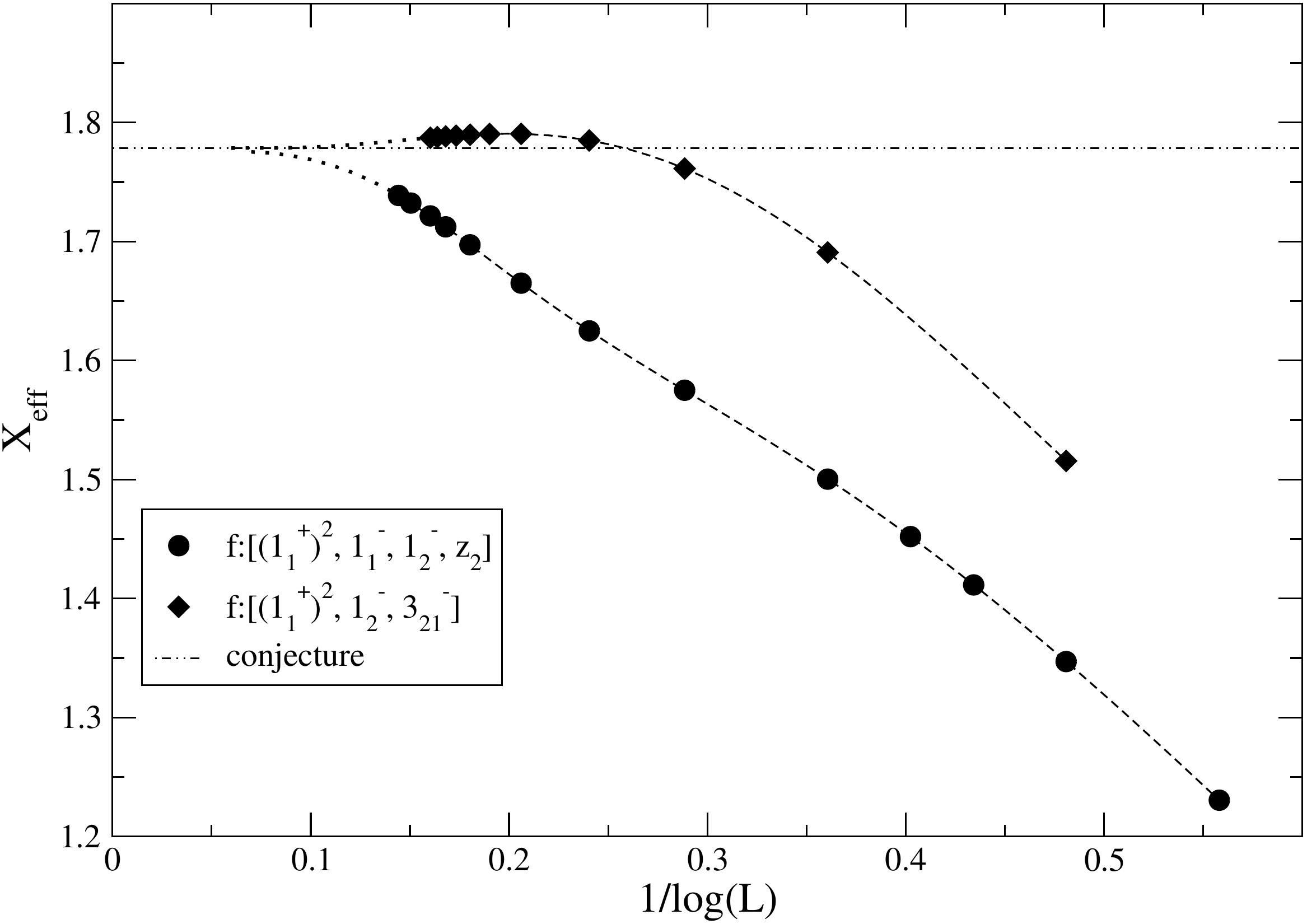}
  \caption{Similar as Fig.~\ref{fig:00_tower-scaling} but for the states in the spectrum of the superspin chain in sector $(1,0)$ extrapolating to \eqref{conj10-s6.9} for $\gamma=2\pi/7$. The dashed-dotted line is our conjecture (\ref{conj10-s6.9}), for this anisotropy.}
  \label{fig:10_6th_9th-scaling}
\end{figure}

Two low energy levels with root configurations $f:[(1_1^+)^2,1_1^-,1_2^-,z_2]$ and 
$f:[(1_1^+)^2,1_2^-,3_{21}^-]$  have spin $s=1$ and $s=0$, respectively.  Their effective scaling dimensions are 
\begin{equation}
  \label{conj10-s6.9}
  X_{1,\frac12}^{0,1} = \Xi_{1,\frac12}^{0,1} - \frac{1}{8} + \frac{1}{2} = -\frac{\gamma}{4\pi} + \frac{\pi}{4(\pi-\gamma)} + \frac{\pi}{4\gamma} + \frac58\,.
\end{equation}
Our finite-size data for these two states can be found in fig~\ref{fig:10_6th_9th-scaling}.

\begin{figure}[ht]
  \includegraphics[width=0.7\textwidth]{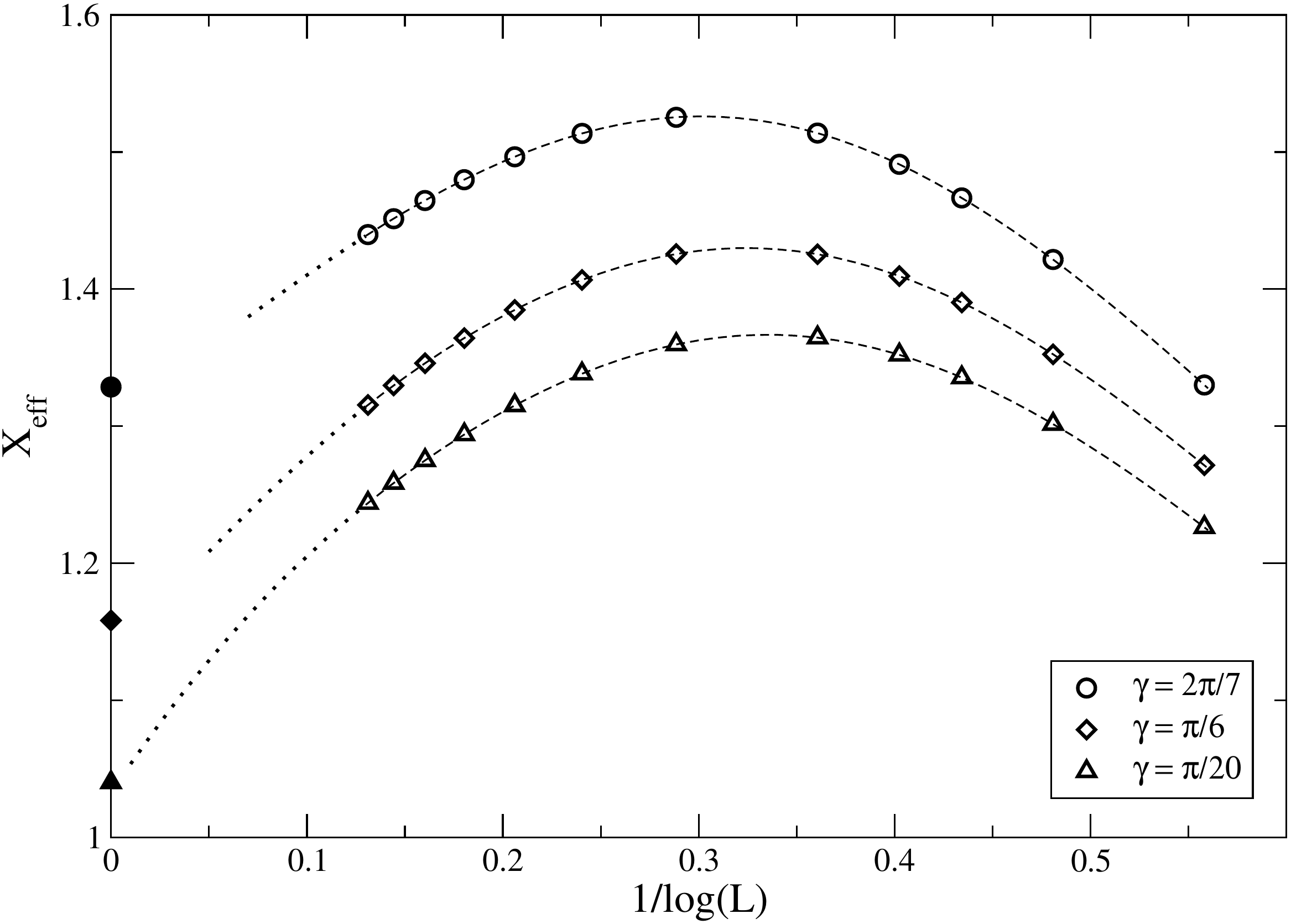}
  \caption{Similar as Fig.~\ref{fig:02_s7-scaling} but for the state in sector $(1,0)$ extrapolating to \eqref{conj10-s8}.}
  \label{fig:10_8th-scaling}
\end{figure}

Another scaling dimension in this sector has been identified from the finite-size scaling of a spin $s=1$ level with root configuration $f:[(1_1^+)^3,1_2^-,2_2^+]$:
\begin{equation}
  \label{conj10-s8}
  X_{1,1}^{0,0} = \Xi_{1,1}^{0,0} - \frac{1}{4} = -\frac{\gamma}{4\pi} + \frac{\pi}{(\pi-\gamma)}\,.
\end{equation}
Our numerical results for this state is shown in fig~\ref{fig:10_8th-scaling}
for several anisotropies.

\begin{figure}[ht]
  \includegraphics[width=0.7\textwidth]{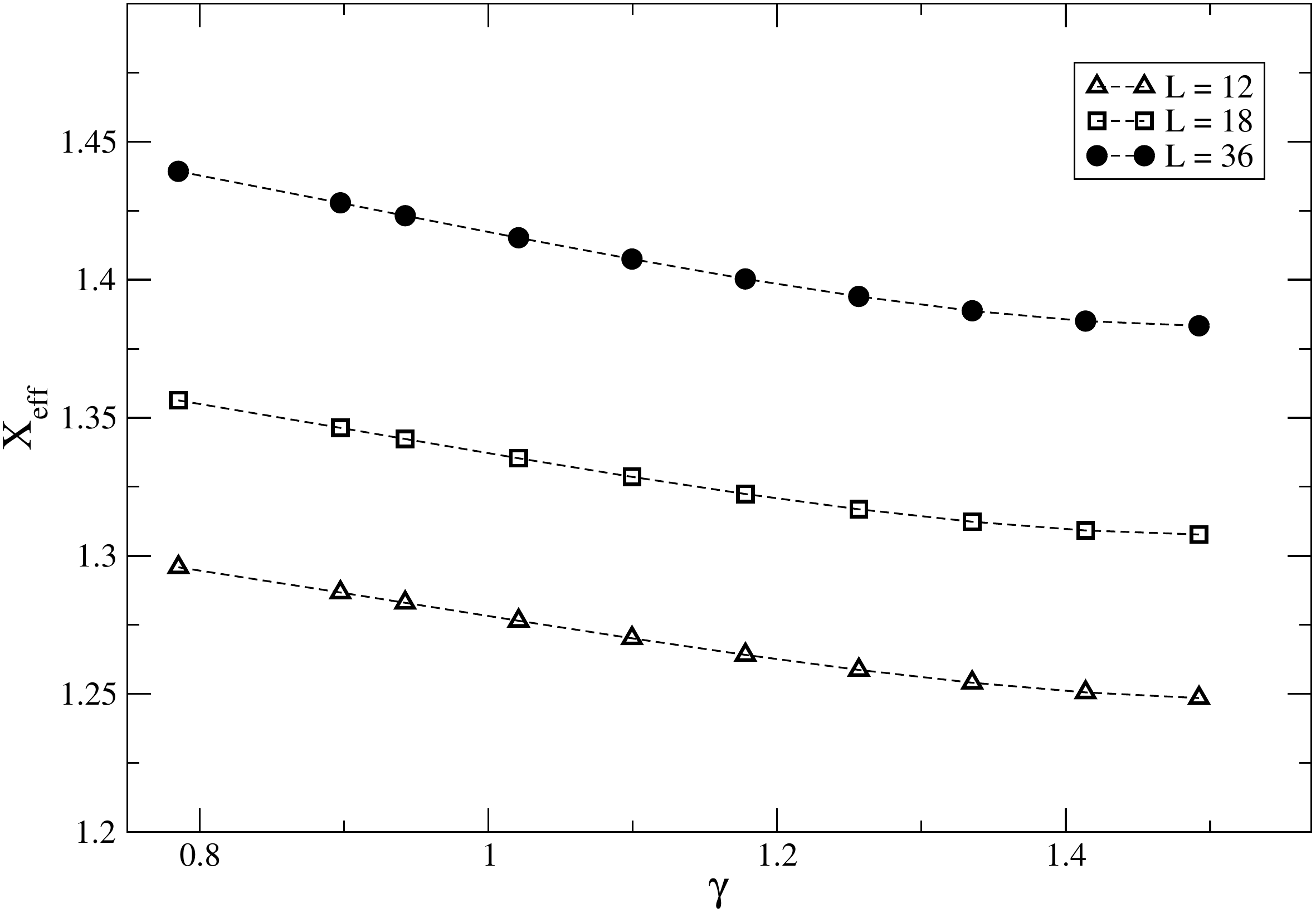}
  \caption{Effective scaling dimension of the state $b:[1_1^-, 1_2^+,3_{21}^+, z_1]$ or equivalently  $f:[3_{12}^+,3_{21}^+]$ in sector $(1,0)$ for small system sizes as a function of $\gamma$.}
  \label{fig:10_3th-xvsg}
\end{figure}

In addition to the levels discussed above there is another state where we have observed a change of the root pattern as the system size changes.  For small $L$ it is described by root configurations $b:[1_1^-,1_2^+,3_{21}^+,z_1]$ and $f:[3_{12}^+,3_{21}^+]$, depending on the grading. As the system size is increased the $bfbfb$ root structure changes when the pair of complex conjugate first level roots degenerates and is replaced by two one-strings with negative parity, i.e.\ $[z_1]\to[(1_1^-)^2]$. For $\gamma=2\pi/7$ this already happens as $L$ grows from $10$ to $12$.  Increasing the system size further, beyond $L=36$ for $\gamma=2\pi/7$, another degeneration is observed, this time affecting the roots on the second level. Again, we present more details in Appendix~\ref{app:roots}.  It has not been possible to follow the evolution of the root configuration beyond the second degeneration. Due to the presence of strong corrections to scaling in this state, see Fig.~\ref{fig:10_3th-xvsg}, these data do not allow for a finite-size analysis.

We finally remark that, for one spin $s=1$ state in the low energy spectrum 
of the superspin chain we 
have been able to identify the root configuration only for $L=8$ 
and $\gamma \leq 2\pi/7$, see Appendix~\ref{app:missing_states}.

As before we present our results for the sector $(n_1,n_2) = (1,0)$ in table~\ref{tab:sec10}.
\begin{table}[t]
\begin{ruledtabular}
\begin{tabular}{l|ccc|ccc|l}
     &  \multicolumn{3}{c|}{$X$} & \multicolumn{3}{c|}{$s$} & \\
 Eq. & $m_1$ & $m_2$ & $x_0$ & total spin & $\sigma_{n_1,m_1}^{n_2,m_2}$ & $s_0$ & remark\\\hline
(\ref{conj10-tower}) & $0$ & $0$ & $-\frac{1}{4}$ & $0$ & $0$ & $0$ & tower \\
(\ref{conj10-s6.9}) & $\frac{1}{2}$ & $1$ & $-\frac{1}{8}+\frac{1}{2}$ & $1,0$ & $ \frac{1}{2}$ & $\pm \frac{1}{2}$ & Ising $(\frac{1}{2}, 0)$, $(0,\frac{1}{2})$\\
(\ref{conj10-s8}) & $1$ & $0$ & $-\frac{1}{4}$ & $ 1$ & $1$ & $0$ & \\
\end{tabular}
\end{ruledtabular}
\caption{\label{tab:sec10}Conformal data for the levels studied in charge sector $(n_1,n_2)=(1,0)$ (see also Table~\ref{tab:sec00}).  We have also observed descendents of (\ref{conj10-tower}), see (\ref{conj10-s2.4}) and (\ref{conj10-s5}).}
\end{table} 

\subsection{Sector $(1,1)$}
The lowest state in this sector in $fbbbf$ grading is described by a root configuration $f:[1_1^+]$, i.e.\ $(L-2)/2$ string complexes (\ref{stringFBBBF}) and one additional root, $\lambda^{(1)}=0$, on the first level.  From our numerical finite-size data for chains with up to $L=2048$ sites we find that the effective scaling dimension of this level is
\begin{equation}
  \label{conj11-tower}
  X_{1,0}^{1,0} = \Xi_{1,0}^{1,0} - \frac{1}{4} = 0\,.
\end{equation}
As in the charge sector $(1,0)$ this state is the lowest in a tower of levels with conformal spin $s=0$ extrapolating to the same dimension.  The root configurations of these excitations are again obtained by breaking string complexes as in (\ref{tower10}), giving $f:[(1_1^+)^{2k+1},(1_2^-)^{2k}]$ with integer $k\geq1$.
\begin{figure}[ht]
  \includegraphics[width=0.7\textwidth]{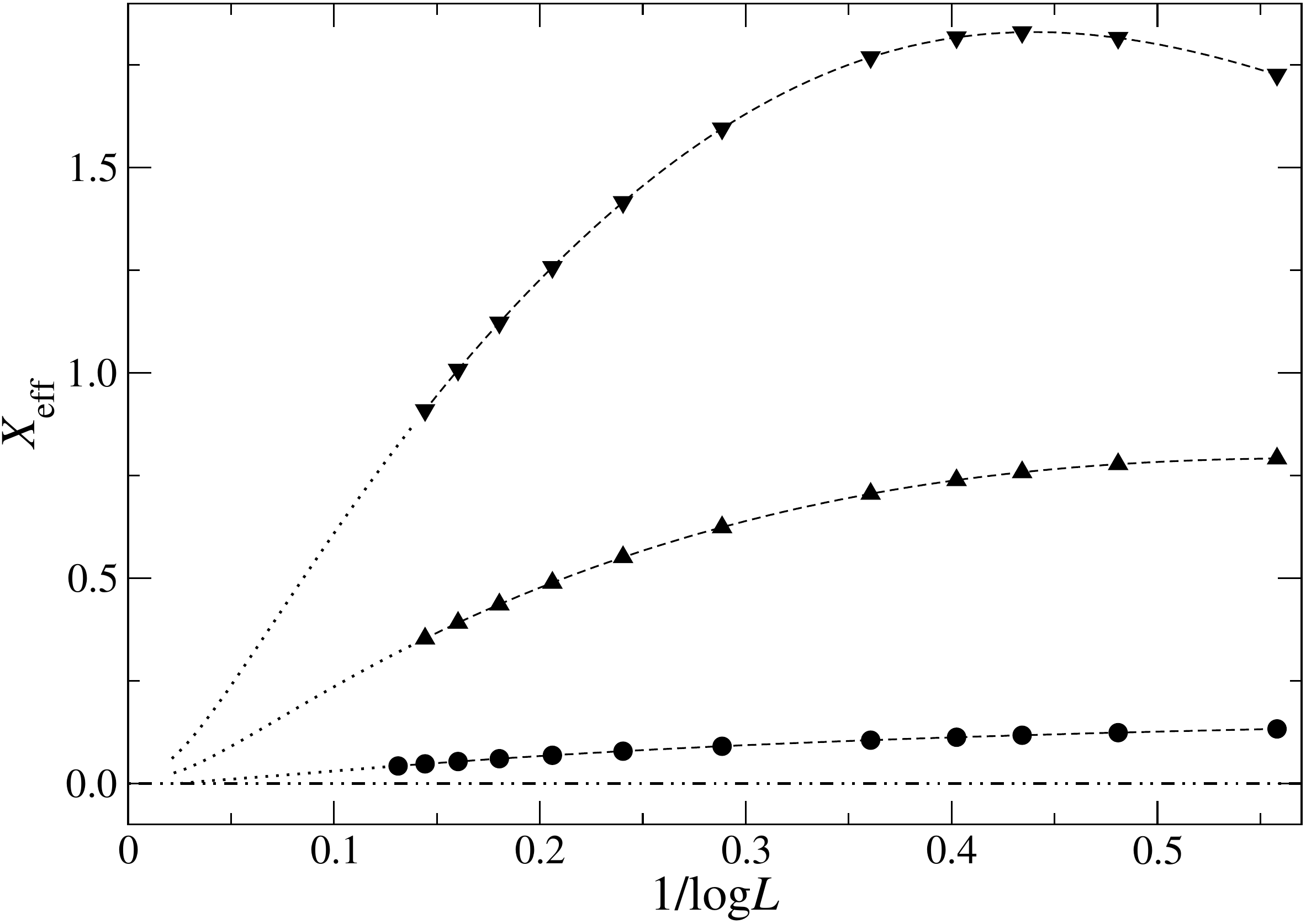}
  \caption{Similar as Fig.~\ref{fig:00_tower-scaling} but for the lowest eigenenergy and the related tower of levels in the spectrum of the superspin chain in sector $(1,1)$ for $\gamma=2\pi/7$. The dashed-dotted line is our conjecture (\ref{conj11-tower}) for this anisotropy.}
  \label{fig:11_tower-scaling}
\end{figure}

\begin{figure}[ht]
  \includegraphics[width=0.7\textwidth]{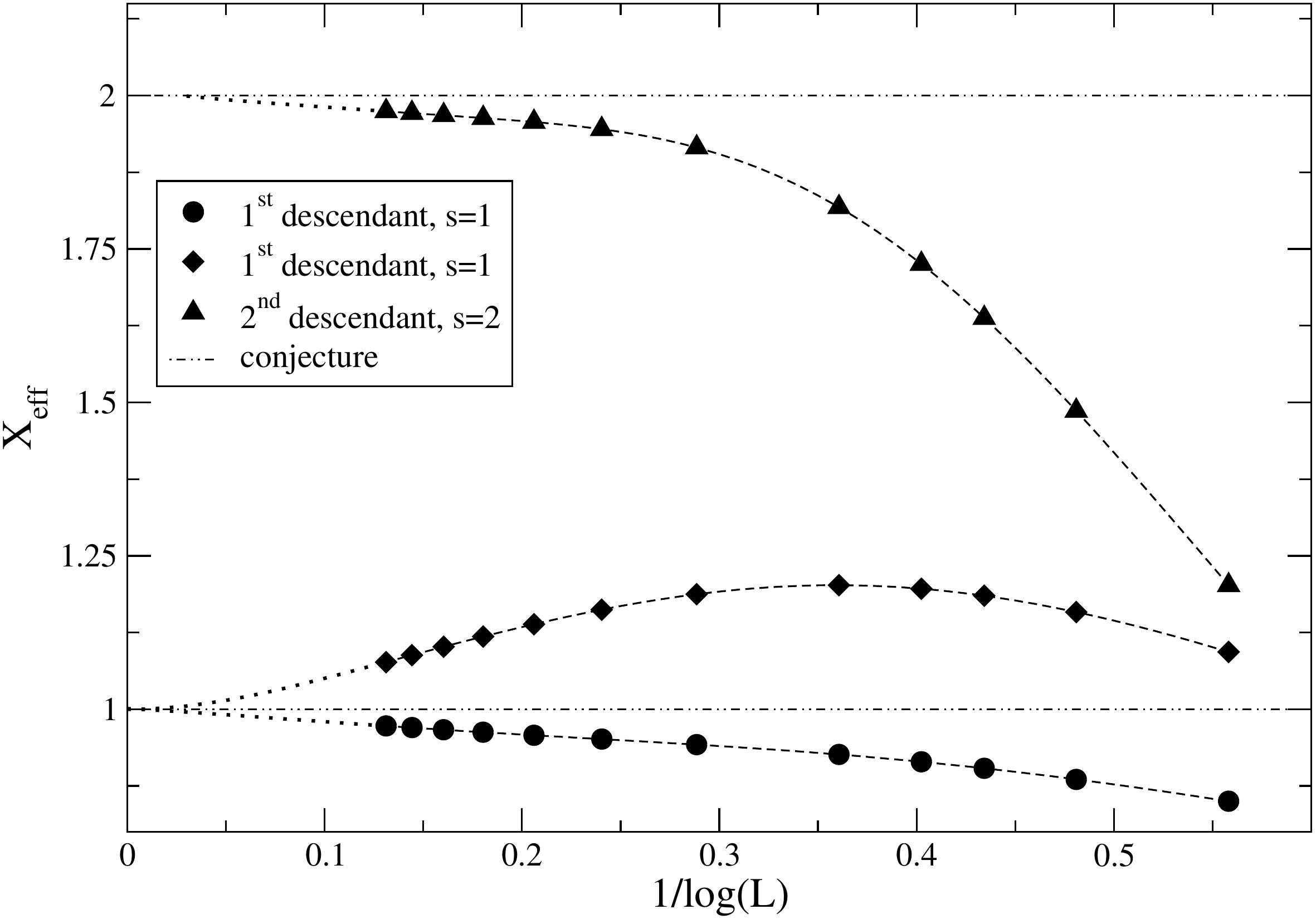}
  \caption{Similar as Fig.~\ref{fig:00_tower-scaling} but for three descendants of the two lowest levels in the spectrum of the superspin chain in sector $(1,1)$ for $\gamma=2\pi/7$. The dashed-dotted lines are our conjectures (\ref{conj11-s3.4}) and \eqref{conj11-s7}.}
  \label{fig:11_3-4-7}
\end{figure}

We have identified descendents of the two lowest levels in this tower. They are parameterized by root configurations $f:[1_1^-]$ and $f:[(1_1^+)^2]\oplus [1_1,2_2^+]_{-\infty}$, respectively, and have conformal spin $s=1$. Our numerical finite-size data extrapolate to
\begin{equation}
  \label{conj11-s3.4}
  X = \Xi_{1,0}^{1,0} - \frac{1}{4} + 1 = 1\,.
\end{equation}
A potential second descendent with scaling dimension
\begin{equation}
  \label{conj11-s7}
  X = \Xi_{1,0}^{1,0} - \frac{1}{4} + 2 = 2\,
\end{equation}
and spin $s=2$ is described by a $b:[1_1^+]$ root configuration.  The scaling behaviour of these descendents is displayed in fig.~\ref{fig:11_3-4-7}.

\begin{figure}[ht]
  \includegraphics[width=0.7\textwidth]{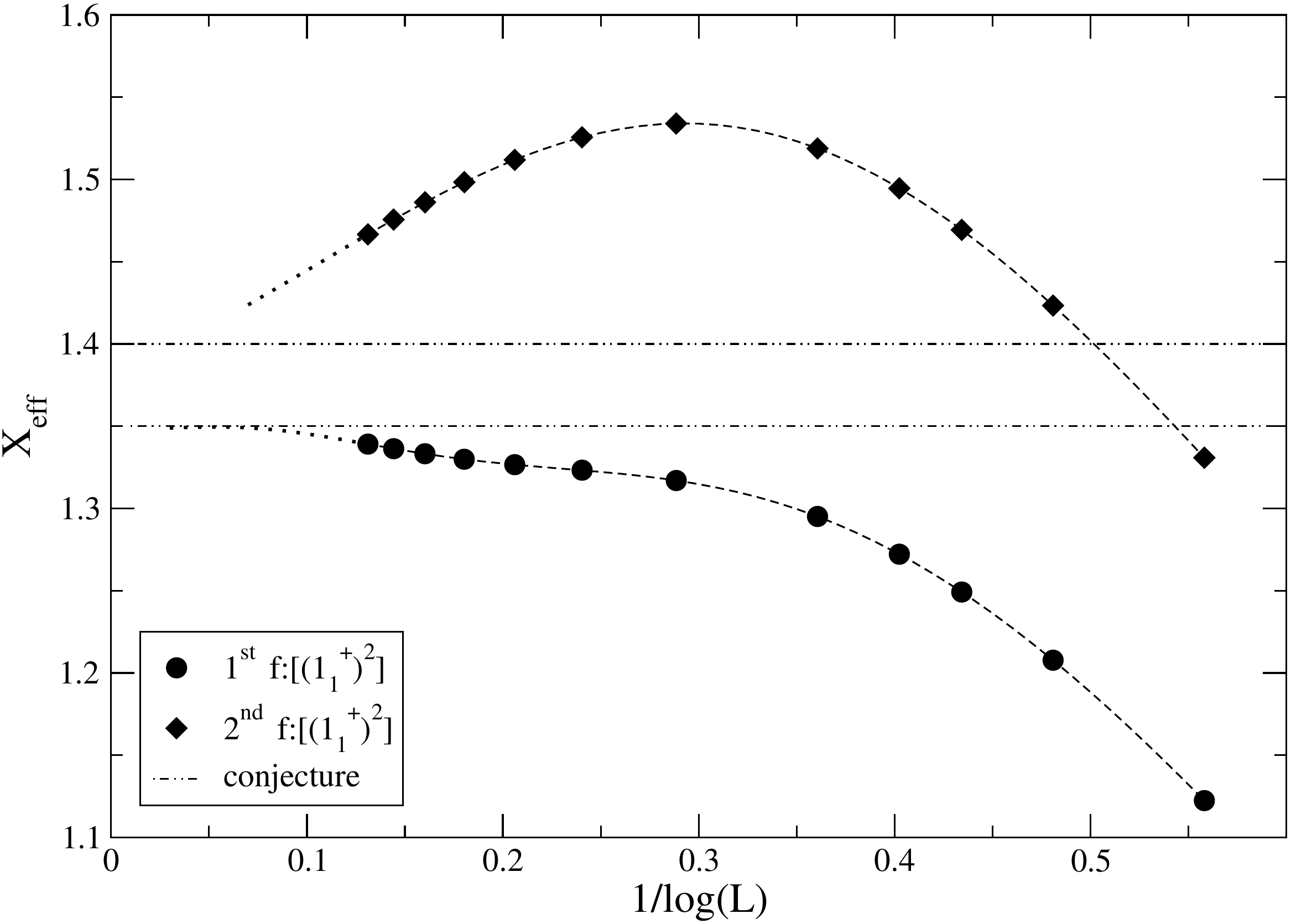}
  \caption{Similar as Fig.~\ref{fig:00_tower-scaling} but for the two states in the spectrum of the superspin chain in sector $(1,1)$ for which the finite part of the root configuration is parametrized by $f:[(1_1^+)^2]$ for $\gamma=2\pi/7$. The dashed-dotted lines are our conjectures (\ref{conj11-s6}) and \eqref{conj11-s9}.}
  \label{fig:11_6-9}
\end{figure}

Among the other low energy states we have identified two spin $s=1$ primaries: one is described by Bethe roots arranged as $f:[(1_1^+)^2] \oplus [1_1,2_2]_{-\infty}$ and scales to 
\begin{equation}
  \label{conj11-s6}
  X_{1,\frac12}^{1,1} = \Xi_{1,\frac12}^{1,1} - \frac{1}{8} = \frac{\pi}{4(\pi-\gamma)} + \frac{\pi}{4\gamma} +\frac18 \,.
\end{equation}
The other one is described by a root configuration $f:[(1_1^+)^2] \oplus [1_1,2_2]_{-\infty}$.  Its scaling dimension is
\begin{equation}
  \label{conj11-s9}
  X_{1,1}^{1,0} =  \Xi_{1,1}^{1,0} - \frac{1}{4} = \frac{\pi}{(\pi-\gamma)} \,.
\end{equation}
The results of our numerical analysis concerning these two states an be found in fig.~\ref{fig:11_6-9}.

\begin{figure}[!ht]
  \includegraphics[width=0.7\textwidth]{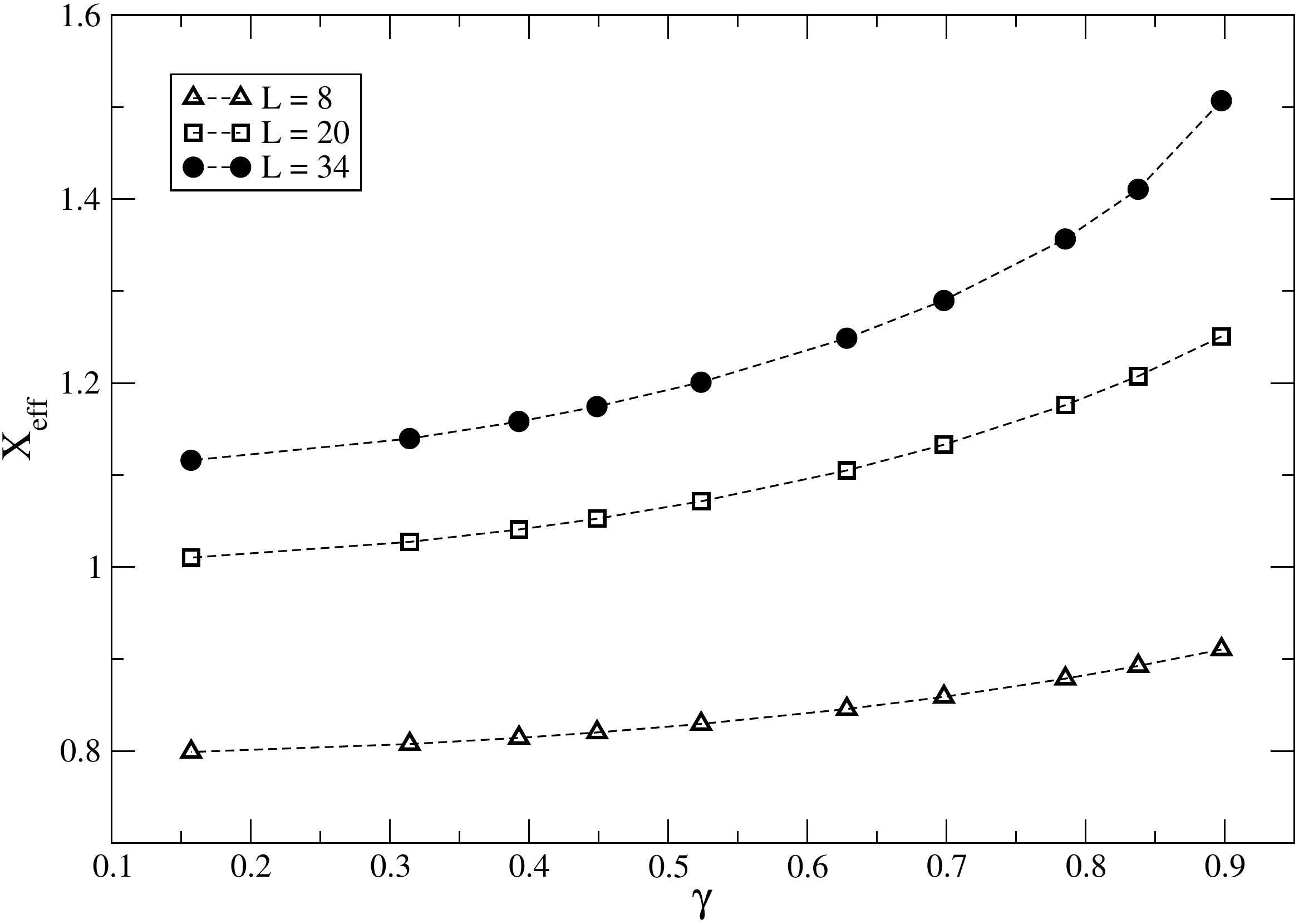}
  \caption{Effective scaling dimension of the state $b:[1_1^+]$ or equivalently $f:[1_1^-]$ in sector $(1,1)$ for small system sizes as a function of $\gamma$.}
  \label{fig:11_2nd-xvsg}
\end{figure}

As in sector $(1,0)$ we have observed degenerations of the root configurations for one spin $s=0$ level: for the smallest system sizes considered its root configurations are $b:[1_1^+]$ or $f:[1_1^-]$. As discussed in Appendix~\ref{app:roots} these patterns change with growing $L$. We have succeeded in following these changes up to, e.g.\ $L=36$ for $\gamma=2\pi/7$. The effective scaling dimensions for this state as obtained from the available finite-size data are shown in Figure~\ref{fig:11_2nd-xvsg}.  Unfortunately, they do not allow for a reliable extrapolation.

There are two low energy states remaining which are present in the spectrum of the superspin chain with lengths accessible to exact diagonalization of the Hamiltonian (\ref{HAM}). Both of these levels have non-zero conformal spin. For one of them we have identified the corresponding Bethe roots for $L=4$, $6$ and anisotropies $\gamma \lesssim \pi/4$, see Appendix~\ref{app:missing_states}, but were not able to go to larger $L$.  For the other one the parameterization in terms of Bethe roots is unkown.

To end the discussion of the sector $(n_1,n_2) = (1,1)$ we present our findings in table~\ref{tab:sec11}.
\begin{table}[t]
\begin{ruledtabular}
\begin{tabular}{l|ccc|ccc|l}
     &  \multicolumn{3}{c|}{$X$} & \multicolumn{3}{c|}{$s$} & \\
 Eq. & $m_1$ & $m_2$ & $x_0$ & total spin & $\sigma_{n_1,m_1}^{n_2,m_2}$ & $s_0$ & remark\\\hline
(\ref{conj11-tower}) & $0$ & $0$ & $-\frac{1}{4}$ & $0$ & $0$ & $0$ & tower\\
(\ref{conj11-s6}) & $\frac{1}{2}$ & $1$ & $-\frac{1}{8}$ & $1$ & $1$ & $0$ &  \\
(\ref{conj11-s9}) & $1$ & $0$ & $-\frac{1}{4}$ & $1$ & $1$ & $0$ & \\
\end{tabular}
\end{ruledtabular}
\caption{\label{tab:sec11}Conformal data for the levels studied in charge sector $(n_1,n_2)=(1,1)$ (see also Table~\ref{tab:sec00}).  We have also observed descendents of (\ref{conj11-tower}), see (\ref{conj11-s3.4}) and (\ref{conj11-s7}).}
\end{table} 

\subsection{Sector $(1,2)$}
\begin{figure}[ht]
  \includegraphics[width=0.7\textwidth]{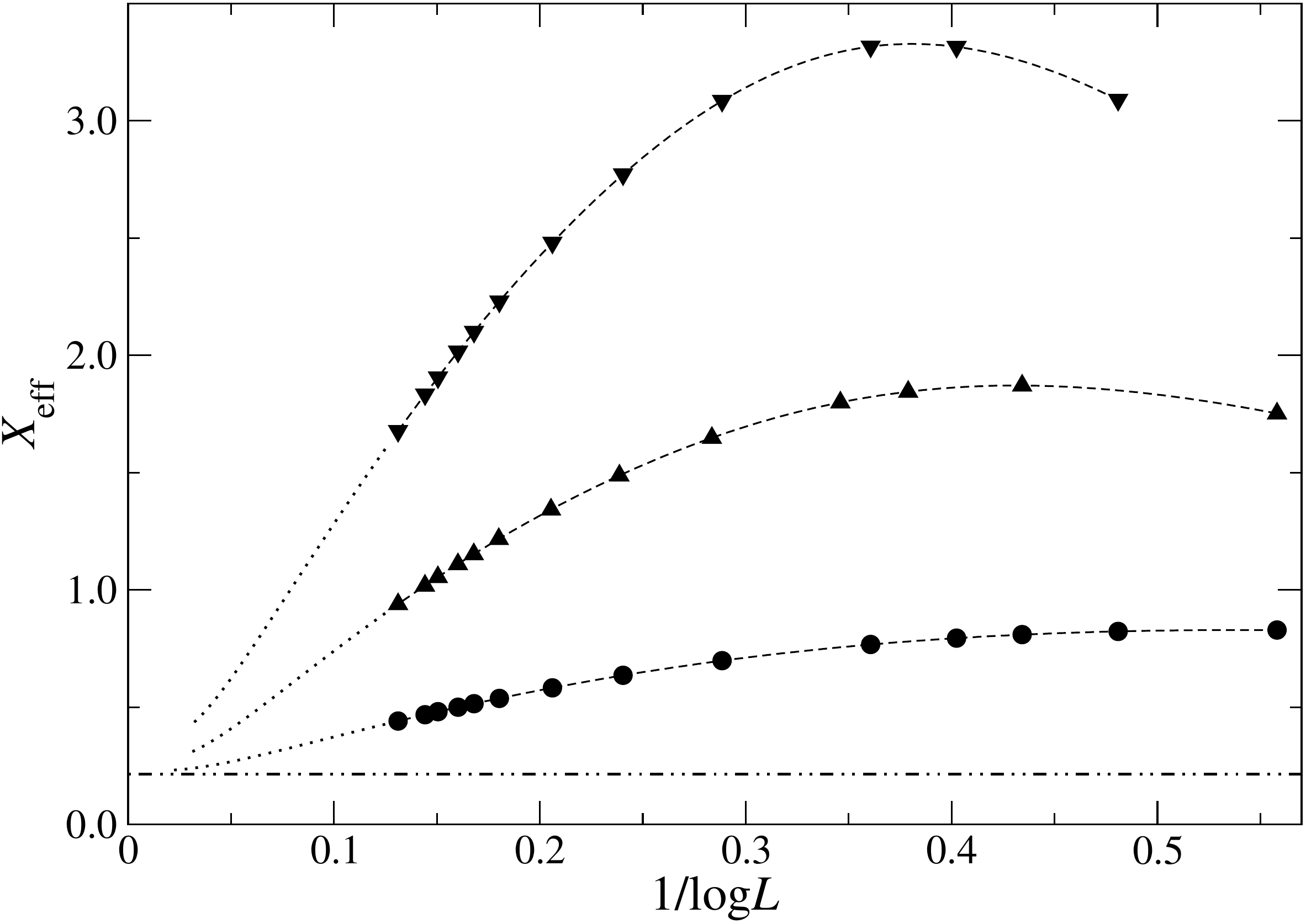}
  \caption{Similar as Fig.~\ref{fig:00_tower-scaling} but for the lowest eigenenergy and the related tower of levels in the spectrum of the superspin chain in sector $(1,2)$ for $\gamma=2\pi/7$. The dashed-dotted line is our conjecture (\ref{conj12-tower}) for this anisotropy.}
  \label{fig:12_tower-scaling}
\end{figure}

The Bethe roots for the lowest state in this sector are arranged in a $b:[(1_1^-)^2,1_2^-]$ configuration, i.e.\ $(L-4)/2$ $bfbfb$ string complexes (\ref{stringBFBFB}) and two (one) additional roots $\lambda^{(1)}$ ($\lambda^{(2)}$) on the line $\mathrm{Im}(\lambda)=\pi/2$.  This states appears in a tower of levels with scaling dimensions extrapolating to, see Figure~\ref{fig:12_tower-scaling},
\begin{equation}
  \label{conj12-tower}
  X_{1,0}^{2,0} = \Xi_{1,0}^{2,0} - \frac{1}{4} = \frac{3\gamma}{4\pi}\,.
\end{equation}
The other members of this tower are described by Bethe root configurations with one or more of the $bfbfb$ string complexes replaced by two $(1_a^-)$-strings
on each level $a=1,2$, i.e.\ $b:[(1_1^-)^{2k+2},(1_2^-)^{2k+1}]$.  The scaling behaviour of the first two excitations in this tower is also exhibited in Figure~\ref{fig:12_tower-scaling}.

\begin{figure}[ht]
  \includegraphics[width=0.7\textwidth]{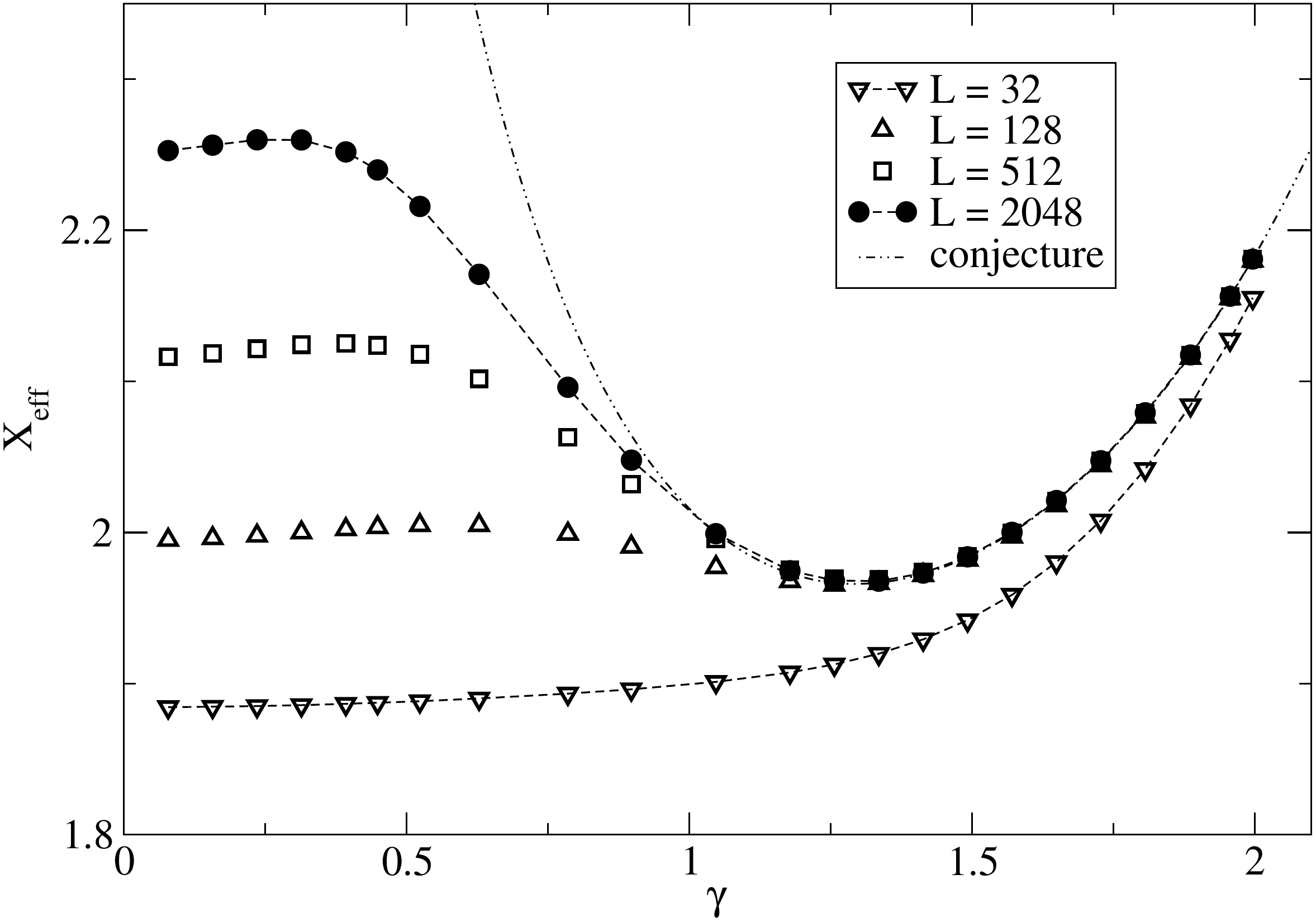}
  \caption{Effective scaling dimension for the charge $(1,2)$ state extrapolating to (\ref{conj12-s2}) as a function of $\gamma$ for various system sizes.}
  \label{fig:12-s2-scaling}
\end{figure}

The next excitation in the charge sector $(1,2)$ which we have analyzed is again described by a root configuration $b:[(1_1^-)^2,1_2^-]$, just as the lowest state in this sector.  The extrapolation of the finite-size effective scaling dimensions gives
\begin{equation}
  \label{conj12-s2}
  X_{1,\frac12}^{2,1} = \Xi_{1,\frac12}^{2,1} - \frac{1}{8} + \frac{1}{2} = \frac{3\gamma}{4\pi} + \frac{\pi}{4(\pi-\gamma)} + \frac{\pi}{4\gamma} +\frac58 \,.
\end{equation}
This state has conformal spin $s=2$ in agreement with (\ref{GAUSconjecture_spin}) in the presence of a chiral Ising contribution. As $\gamma\to\pi/2$ the subleading corrections to scaling become small, for $\gamma\to0$ the level disappears from the low energy spectrum, see Figure~\ref{fig:12-s2-scaling}. 

\begin{figure}[ht]
  \includegraphics[width=0.7\textwidth]{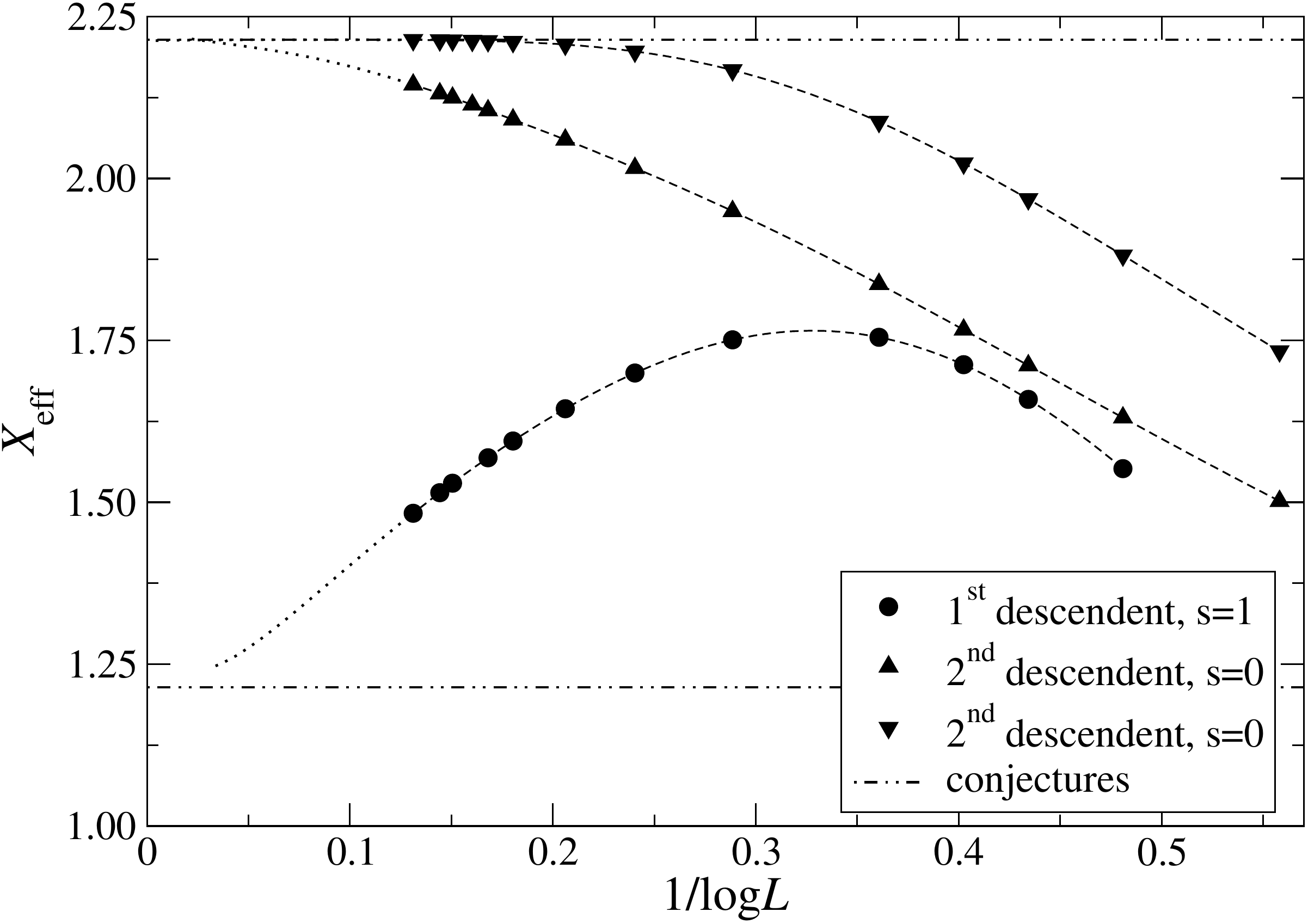}
  \caption{Similar as Fig.~\ref{fig:00_tower-scaling} but for the descendents of the tower states in sector $(1,2)$ extrapolating to (\ref{conj12-s3.4.6}) for $\gamma=2\pi/7$. The dashed-dotted line are our conjectured values for $L\to\infty$ for this anisotropy.}
  \label{fig:12_tower-extra}
\end{figure}

Three additional excitations in this sector are identified as descendents of the tower states (\ref{conj12-tower}).  They are described by root configurations $b:[(1_1^-)^2,1_2^-)]$, $b:[(1_1^+)^2,1_2^-]$, and $b:[(1_1^+)^2,1_2^+]$, respectively.  The first of these levels has conformal spin $1$, the others $s=0$.  Their effective scaling dimensions extrapolate to 
\begin{equation}
  \label{conj12-s3.4.6}
  X = \Xi_{1,0}^{2,0} - \frac{1}{4} + n = \frac{3\gamma}{4\pi} +n \,,\quad n=1,2\,,
\end{equation}
with $n=1$ ($2$) for the spin $s=1$ ($0$) states.  The $L$-dependence of the  corrections to scaling for $\gamma=2\pi/7$ is displayed in Figure~\ref{fig:12_tower-extra}.

\begin{figure}[ht]
  \begin{minipage}{0.48\textwidth}
    \subfigure[]{\includegraphics[width=\textwidth]{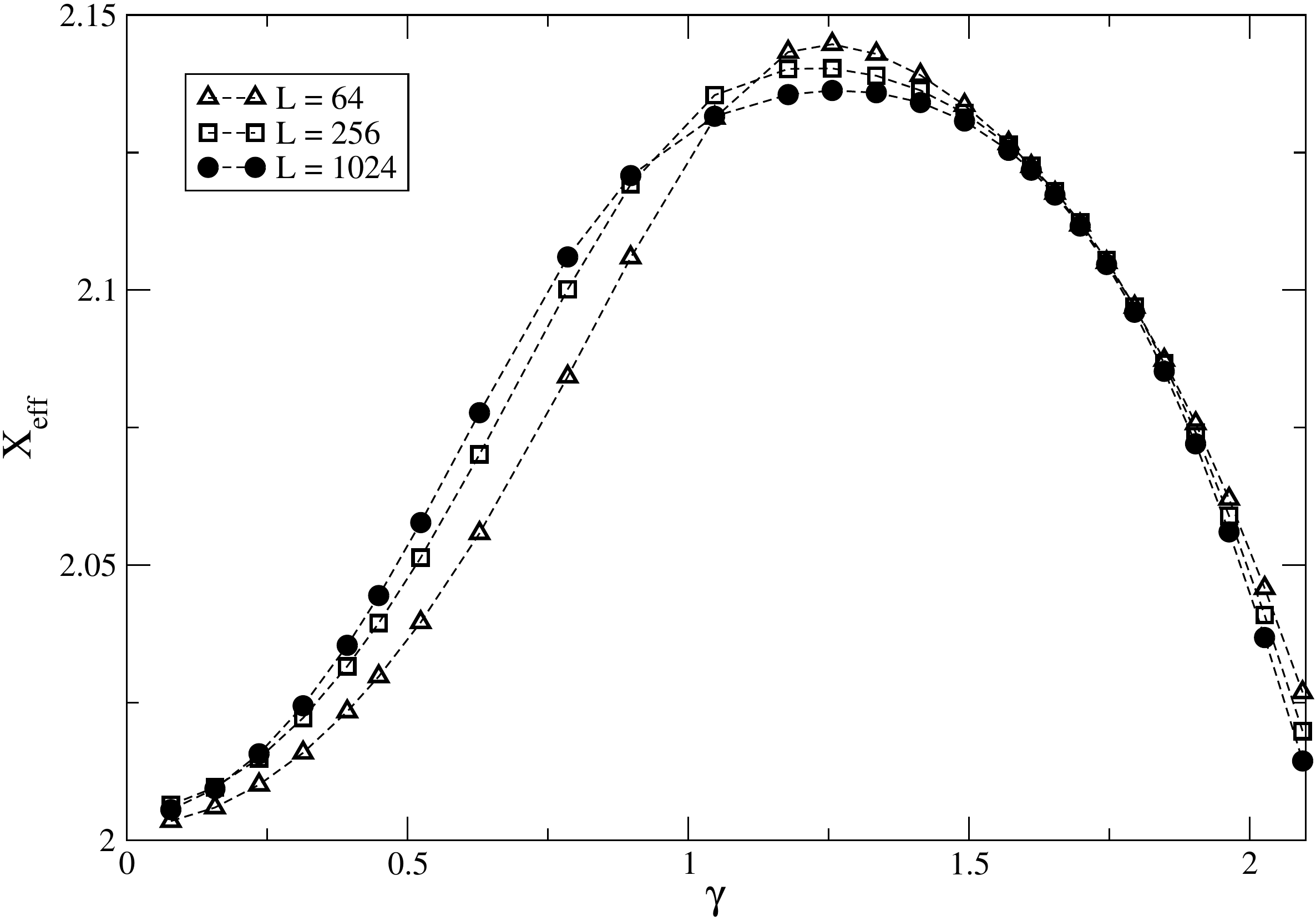}}
  \end{minipage}
  \begin{minipage}{0.48\textwidth}
    \subfigure[]{\includegraphics[width=\textwidth]{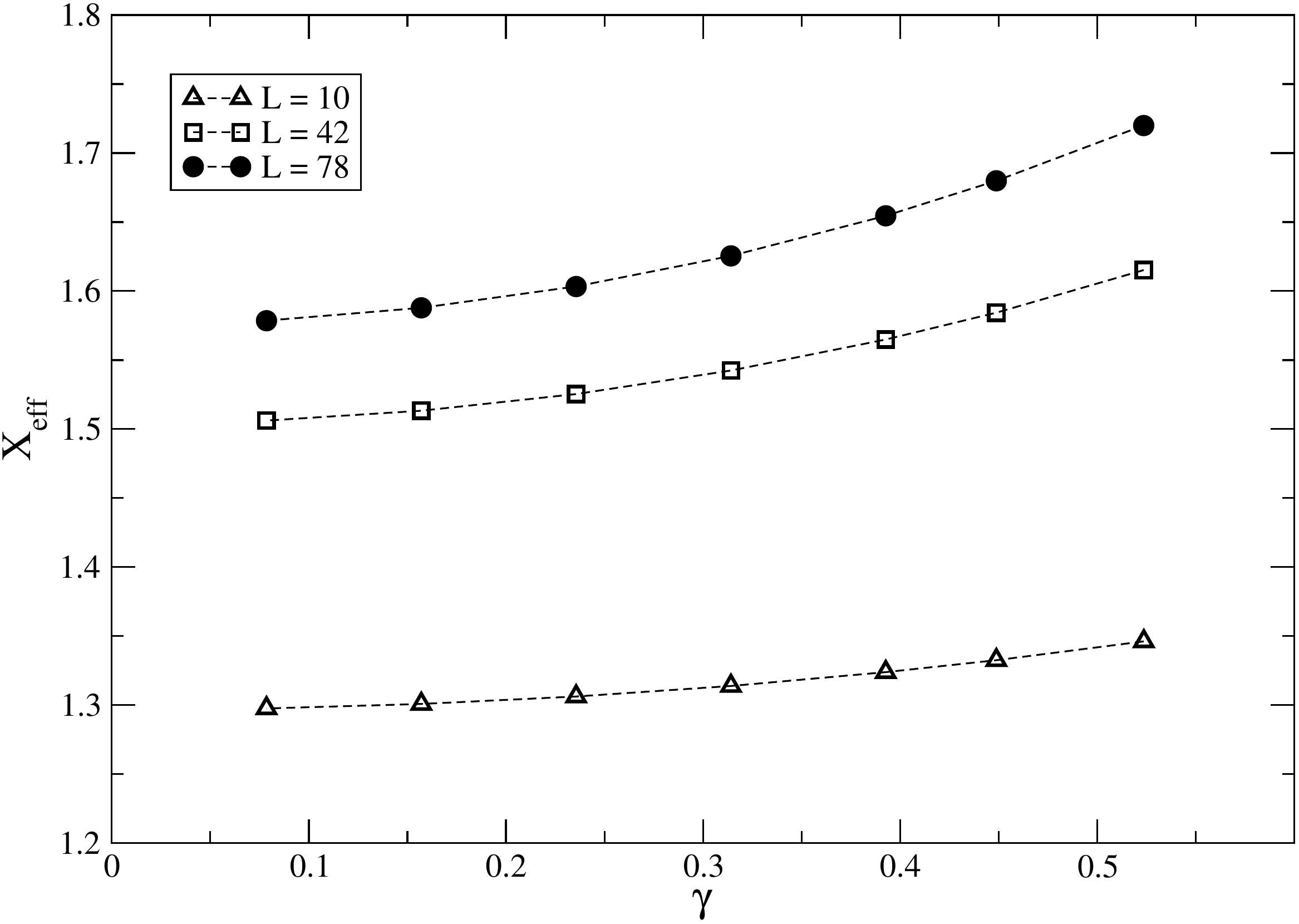}}
  \end{minipage}
  \caption{Effective scaling dimension for the charge $(1,2)$ state with root configuration (a) $f:[(1_1^+)^2,1_1^-,1_2^+]$ and (b) $f:[(1_1^+)^2,1_1^-,1_2^+]$ or equivalently $b:[3_{12}^+]$ as a function of $\gamma$ for various system sizes.}
  \label{fig:12-s5-s1-scaling}
\end{figure}

For another excitation we find that the Bethe roots are arranged according to
$f:[(1_1^+)^2,1_1^-,1_2^+]$. The spin of this excitation is
$s=2$. Finite-size data for the effective scaling dimension are available
  for systems with up to $1024$ lattice sites and anisotropies
  $0\le\gamma\le2\pi/3$, see Fig.~\ref{fig:12-s5-s1-scaling}(a).  Although the
  corrections to scaling appear to be small, in particular for $\gamma\gtrsim
  \pi/2$, we did not manage to describe the effective scaling dimensions in
  terms of our scheme (\ref{conjecture_full}).  A possible explanation for this problem might be a crossing 
between two levels with similar root configurations for some $\gamma<\pi/2$ which 
we have not resolved properly.  In this situation the data displayed in Fig.~\ref{fig:12-s5-s1-scaling}(a) 
would correspond to two different operators. For $\gamma\to 0$ a possible candidate would have, e.g., 
scaling dimension $X=\Xi_{1,0}^{2,0}-1/4+2=(3\gamma/4\pi)+2$. Around $\pi/2$ the data might correspond 
to an operator with $m_2=1$.  If this is the case, however, the crossing would come along with 
huge corrections to scaling which cannot be handled with the available data.

Again, there is one low energy state present in this charge sector where changes in the root configuration with the system size aggravate the solution of the corresponding Bethe equations.  This level has conformal spin $s=0$ and is parameterized by roots arranged as $f:[(1_1^+)^2,1_1^-,1_2^+]$ or $b:[3_{12}^+]$ for small system sizes.  Similarly as in the $(n_1,n_2)=(0,0)$ example discussed in Appendix~\ref{app:roots} the $1_1^+$ diverge at some finite value of the system size so that we cannot determine the scaling dimension from the available finite-size data. Our numerical results for small system sizes however are presented in Figure~\ref{fig:12-s5-s1-scaling}(b).

Again we summarize the study of the sector $(n_1,n_2) = (1,2)$ in table~\ref{tab:sec12}.
\begin{table}[t]
\begin{ruledtabular}
\begin{tabular}{l|ccc|ccc|l}
     &  \multicolumn{3}{c|}{$X$} & \multicolumn{3}{c|}{$s$} & \\
 Eq. & $m_1$  & $m_2$ & $x_0$ & total spin & $\sigma_{n_1,m_1}^{n_2,m_2}$ & $s_0$ & remark\\\hline
(\ref{conj12-tower}) & $0$ & $0$ & $-\frac{1}{4}$ & $0$ & $0$ & $0$ &  tower\\
(\ref{conj12-s2}) & $\frac{1}{2}$ & $1$ & $-\frac18 + \frac{1}{2}$ & $2$ & $\frac{3}{2}$ & $\frac{1}{2}$ & Ising $(\frac{1}{2},0)$\\
\end{tabular}
\end{ruledtabular}
\caption{\label{tab:sec12}Conformal data for the levels studied in charge sector $(n_1,n_2)=(1,2)$ (see also Table~\ref{tab:sec00}). We have also observed descendents of (\ref{conj12-tower}), see (\ref{conj12-s3.4.6}) .}
\end{table}

\subsection{Sector $(2,0)$}
\begin{figure}[ht]
  \begin{minipage}{0.48\textwidth}
    \subfigure[]{\includegraphics[width=\textwidth]{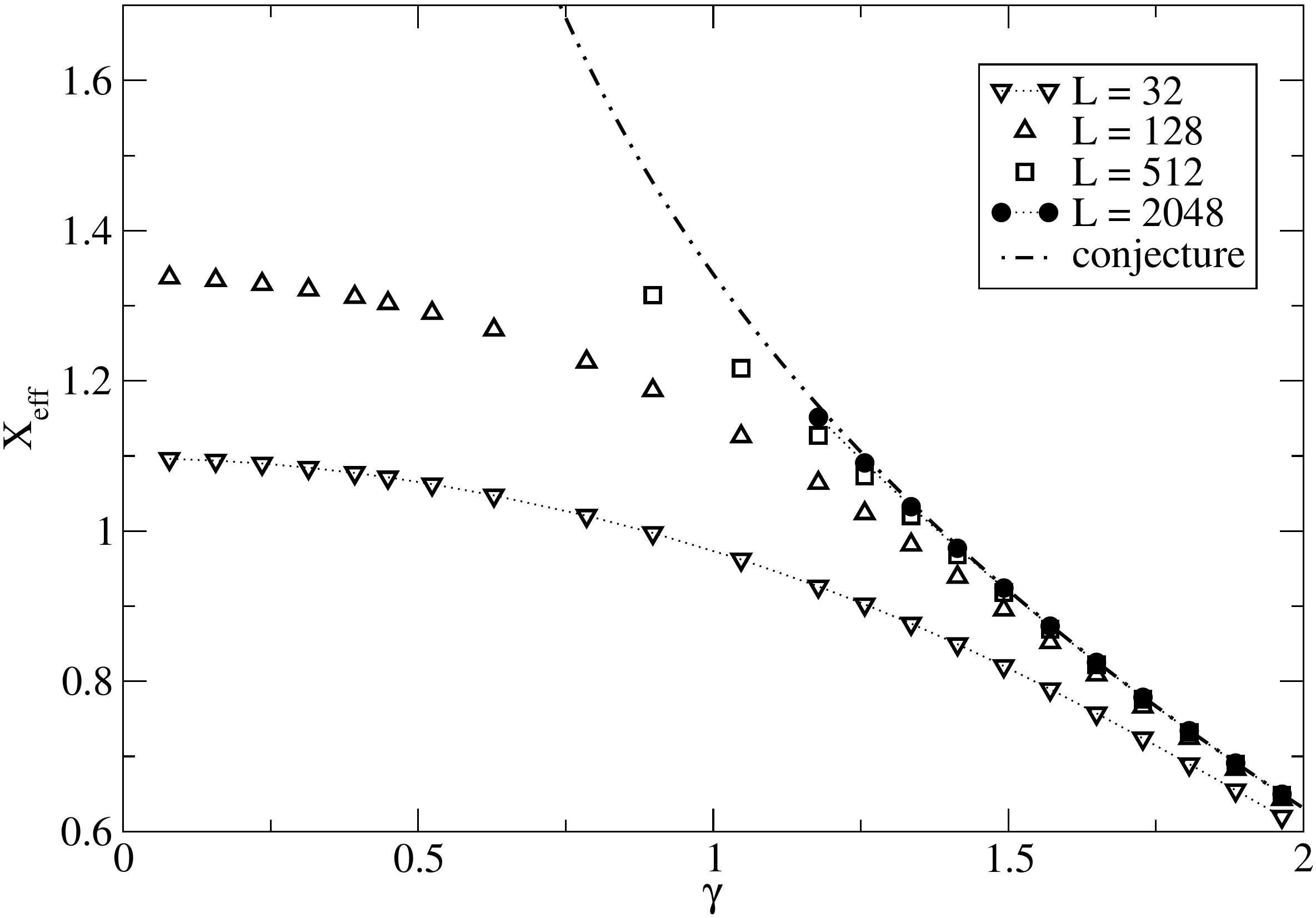}}
  \end{minipage}
    \begin{minipage}{0.48\textwidth}
    \subfigure[]{\includegraphics[width=\textwidth]{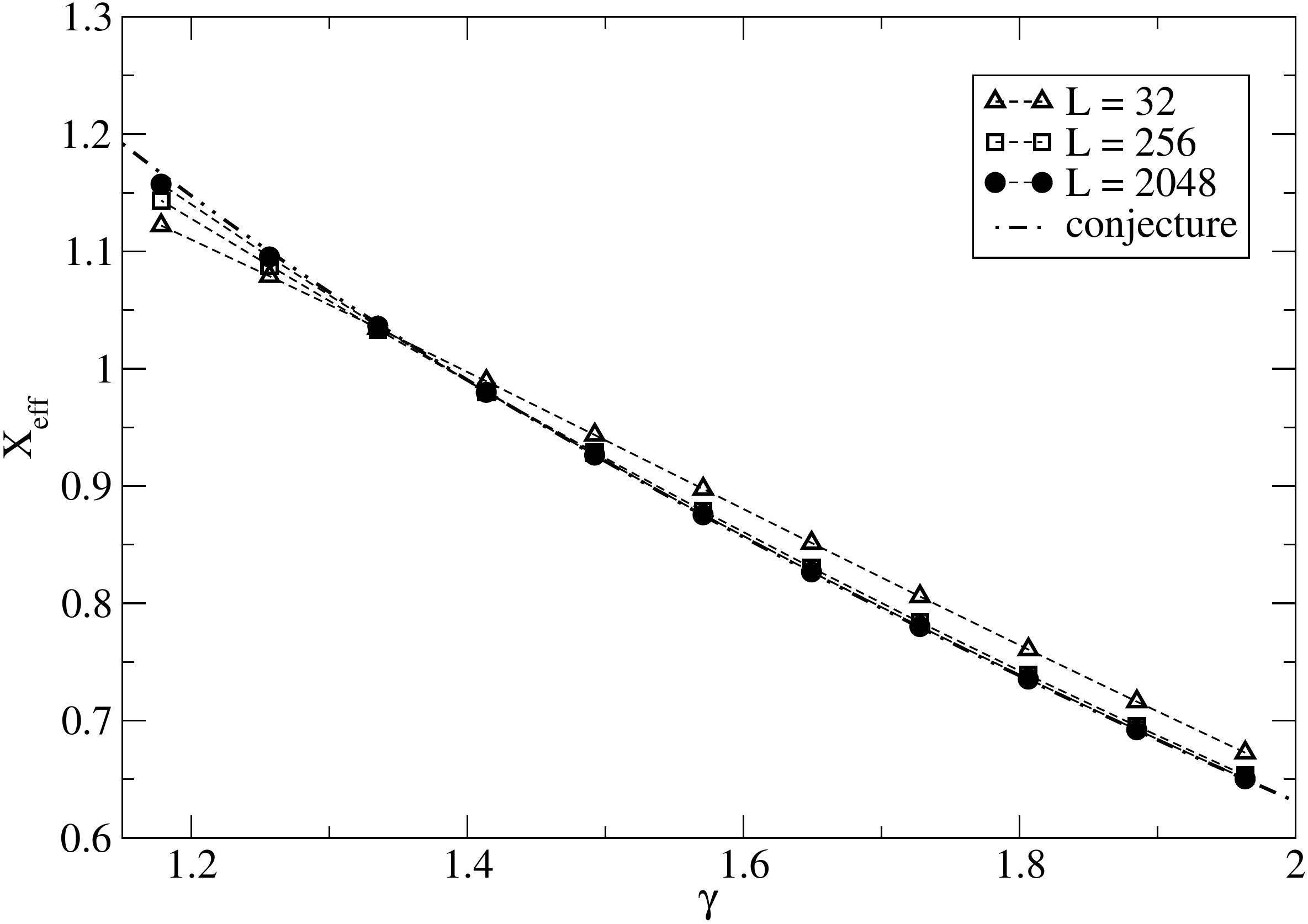}}
  \end{minipage}
  \caption{Effective scaling dimension for the state in sector $(2,0)$ with root configuration (a) $f:[(1_1^+)^2,z_2]$ (Note, this is the lowest state in this charge sector) and (b) $f:[(1_1^+)^2,1_2^+,1_2^-]$ as a function of $\gamma$ for various system sizes. The dashed-dotted lines are our conjectures (\ref{conj20-s0.2}) for these levels.}
  \label{fig:20-s0-s2-scaling}
\end{figure}

For $L=6$ the ground state in charge sector $(2,0)$ is parameterized by two $fbbbf$ string complexes (\ref{stringFBBBF}).  As the system size is increased this configuration degenerates. For example, at $\gamma=2\pi/7$ this happens at $L_*=10$. Beyond this $L_*$ the root configuration consists of $(L-4)/2$ $fbbbf$ string complexes and, in addition, $\lambda_\pm^{(1)}=\pm \xi\in\mathbb{R}$ and $\lambda_\pm^{(2)}\simeq \pm i\eta$, i.e. $f:[(1_1^+)^2,z_2]$.  At least for deformation parameters $\gamma$ close to $\pi/2$ no further degeneration occurs so that we have been able to study the effective scaling dimensions of this level in this regime.  From our data we conclude that this level has conformal spin $s=0$ and that the finite-size data extrapolate to
\begin{equation}
  \label{conj20-s0.2}
  X_{2,0}^{0,1} = X_{2,0}^{0,1} - \frac{1}{8} = 1-\frac{\gamma}{\pi} + \frac{\pi}{4\gamma} -\frac18\,,
\end{equation}
see Figure\ref{fig:20-s0-s2-scaling}(a).
There exists a second $s=0$ state with this scaling dimension with root configuration $f:[(1_1^+)^2,1_2^+,1_2^-]$.  Here we had to study deformation parameters from $3\pi/8\leq\gamma\leq5\pi/8$ for data with sufficiently large $L$ for the extrapolation. The results of our numerical analysis for this state is shown in Figure~\ref{fig:20-s0-s2-scaling}(b). Note that the corrections to scaling for both states extrapolating to \eqref{conj20-s0.2} are small for anisotropies $\gamma \geq 3\pi/8$.

\begin{figure}[ht]
  \includegraphics[width=0.7\textwidth]{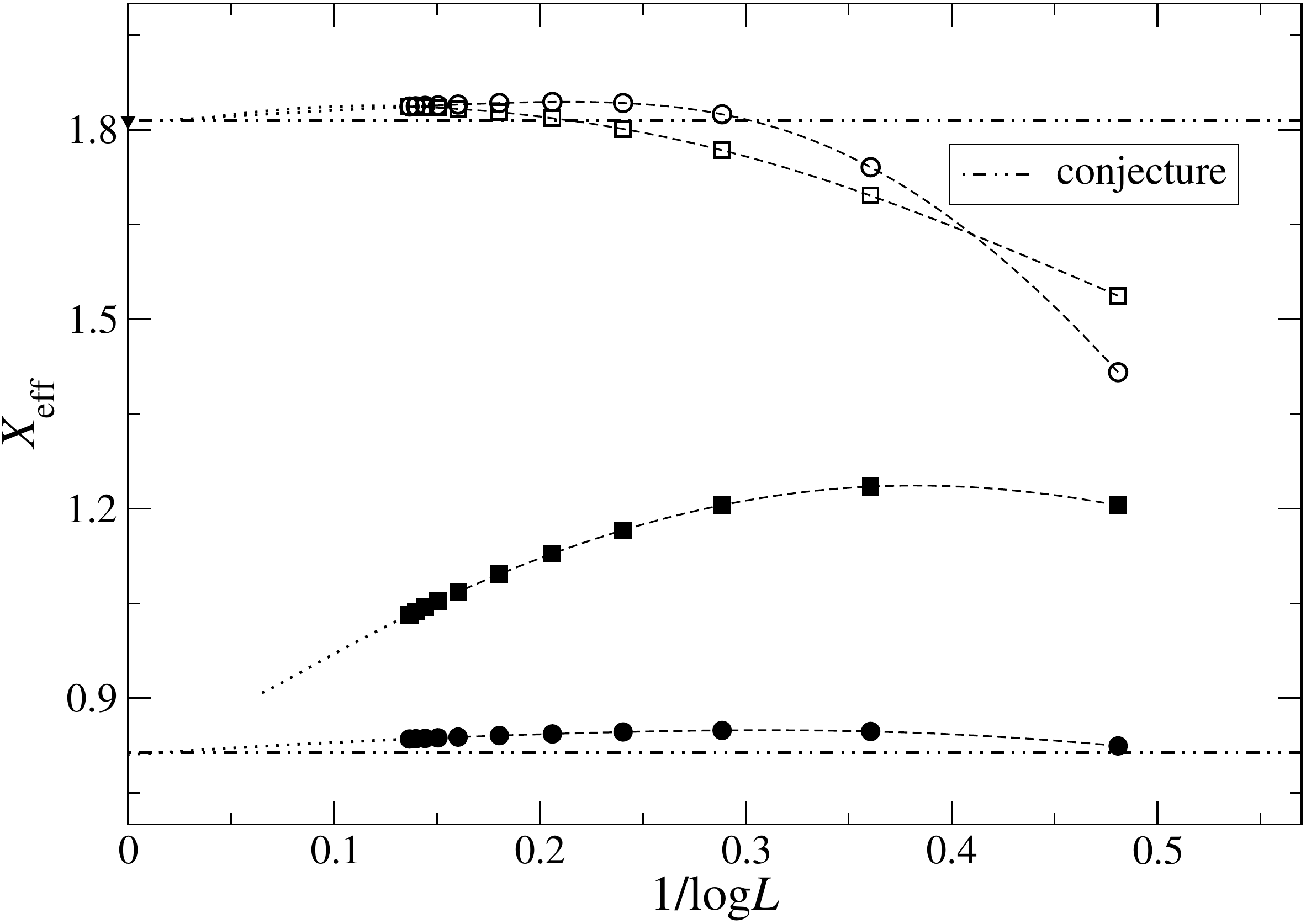}
  \caption{Similar as Fig.~\ref{fig:00_tower-scaling} but for the states in sector $(2,0)$ extrapolating to (\ref{conj20-s1.3.4.8}) for $\gamma=2\pi/7$. The filled symbols denote states with conformal spin $s=1$, the open ones are levels with $s=2$ and $s=0$, respectively. The dashed-dotted lines are our conjectured values for $L\to\infty$ for this anisotropy.}
  \label{fig:20_other}
\end{figure}

Among the next set of low energy levels we have identified in this sector two states with root configurations consisting only of $(L-2)/2$ $fbbbf$ string complexes (\ref{stringFBBBF}).  Their effective scaling dimensions extrapolate to
\begin{equation}
  \label{conj20-s1.3.4.8}
  X = \Xi_{2,\frac12}^{0,0} - \frac{1}{4} + n = 1-\frac{\gamma}{\pi} + \frac{\pi}{4(\pi-\gamma)} -\frac14 + n\,,
  \quad n=0,1\,,
\end{equation}
and have conformal spin $s=1$ ($2$) for $n=0$ ($1$).  Their scaling behaviour is shown in Figure~\ref{fig:20_other}.
In addition there is a pair of excitations combining $(L-4)/2$ $fbbbf$ string complexes with different structures for the additional roots.  One of the excitations has a root configuration $f:[(1_1^+)^2,(1_2^-)^2]$, the other one is described by a pattern $f:[1_1^+,1_1^-,z_2]$ where the roots in the pair $[z_2]$  appear to approach $\lambda^{(2)}_{1,2}\simeq \pm i\pi/4$ for large $L$.  Their scaling dimensions also extrapolate to (\ref{conj20-s1.3.4.8}) while their have conformal spins are $s=1$ ($0$) for $n=0$ ($1$).  The spin $1$ of the lower levels is expected from Eq.~(\ref{GAUSconjecture_spin}) for the $(n_1,m_1)=(2,1/2)$ primary field, the $s=0$ and $2$ levels are $n=1$ descendents.
Their finite-size scaling is displayed in Figure~\ref{fig:20_other}.

\begin{figure}[ht]
  \begin{minipage}{0.48\textwidth}
    \subfigure[]{\includegraphics[width=\textwidth]{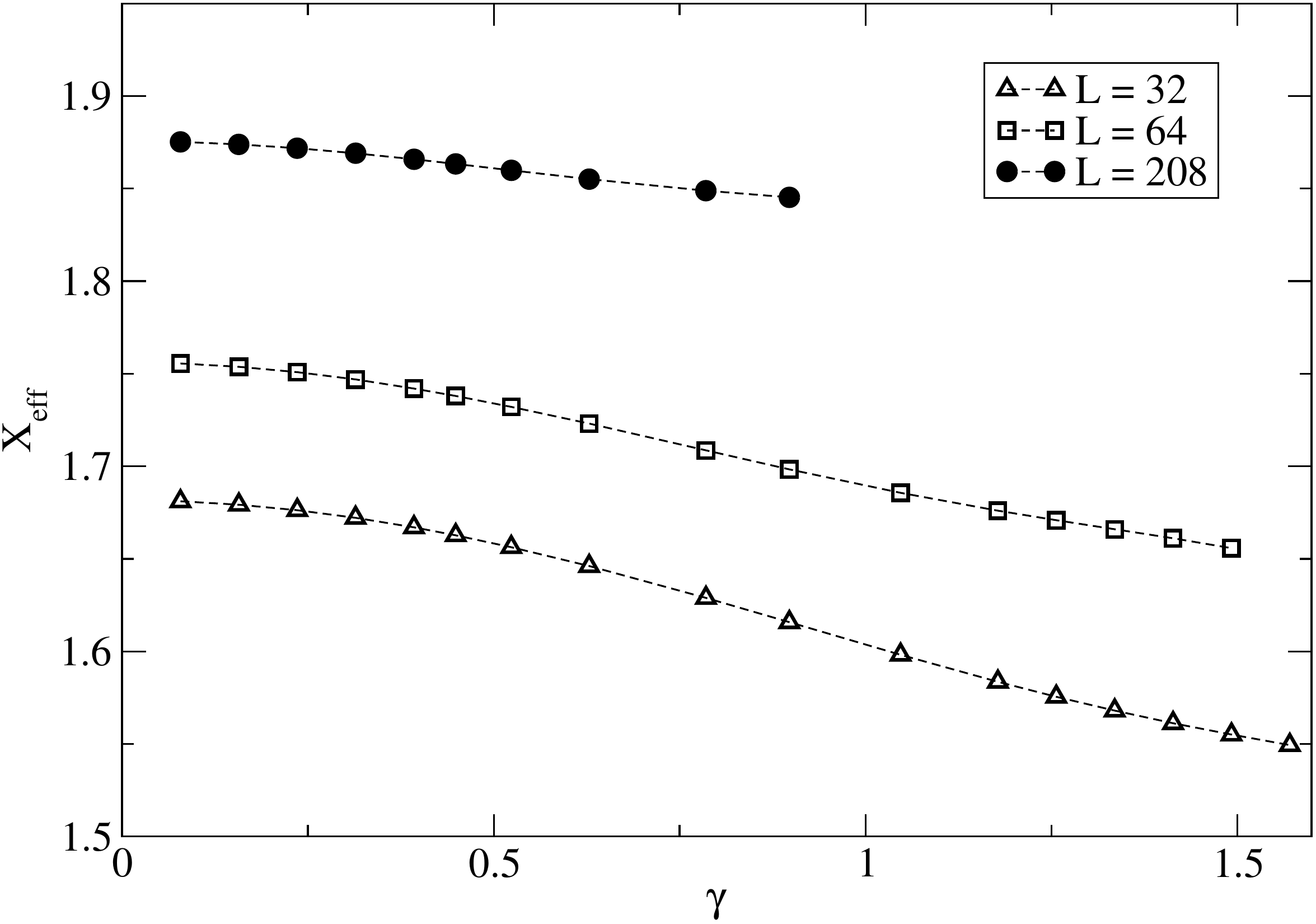}}
  \end{minipage}
    \begin{minipage}{0.48\textwidth}
    \subfigure[]{\includegraphics[width=\textwidth]{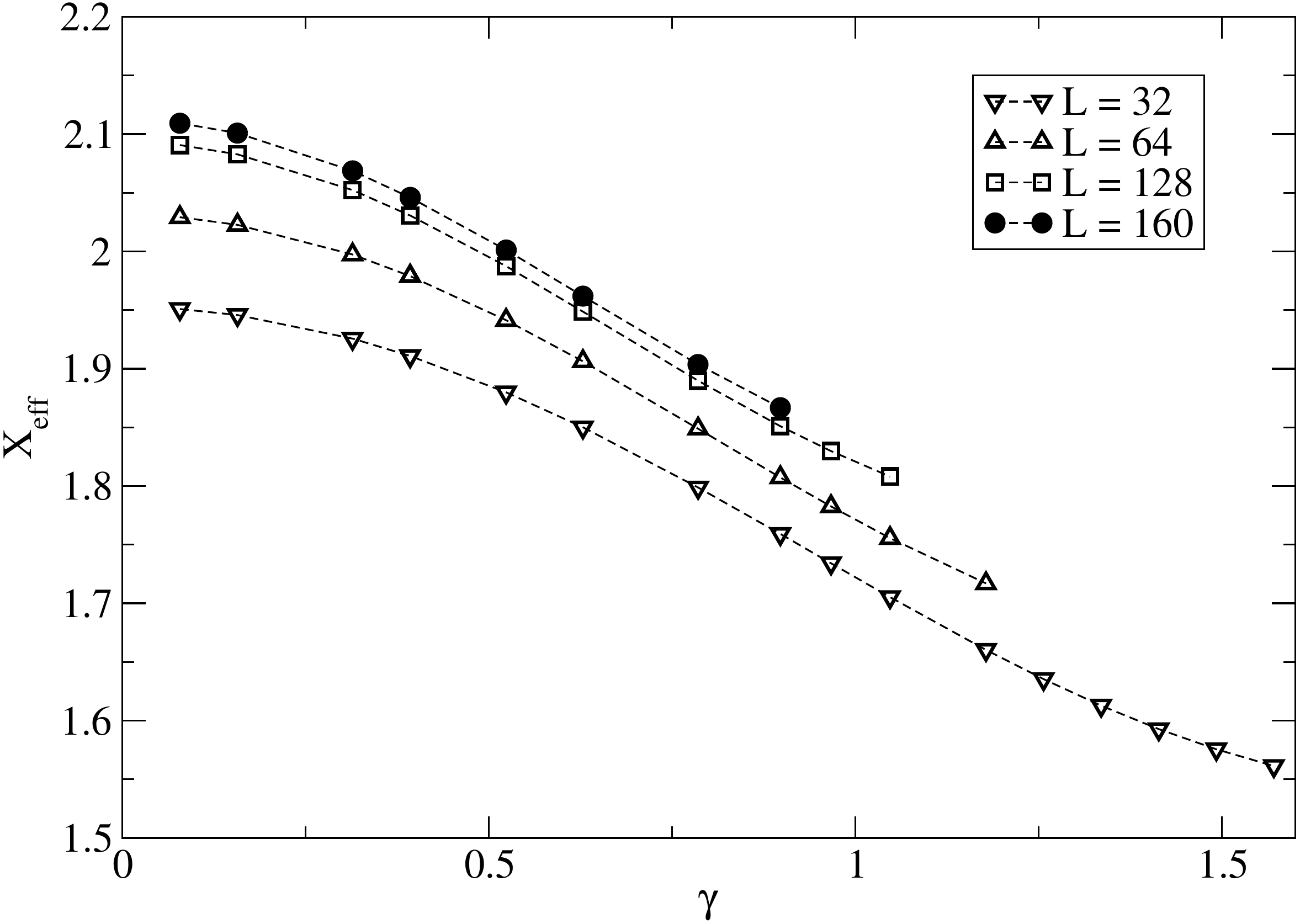}}
  \end{minipage}
  \caption{Effective scaling dimension for the states in sector $(2,0)$ with root configuration $b:[1_1^+,(1_1^-)^3,(1_2^-)^2]$ or equivalently $f:[(1_1^+)^2,(1_2^-)^2]$ and conformal spin (a) $s=1$ and (b) $s=0$ as a function of $\gamma$ for various system sizes.}
  \label{fig:20-s5-s6-scaling}
\end{figure}

For two excitations the root configurations are $b:[1_1^+,(1_1^-)^3,(1_2^-)^2]$ or $f:[(1_1^+)^2,(1_2^-)^2]$ for not too large $L$ but the presence of Bethe roots diverging as the system size grows prevented us from getting finite-size data for $L\gtrsim200$. They have conformal spin $s=1$ and $s=0$, respectively. The results of our numerical study up to $L = 208~(160)$ for these two states is shown in fig.~\ref{fig:20-s5-s6-scaling}.

\begin{figure}[ht]
  \includegraphics[width=0.7\textwidth]{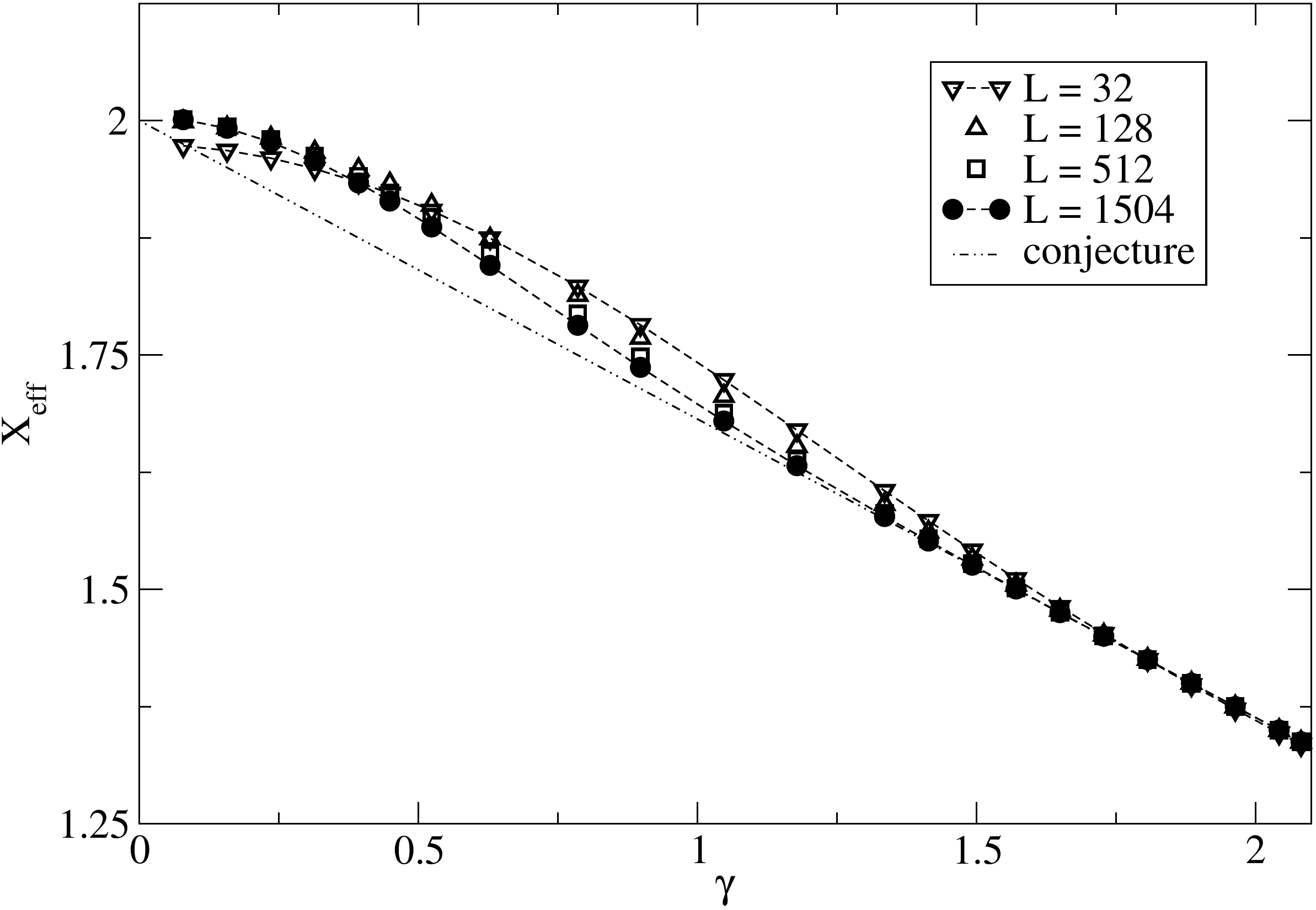}
  \caption{Effective scaling dimension for the state in sector $(2,0)$ conjectured to extrapolate to (\ref{conj20-s10}) as a function of $\gamma$ for various system sizes.}
  \label{fig:20-s10-scaling}
\end{figure}

Finally, we have studied the scaling of a spin $s=0$ level with $(L-4)/2$ $fbbbf$ string complexes  (\ref{stringFBBBF}) and extra roots $\lambda_{\pm}^{(1)}=\pm i\gamma/2$ and two degenerate second level roots at $\lambda^{(2)}=0$, i.e.\ $f:[1_2^+,3_{12}^+]$. The effective scaling dimension extrapolate to
\begin{equation}
  \label{conj20-s10}
  X = \Xi_{2,0}^{0,0} + 1 = -\frac{\gamma}{\pi} + 2 \,,
\end{equation}
see Fig.~\ref{fig:20-s10-scaling}.

Again, there is one low energy excitation present in the spectrum of small systems for which no Bethe ansatz solution has been found.

A summary of the numerical study of the sector $(n_1,n_2) = (2,0)$ is given in table~\ref{tab:sec20}.
\begin{table}[t]
\begin{ruledtabular}
\begin{tabular}{l|ccc|ccc|l}
     &  \multicolumn{3}{c|}{$X$} & \multicolumn{3}{c|}{$s$} & \\
 Eq. & $m_1$ & $m_2$ & $x_0$ & total spin & $\sigma_{n_1,m_1}^{n_2,m_2}$ & $s_0$ & remark\\\hline
(\ref{conj20-s0.2}) & $0$ & $1$ & $-\frac{1}{8}$ & $0$ & $0$ & $0$ & \\ 
(\ref{conj20-s1.3.4.8}) & $\frac{1}{2}$ & $0$ & $-\frac14$ & $1$ & $1$ & $0$ &\\ 
(\ref{conj20-s10}) & $0$ & $0$ & $1 $ & $0$ & $0$ & $0$ & Ising $(\frac{1}{2},\frac{1}{2})$ \\
\end{tabular}
\end{ruledtabular}
\caption{\label{tab:sec20}Conformal data for the levels studied in charge sector $(n_1,n_2)=(2,0)$ (see also Table~\ref{tab:sec00}). 
We have also observed descendents of (\ref{conj20-s1.3.4.8}), see Fig.~\ref{fig:20_other}.}
\end{table} 

\subsection{Sector $(2,1)$}
\begin{figure}[ht]
  \includegraphics[width=0.7\textwidth]{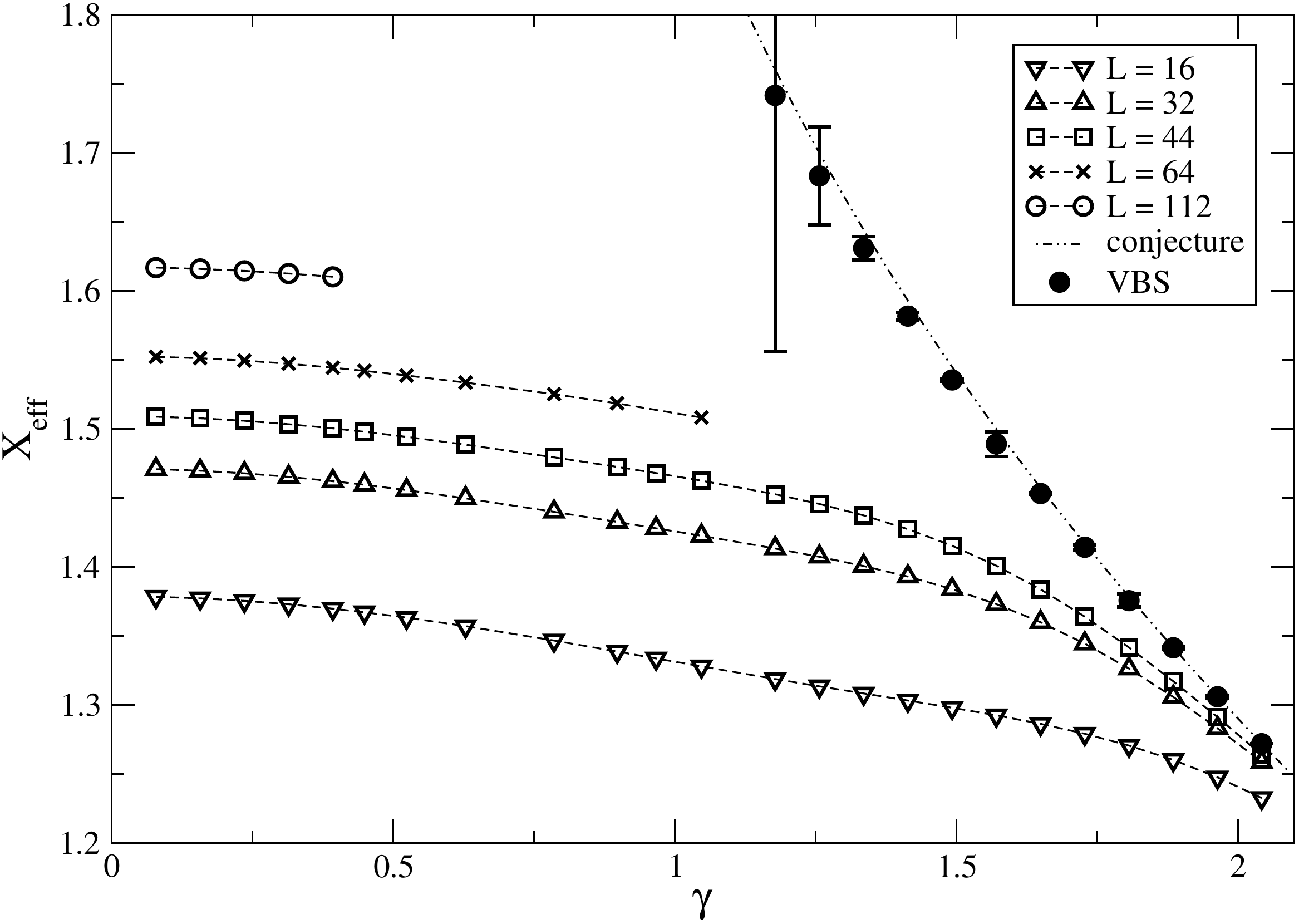}
  \caption{Effective scaling dimension for the lowest state in sector $(2,1)$ as a function of $\gamma$ for various system sizes.}
  \label{fig:21-s0}
\end{figure}

The lowest level in this sector is described by root configurations $b:[1_1^-,3_{12}^+]$ or $f:[(1_1^+)^2,1_2^+]$ for small system sizes.  Since this configuration degenerates at intermediate $L$ we were able to solve the Bethe ansatz equations only for $L \leq 128$, depending on the anisotropy. However, based on a VBS analysis of our data for $3\pi/8 \leq \gamma \leq 2\pi/3$ we conjecture that its scaling dimension extrapolates to 
\begin{equation}
  \label{conj21-s0.2.3}
  X_{2,0}^{1,1} = \Xi_{2,0}^{1,1} -\frac{1}{8} + \frac12 = 1-\frac{3\gamma}{4\pi} + \frac{\pi}{4\gamma} +\frac{3}{8}\,.
\end{equation}
It has conformal spin $s=0$. Our finite-size results for small system sizes are depicted in Fig.~\ref{fig:21-s0}.

\begin{figure}[ht]
  \includegraphics[width=0.7\textwidth]{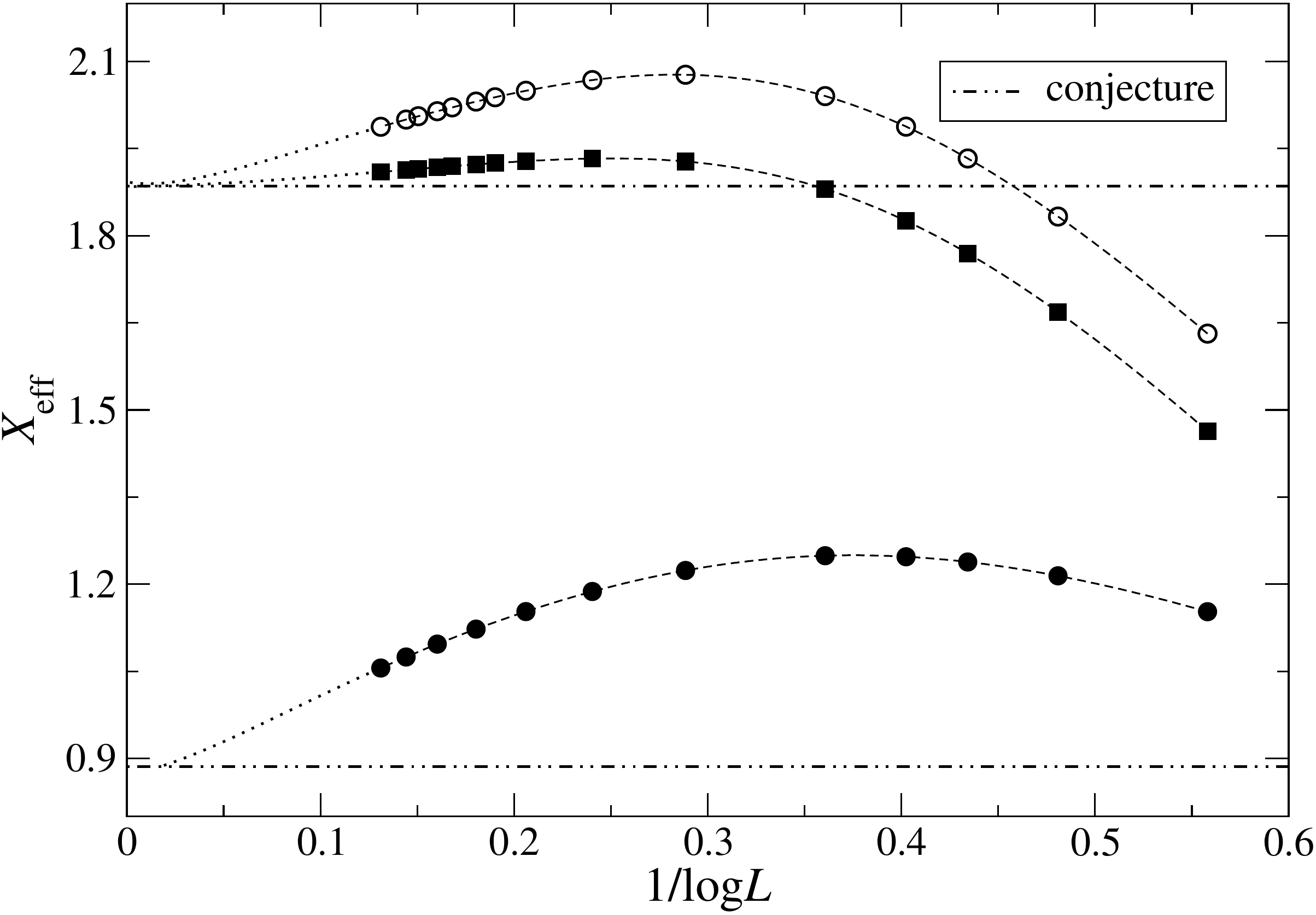}
  \caption{Similar as Fig.~\ref{fig:00_tower-scaling} for the first excitation with spin $s=1$ in sector $(2,1)$ extrapolating to $X$ given by (\ref{conj21-s1}) and two higher $s=2$ levels extrapolating to $X+1$. The data are for $\gamma=2\pi/7$.}
  \label{fig:21_other}
\end{figure}

The first excitation is described by an $f:[(1_1^+)^2, 1_2^-]$ configuration, i.e.\ it has $(L-4)/2$ $fbbbf$ string complexes (\ref{stringFBBBF}), two additional real roots on the first level, and a single root with $\mathrm{Im}(\lambda^{(2)})=\pi/2$ on the second. The conformal spin of this excitation is $s=n_1 m_1=1$. For $L\to\infty$ its effective scaling dimension is
\begin{equation}
  \label{conj21-s1}
  X_{2,\frac12}^{1,0} = \Xi_{2,\frac12}^{1,0} - \frac{1}{4} = 1-\frac{3\gamma}{4\pi} + \frac{\pi}{4(\pi-\gamma)} -\frac14 \,
\end{equation}
with strong subleading corrections, see Figure~\ref{fig:21_other}.
Also shown in this Figure are two higher excitations with dimension $X = \Xi_{2,\frac12}^{1,0} - 1/4 + 1$ and spin $s=2$.  Their root configurations are best described in $bfbfb$ grading where both contain $(L-4)/2$ string complexes (\ref{stringBFBFB}).  
The full configurations are $b:[1_1^-,3_{12}^+]$ and$b:[1_1^+,(1_1^-)^2,1_2^-]$, respectively.
  
\begin{figure}[ht]
  \begin{minipage}{0.48\textwidth}
    \subfigure[]{\includegraphics[width=\textwidth]{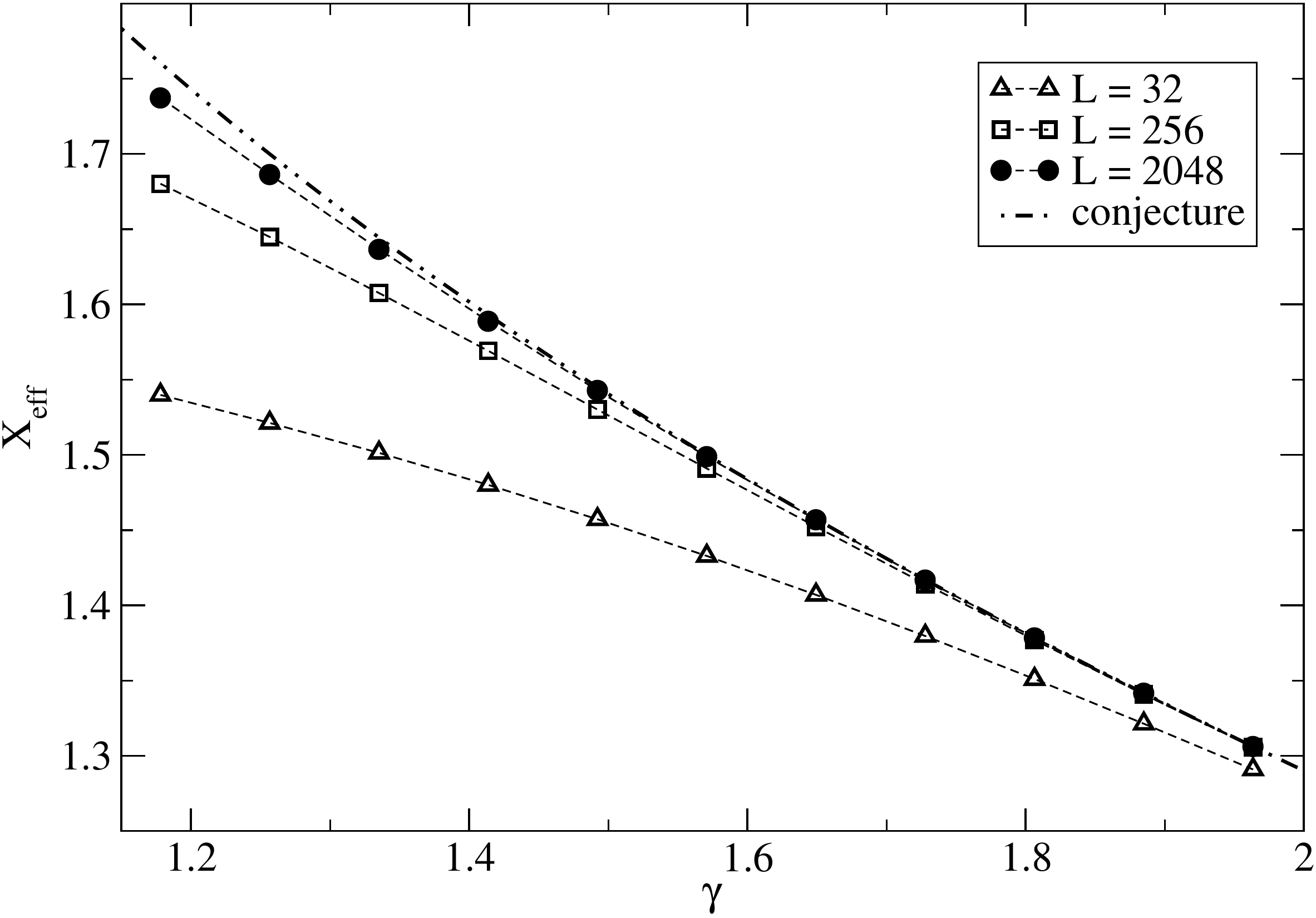}}
  \end{minipage}
  \begin{minipage}{0.48\textwidth}
    \subfigure[]{\includegraphics[width=\textwidth]{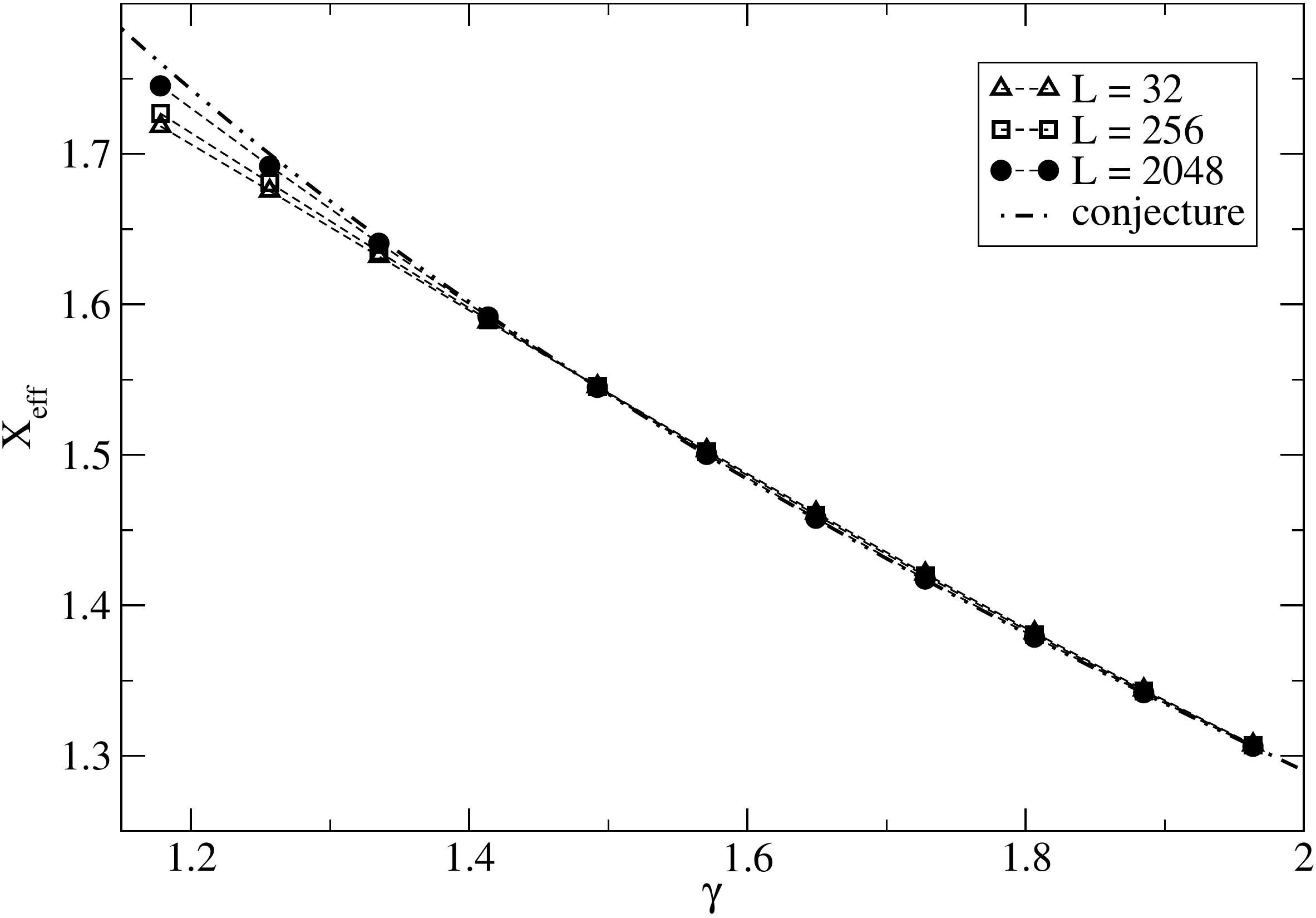}}
  \end{minipage}
  \caption{Effective scaling dimension for the states in sector $(2,1)$ conjectured to extrapolate to (\ref{conj21-s0.2.3}) with conformal spin (a) $s=1$ and (b) $s=0$ as a function of $\gamma$ for various system sizes.}
  \label{fig:21-s2+3}
\end{figure}

For two other excitations with root configurations $b:[1_1^+,(1_1^-)^2,1_2^-]$ we had to solve the Bethe equations for $3\pi/8 \leq \gamma \leq 5\pi/8$ to get the energies for sufficient large systems.  Based on these data, see Fig.~\ref{fig:21-s2+3}, we propose that their scaling dimensions extrapolate to \eqref{conj21-s0.2.3}. Their conformal spin is $s=1$ and $s=0$, respectively.  Given that there is a contribution $n_2 m_2/2=1/2$ from the $\gamma$ dependent part of the of the conformal weights we argue that their degeneracy is a consequence of the combination with an Ising energy operator, similar  as in (\ref{conj01-s1.5.7}).  Since $m_2\ne0$ these levels disappear from the low energy spectrum for $\gamma\to0$. 

\begin{figure}[ht]
  \includegraphics[width=0.7\textwidth]{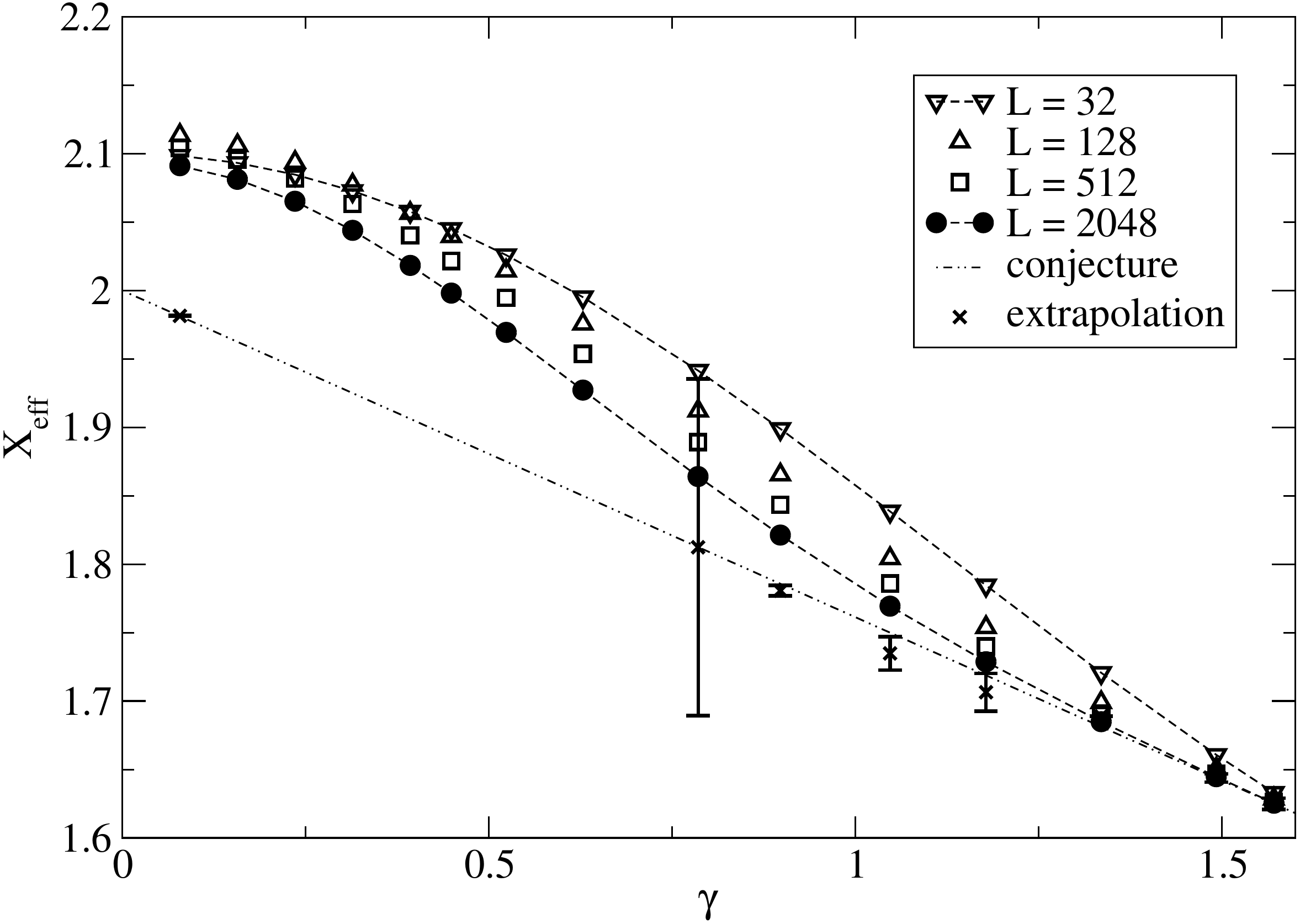}
  \caption{Effective scaling dimension for the state in sector $(2,1)$ state conjectured to extrapolate to (\ref{conj21-s6}) as a function of $\gamma$ for various system sizes.}
  \label{fig:21-s6-scaling}
\end{figure}

The last state we have studied in this sector has a $b:[(1_1^+)^2,1_1^-,1_2^-]$ root configuration.
This state has conformal spin $s=0$ and its scaling dimension extrapolates to
\begin{equation}
  \label{conj21-s6}
  X = \Xi_{2,0}^{1,0} + 1 = -\frac{3\gamma}{4\pi} + 2\,,
\end{equation}
see Figure~\ref{fig:21-s6-scaling}.

There is one remaining low energy excitation present in the spectrum of this charge sector for which the corresponding solution to the Bethe equations has been found only for $L\le16$, see Appendix~\ref{app:missing_states}.
In table~\ref{tab:sec21} we present a summary of our findings in the sector $(n_1,n_2) = (2,1)$.
\begin{table}[t]
\begin{ruledtabular}
\begin{tabular}{l|ccc|ccc|l}
     &  \multicolumn{3}{c|}{$X$} & \multicolumn{3}{c|}{$s$} & \\
 Eq. & $m_1$ & $m_2$ & $x_0$ & total spin & $\sigma_{n_1,m_1}^{n_2,m_2}$ & $s_0$ & remark\\\hline
(\ref{conj21-s0.2.3}) & $0$ & $1$ & $-\frac{1}{8} + \frac{1}{2}$ & $1,0$ & $\frac{1}{2}$ & $\pm \frac{1}{2}$ & Ising $(\frac{1}{2},0)$, $(0,\frac{1}{2})$\\
(\ref{conj21-s1}) & $\frac{1}{2}$ & $0$ & $-\frac14$ & $1$ & $1$ & $0$ &\\
(\ref{conj21-s6}) & $0$ & $0$ & $1 $ & $0$ & $0$ & $0$ & Ising $(\frac{1}{2}, \frac{1}{2})$\\
\end{tabular}
\end{ruledtabular}
\caption{\label{tab:sec21}Conformal data for the levels studied in charge sector $(n_1,n_2)=(2,1)$ (see also Table~\ref{tab:sec00}). 
We have also observed descendents of (\ref{conj21-s1}), see Fig.~\ref{fig:21_other}.}
\end{table}
\subsection{Sector $(2,2)$}
\label{sec:sub22}
Before we consider the low lying states in this sector let us recall our discussion at the end of Section~\ref{sec:sub00}: although there exist exceptions in the spectra obtained from exact diagonalization of small systems with $L\geq6$ sites we observe that many of the eigenenergies in sector $(2,2)$ appear also in the zero charges sector.  In fact this includes all of the levels discussed below.  We expect that the formation of such multiplets can be understood in the context of the $U_q[OSp(3|2)]$ symmetry in the presence of periodic boundary conditions.  This, however, is beyond the scope of this work.

We note that, in addition to the numerical evidence, this spectral inclusion is compatible with our hypothesis (\ref{conjecture_full}) for the effective scaling dimensions.  The latter implies
\begin{equation}
    X_{2,m_1}^{2,m_2} = X_{0,m_1}^{0,m_2} + 1\,.
\end{equation}
Similarly, the conformal spins according to (\ref{GAUSconjecture_spin}) are related as $s_{2,m_1}^{2,m_2} = s_{0,m_1}^{0,m_2} + 2m_1+m_2$.
Therefore, levels with $2m_1+m_2=1$ considered in this section may be considered either as primaries in charge sector $(2,2)$ or, alternatively, as descendents of a lower energy state with spin $s=0$ in sector $(0,0)$.

\begin{figure}[ht]
    \centering
    \includegraphics[width=0.7\textwidth]{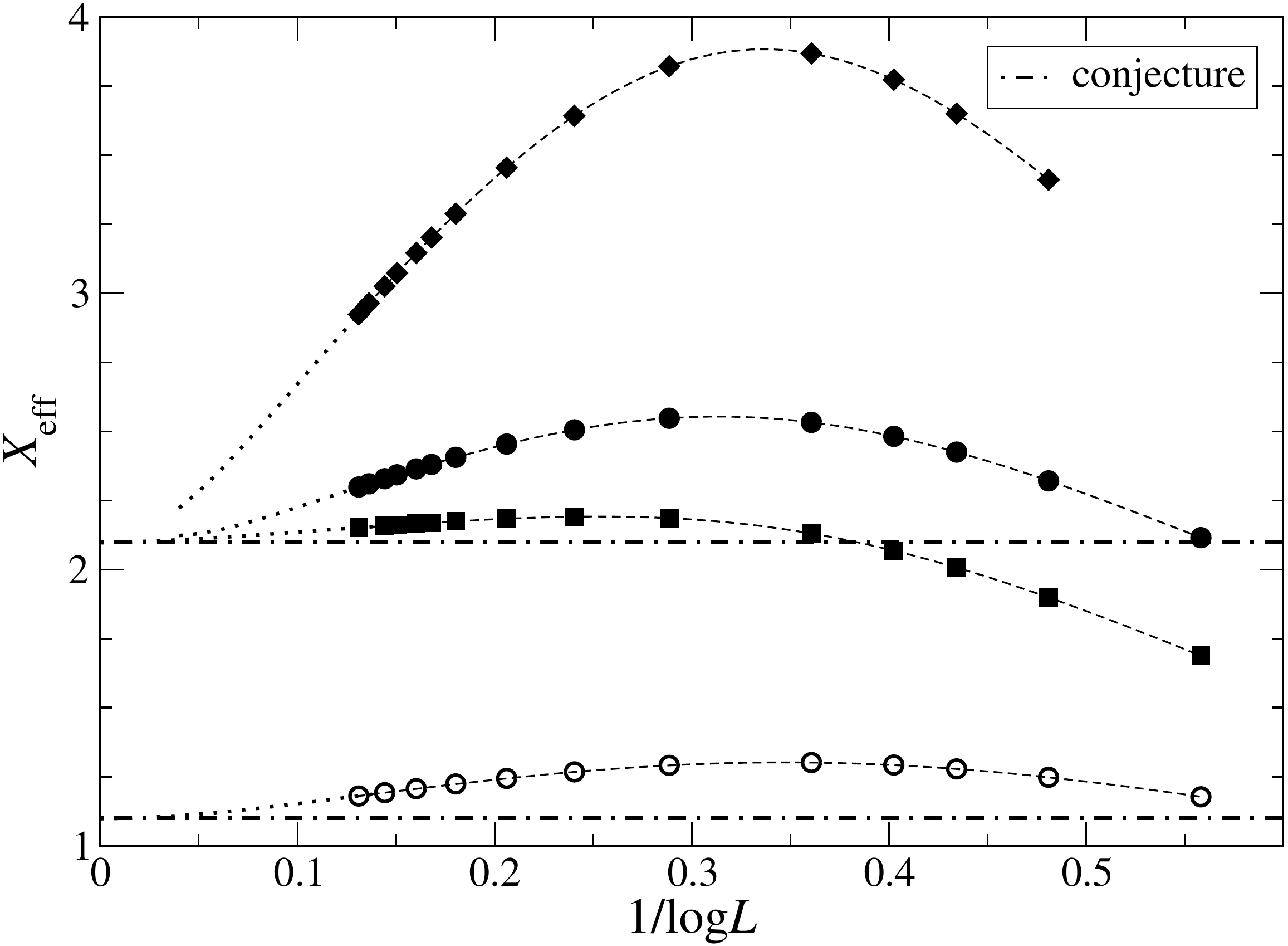}
    \caption{Similar as Fig.~\ref{fig:00_tower-scaling} for the lowest state in sector $(2,2)$ extrapolating to $X_{2,\frac12}^{2,0}$, Eq.~(\ref{conj22-s0}), and three higher levels extrapolating to $X=X_{2,\frac12}^{2,0}+1$, Eq.~(\ref{conj22-s5.X.X2}). The data shown are for $\gamma=2\pi/7$.}
    \label{fig:22-0th}
\end{figure}

\begin{figure}[p]
\centering
  \begin{minipage}{0.48\textwidth}
    \subfigure[]{\includegraphics[width=\textwidth]{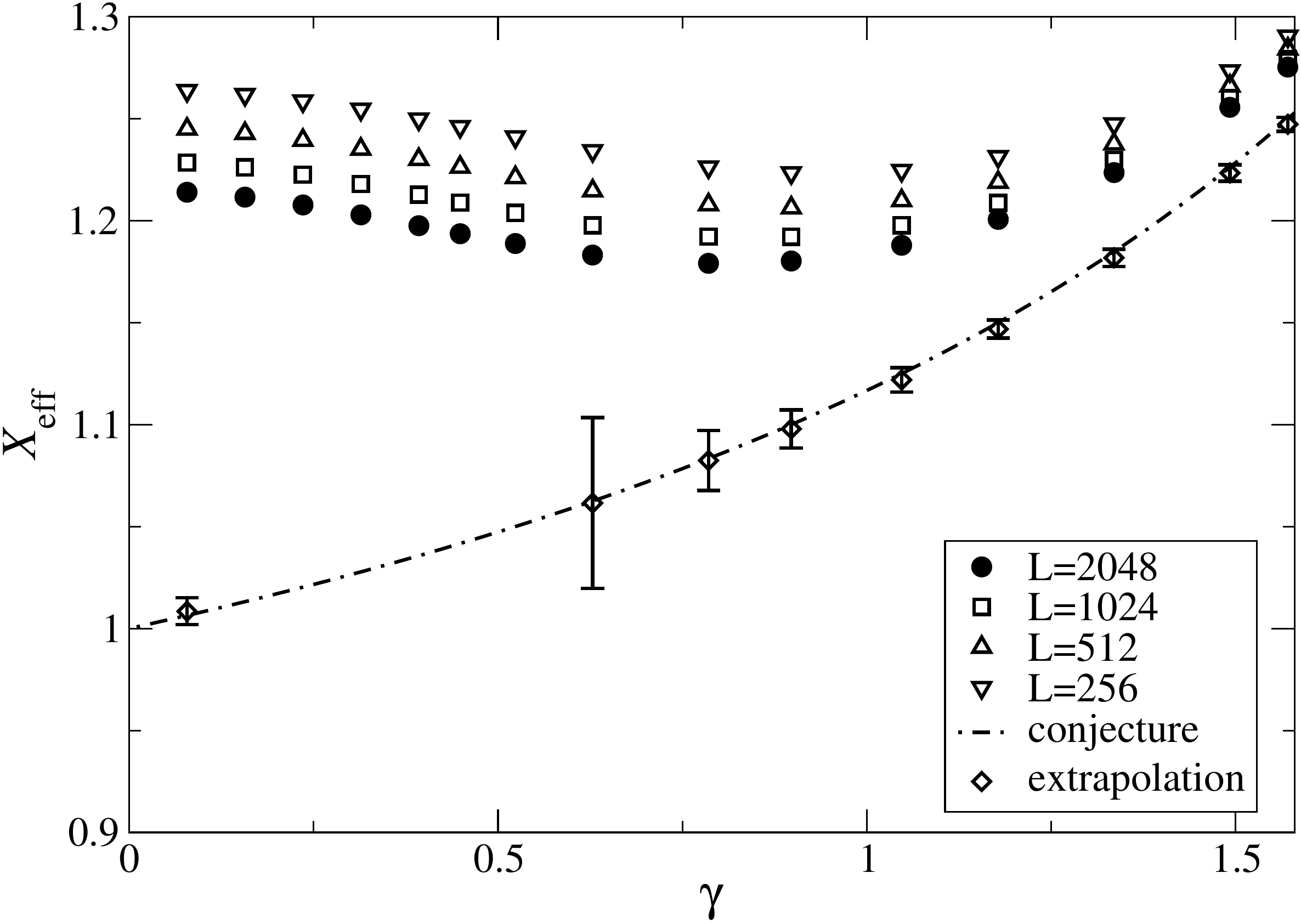}}
  \end{minipage}
  \addtocounter{subfigure}{2}
  \begin{minipage}{0.48\textwidth}
    \subfigure[]{\includegraphics[width=\textwidth]{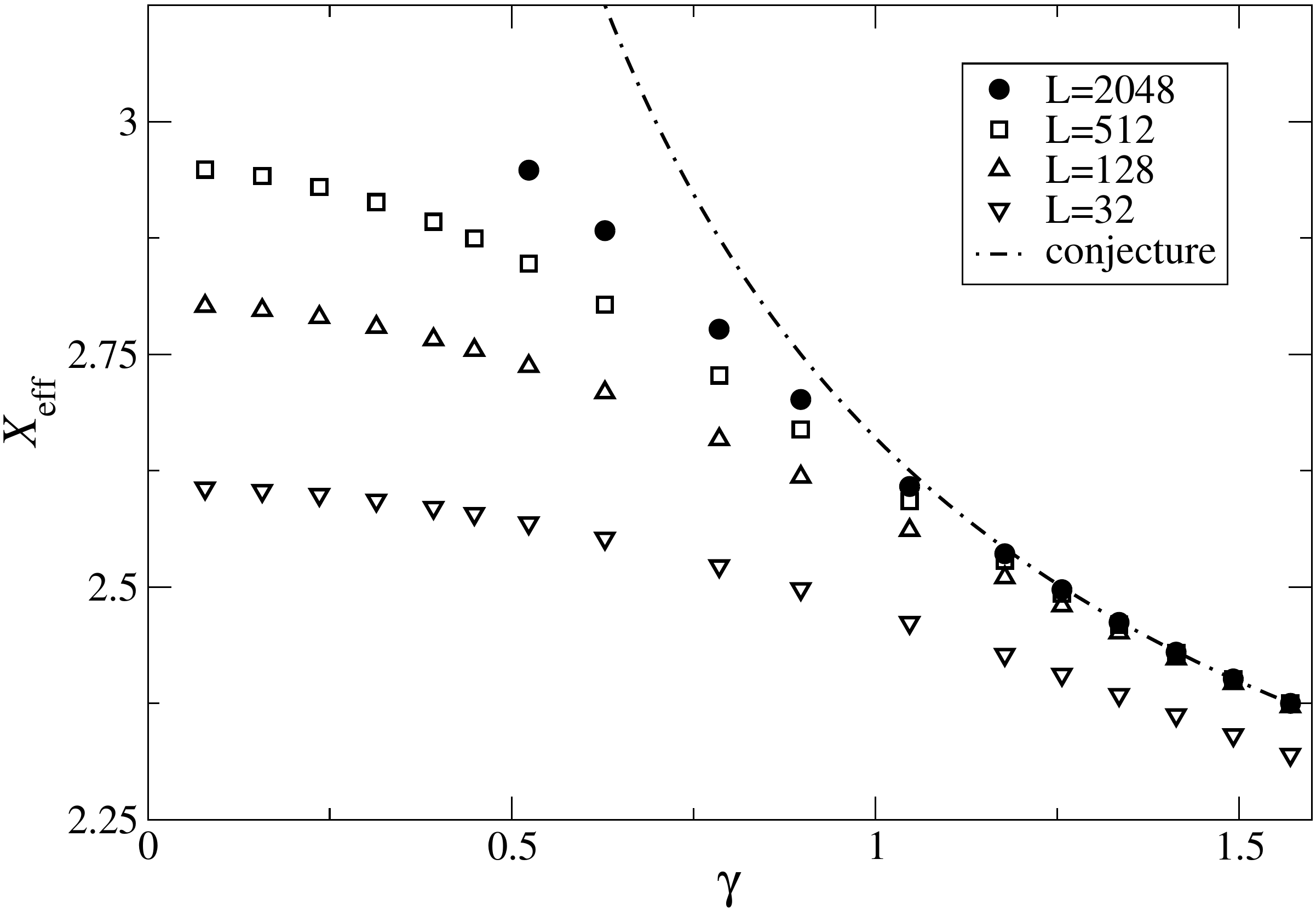}}
  \end{minipage}
  \addtocounter{subfigure}{-3}
  \begin{minipage}{0.48\textwidth}
    \subfigure[]{\includegraphics[width=\textwidth]{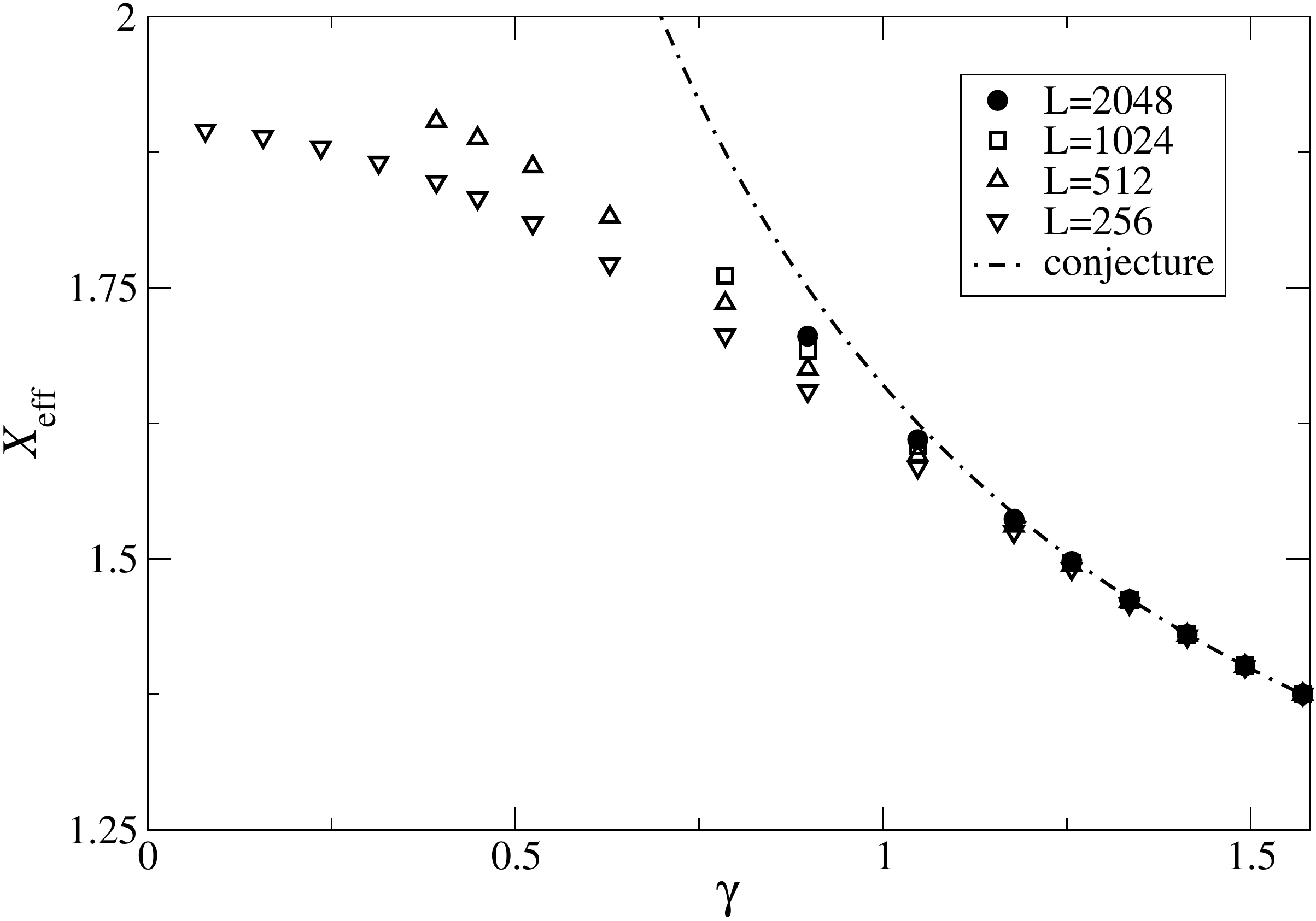}} 
  \end{minipage}
  \addtocounter{subfigure}{2}
  \begin{minipage}{0.48\textwidth}
    \subfigure[]{\includegraphics[width=\textwidth]{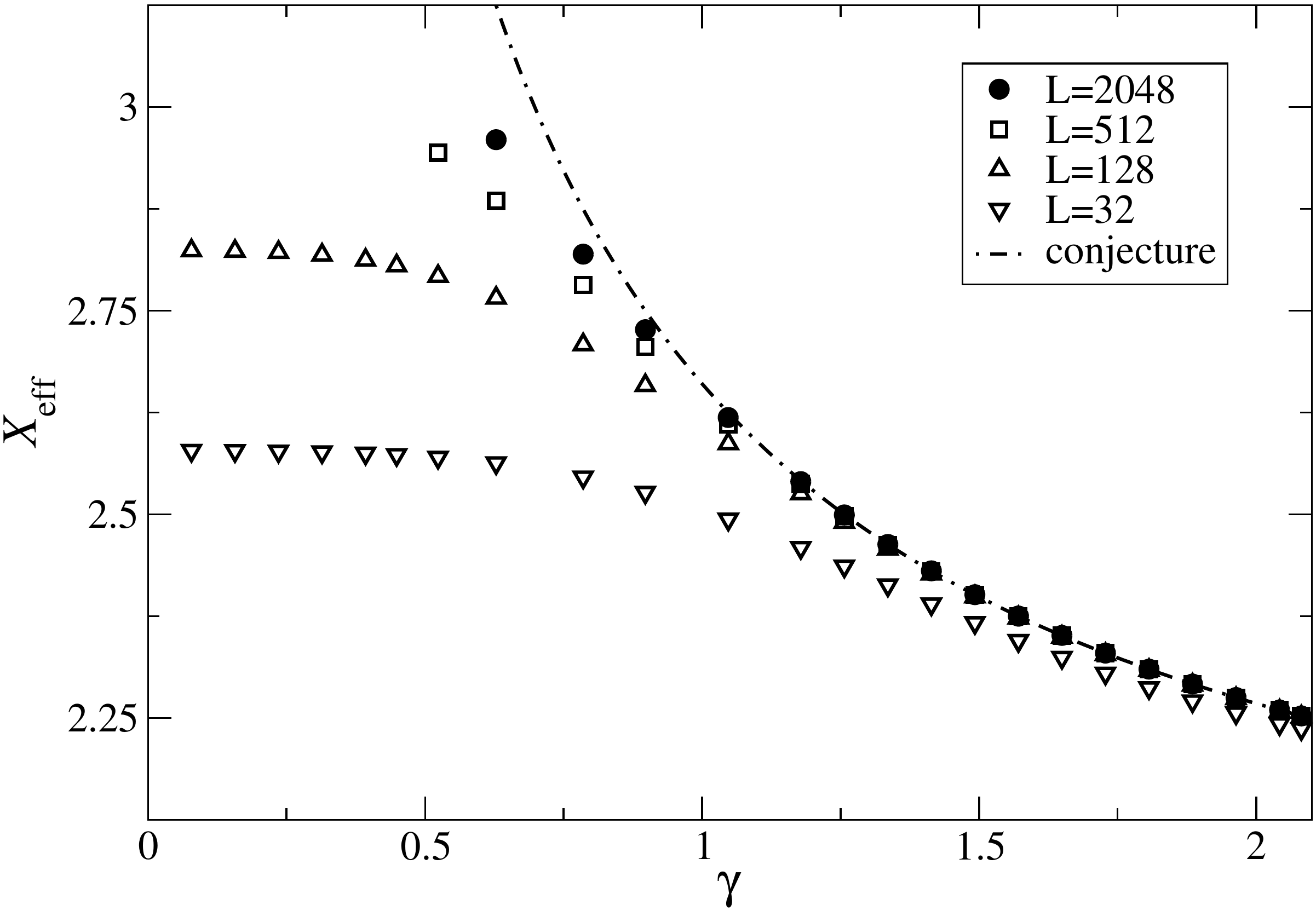}} 
  \end{minipage}
  \addtocounter{subfigure}{-3}
  \begin{minipage}{0.48\textwidth}
    \subfigure[]{\includegraphics[width=\textwidth]{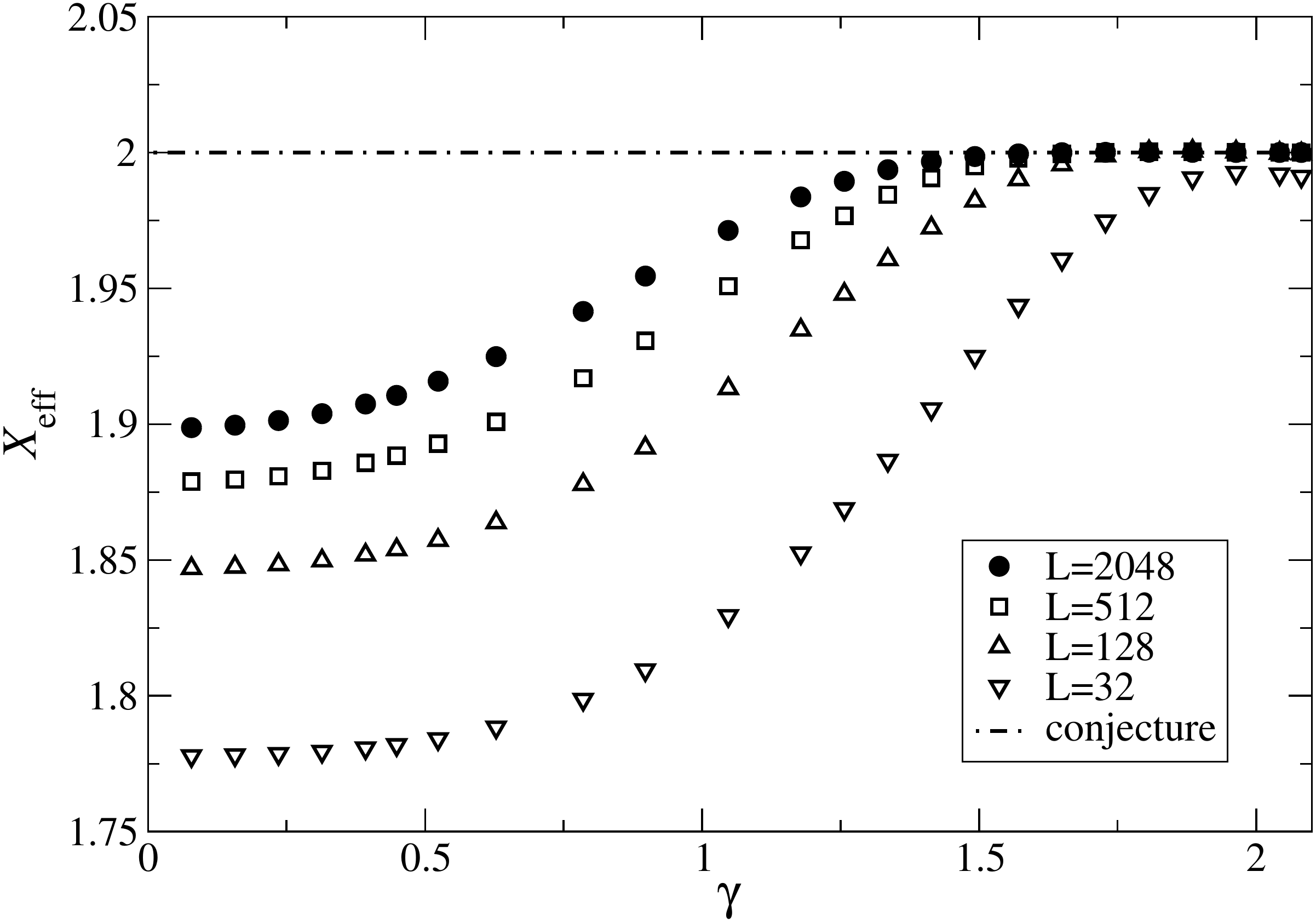}} 
  \end{minipage}
  \addtocounter{subfigure}{2}
  \begin{minipage}{0.48\textwidth}
    \subfigure[]{\includegraphics[width=\textwidth]{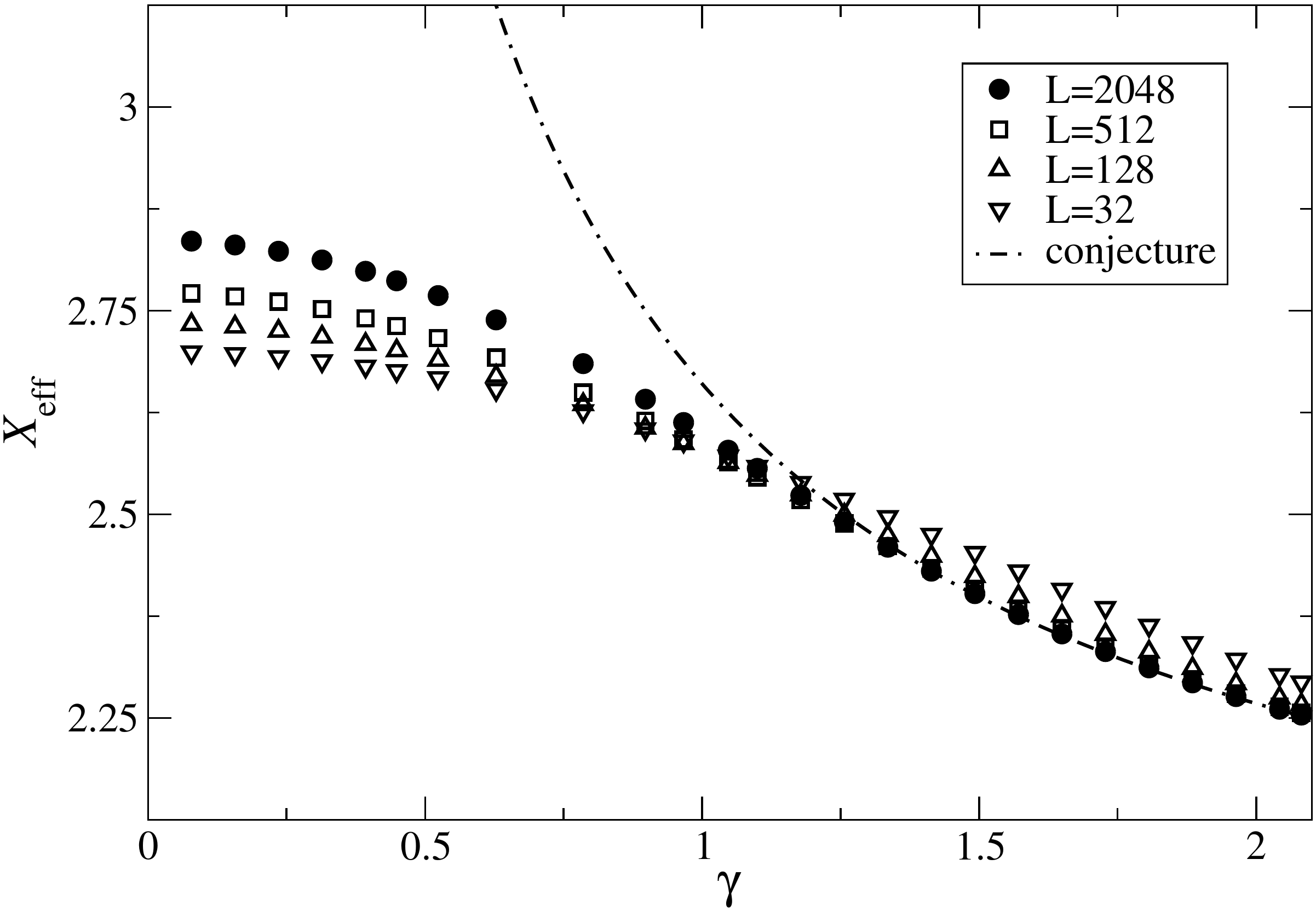}} 
  \end{minipage}
  \caption{Effective scaling dimensions of several low energy states in sector $(2,2)$ as a function of $\gamma$ for various system sizes: displayed in the left panel are (a) the spin $s=1$ ground state with effective scaling dimension extrapolating to $X^{2,0}_{2,1/2}$, Eq.~(\ref{conj22-s0}), in (b) the spin $s=1$ level extrapolating to $X^{2,1}_{2,0}+1$, Eq.~(\ref{conj22-s1}) and in (c) the $s=0$ level extrapolating to $X=2$, Eq.~\eqref{conj22-s2}. In the right panel the effective scaling dimension of three descendants of (b), Eq.~(\ref{conj22-s4.6.Z}) with spin (d) and (e) $s=2$ and (f) $s=0$, are shown. Dashed-dotted lines show the conjectured $\gamma$-dependence. In (a), the extrapolated data were calculated assuming a rational dependence of the finite-size data on $x=1/\log L$. Since the finite-size corrections become larger the extrapolation starts to fail for small $\gamma$.}
  \label{fig:22-xvsg}
\end{figure}

As mentioned in Section~\ref{sec:sub00} above, the lowest level in charge sector $(2,2)$ is part of a multiplet which also appears as an excitation in sector $(0,0)$.  Here its root configuration is $f:[(1_1^+)^2]$. It has conformal spin $s=1$ in agreement with (\ref{GAUSconjecture_spin}) and its effective scaling dimension extrapolates to
\begin{equation}
  \label{conj22-s0}
  X_{2,\frac12}^{2,0} = \Xi_{2,\frac12}^{2,0} - \frac{1}{4} = 1+ \frac{\pi}{4(\pi-\gamma)} -\frac14\,,
\end{equation}
see Figures~\ref{fig:22-0th} and \ref{fig:22-xvsg} (a).

Possible descendents of this level are described by root configurations $b:[1_1^+,1_1^-]$,  $f:[(1_1^+)^3,1_1^-,z_2]$, and $f:[(1_1^+)^5,(1_2^-)^2,3_{21}^-]$.  They have conformal spin $2$, $0$, and $0$, respectively, and their scaling dimensions extrapolate to
\begin{equation}
  \label{conj22-s5.X.X2}
  X = \Xi_{2,\frac12}^{2,0} - \frac{1}{4} +  1 = 1+\frac{\pi}{4(\pi-\gamma)} -\frac14 + 1\,.
\end{equation}
Their scaling behaviour is also shown in Figure~\ref{fig:22-0th}.

Another $s=1$ level described by a root configuration is $f:[(1_1^+)^2]$ gives an effective scaling dimension:
\begin{equation}
  \label{conj22-s1}
  X_{2,0}^{2,1} = \Xi_{2,0}^{2,1} -\frac{1}{8} = 1+ \frac{\pi}{4\gamma} -\frac18\,,
\end{equation}
see Figure~\ref{fig:22-xvsg}(b).
Again, we have identified three possible descendents of this level: their root configurations are $b:[1_1^+,1_1^-]$, $b:[(1_1^+)^2]$, and $f:[(1_1^+)^4,1_2^+,1_2^-]$.  The conformal spin of these descendents are $s=2$, $2$, and $0$, respectively and their scaling dimension is
\begin{equation}
  \label{conj22-s4.6.Z}
  X = \Xi_{2,0}^{2,1} - \frac{1}{8} + 1 = 1+\frac{\pi}{4\gamma} -\frac18 + 1\,,
\end{equation}
see also Figure~\ref{fig:22-xvsg} (d), (e), and (f).

Also in this sector we found a level corresponding to an operator with spin $s=0$ and scaling dimension 
\begin{equation}
  \label{conj22-s2}
  X = \Xi_{2,0}^{2,0} +1 = 2\,,
\end{equation}
see Figure~\ref{fig:22-xvsg}(c).
Its root configuration is $f:[1_1^+,1_1^-]$.

Apart from these states we have identified the Bethe roots for two additional levels: a spin $s=0$ excitation with root configuration $b:[1_1^+,1_1^-]$ or $f:[(1_1^+)^2]$ and a $s=1$ excitation described by roots $f:[1_1^+,1_1^-]$ or $b:[(1_1^+)^2]$. As the system size increases roots in these configurations degenerate, therefore we do not have conjecture for their finite-size scaling.

To conclude our investigation of the sector $(n_1,n_2) = (2,2)$ we show our results in table~\ref{tab:sec22}.
\begin{table}[t]
\begin{ruledtabular}
\begin{tabular}{l|ccc|ccc|l}
     &  \multicolumn{3}{c|}{$X$} & \multicolumn{3}{c|}{$s$} & \\
 Eq. & $m_1$ & $m_2$ & $x_0$ & total spin & $\sigma_{n_1,m_1}^{n_2,m_2}$ & $s_0$ & remark\\\hline
(\ref{conj22-s0}) & $\frac{1}{2}$ & $0$ & $-\frac{1}{4}$ & $1$ & $1$ & $0$ & \\
(\ref{conj22-s1}) & $0$ & $1$ & $-\frac{1}{8} $ & $1$ & $1$ & $0$ & \\
(\ref{conj22-s2}) & $0$ & $0$ & $1$ & $0$ & $0$ & $0$ & Ising $(\frac{1}{2}, \frac{1}{2})$\\
\end{tabular}
\end{ruledtabular}
\caption{\label{tab:sec22}Conformal data for the levels studied in charge sector $(n_1,n_2)=(2,2)$ (see also Table~\ref{tab:sec00}). We have also observed descendents of (\ref{conj22-s0}), see (\ref{conj22-s5.X.X2}), and descendants of (\ref{conj22-s1}), see (\ref{conj22-s4.6.Z}).}
\end{table} 

\section{Summary and Outlook}
\begin{figure}[ht]
  \includegraphics[width=0.7\textwidth]{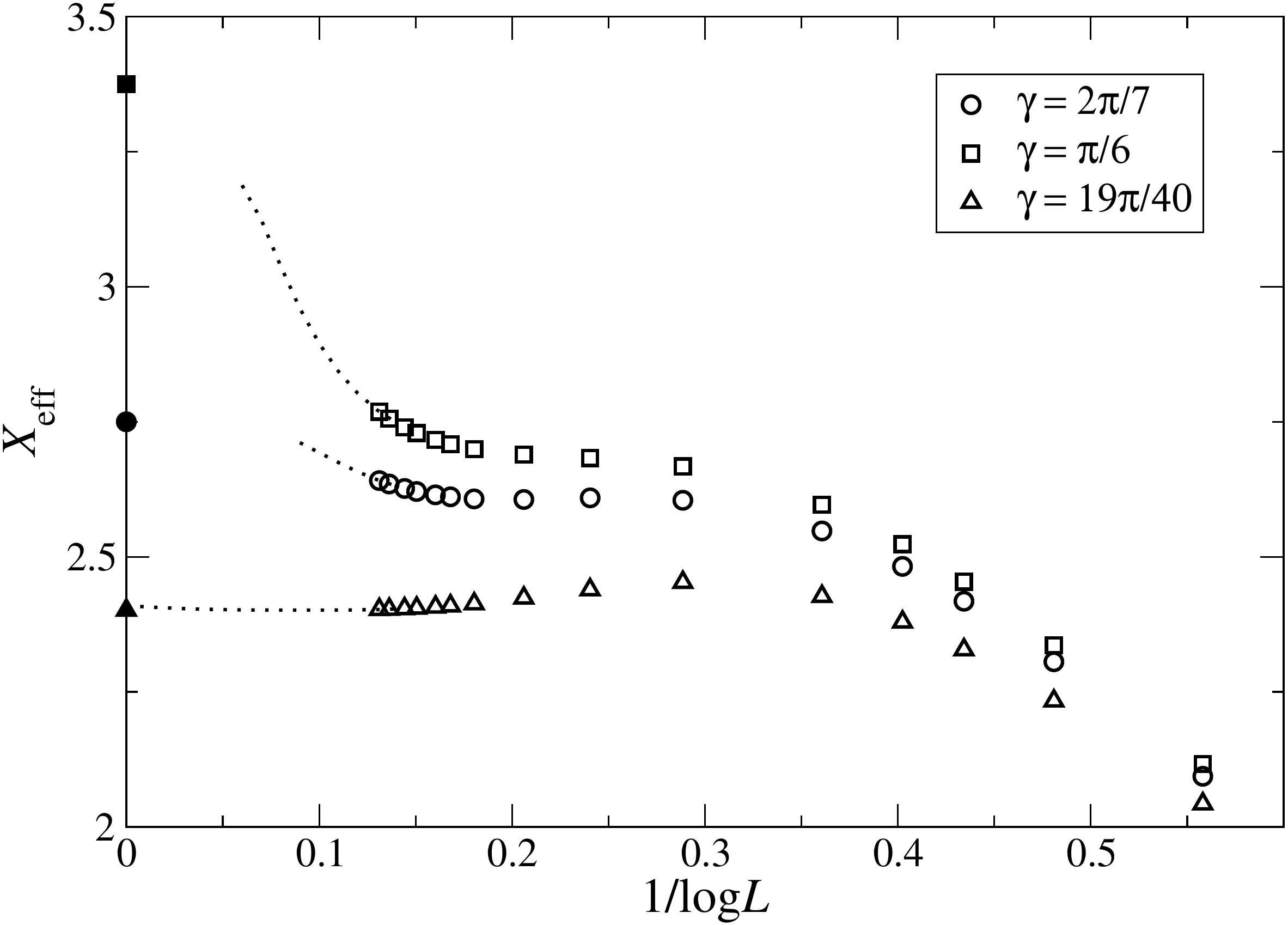}
  \caption{The difficulties with numerical extrapolation based on finite-size
    data in the presence of the strong corrections to scaling observed in some
    eigenenergies of the $U_q[OSp(3|2)]$ are evident in the scaling behaviour
    of one of the three descendants \eqref{conj22-s4.6.Z}, see fig.~\ref{fig:22-xvsg} (f): open
    symbols are data from the numerical solution of the Bethe equations for
    various values of $\gamma$ and system sizes up to $L=2048$. The dotted
    lines are extrapolations assuming a rational dependence of
    $X_{\text{eff}}$ on $1/\log L$.  While the radius of convergence of the
    latter may not be sufficient to read off the effective scaling dimensions
    for $L\to\infty$ the extrapolation is consistent with the conjectured
    values (filled symbols).  This picture indicates the difficulties in using
    standard numerical extrapolation techniques for the calculation of
    effective scaling dimensions for this model.}
  \label{fig:22-ZZth}
\end{figure}

In this paper we have reported the results obtained in a comprehensive finite-size study of the deformed $OSp(3|2)$ superspin chain.  We have identified the configurations of roots to the Bethe equations (\ref{betheBFBFB}) and (\ref{betheFBBBF}) for most of the lowest energy states. Taking these configurations as an input we have computed the corresponding eigenenergies as a function of the system size.  With data available for lattices up to several thousand sites, combined with insights from the root density approach or at $\gamma=\pi/2$ as discussed in Section~\ref{secFINITESIZE}, this allowed to compute the effective scaling dimensions even in the presence of very strong corrections to scaling, see e.g.\ Figure~\ref{fig:22-ZZth}.
There exist a few states where our solution of the Bethe equations has been
limited to several tens or a few hundreds of sites usually due to changes or
degenerations of the corresponding root configurations when the system size
was varied, see e.g.\ Appendix~\ref{app:roots}.  In these cases a reliable
extrapolation has not been possible.

For the majority of states, however, we have been able to extrapolate the
numerical data and found that the scaling dimensions can be described by our proposal, see (\ref{conjecture_full}),
\begin{equation}
  \label{conjecture_summary}
  X_{n_1,m_1}^{n_2,m_2} =
  n_1^2\,\frac{\pi-\gamma}{4\pi} + m_1^2\, \frac{\pi}{\pi-\gamma} 
  + n_2^2\,\frac{\gamma}{4\pi}   + m_2^2\, \frac{\pi}{4\gamma}
  +x_0\,.
\end{equation}
We note that modes with $m_2\ne0$ disappear from the low energy spectrum in
the isotropic limit $\gamma\to0$.  Such a behaviour has also been observed in
other superspin chains based on deformations of orthosymplectic superalgebras,
see e.g.\ \cite{GaMa07}.

Based on the finite size scaling of the states studied in this paper we find that $x_0$ takes values from a discrete set depending on the
quantum numbers $n_1$, $m_1$, $n_2$, and $m_2$ of the corresponding level.
The formulation of the general pattern of such constraints has 
eluded us so far, but we note that they should at least include the following rules (for even $L$)
  \begin{equation}
    \label{conj_x0}
    \begin{aligned}
      &\text{for~} n_1+2m_1 \text{~odd and~}\, m_2=0:\, && x_0=-\frac14\,,\\
      &\text{for~} n_1=0 \text{~and~}\, m_1=m_2=0:\, && x_0= 0\,,\\
      &\text{for~} n_1+n_2 \text{~even and~} m_2=1:\, && x_0 = -\frac18\,,\\
      &\text{for~} n_1+n_2 \text{~odd and~} m_2=1:\, && x_0 =  \frac38 \,.
    \end{aligned}
  \end{equation}
We recall that the rules for $m_2=0$ are consistent to what 
is expected for the
conformal spectrum of the isotropic $OSp(3|2)$ superspin chain \cite{FrMa15}.
The two possible values of $x_0$ observed for $m_2=1$ provide a hint
that the fields in the low energy effective continuum description of the model
are composites of Gaussian fields and an Ising operator.  The connection to
the integrable spin $S=1$ XXZ chain provides additional support for this
interpretation. At present we don't have a complete understanding of this feature
and it may require further studies beyond the scope of this paper.

Another characteristic feature of the conformal spectrum of the isotropic
model are macroscopic degeneracies in the thermodynamic limit $L\to\infty$.
Their presence appears to be a general feature of spin chains invariant under
the superalgebras $OSp(n|2m)$ \cite{FrMa18} and is consistent with the
expected low temperature behaviour of the related intersecting loop models
\cite{MaNR98,JaRS03}.
Here, i.e.\ in the anisotropic deformation of the $OSp(3|2)$ superspin chain
with general values of $\gamma$, we observe a similar feature:
in each of the charge sectors $(n_1,n_2)$ studied in this paper we have
identified groups of levels extrapolating to the same effective scaling
dimension. They are subject to strong corrections to scaling which vanish as a
function of $1/\log L$. In this work such
towers of levels have been found to appear on top of the dimensions:
\begin{equation}
  \label{conj_tower}
  X_{\text{tower}} = X_{n_1,m_1}^{n_2,0} =
  n_1^2\,\frac{\pi-\gamma}{4\pi} + m_1^2\, \frac{\pi}{\pi-\gamma} 
  + n_2^2\,\frac{\gamma}{4\pi} - \frac14\,
\end{equation}
for $(n_1,m_1)=(0,1/2)$, see Figures~\ref{fig:00_tower-scaling},
\ref{fig:01_tower-scaling}, \ref{fig:02_tower-scaling}, and $(n_1,m_1)=(1,0)$ as shown
in Figures \ref{fig:10_tower-scaling}, \ref{fig:11_tower-scaling} and
\ref{fig:12_tower-scaling}.

With these results we provide a first phenomenological picture of the finite
size spectrum of the deformed $OSp(3|2)$ superspin chain.
We emphasize that although most of
our numerical data are for anisotropies in the interval $0\le\gamma\le\pi/2$ we expect that
the proposal (\ref{conjecture_full}) also captures the behaviour of the conformal dimensions in the complementary region $\pi/2<\gamma<\pi$.  The confirmation of this expectation, however, requires a large amount of additional numerical work which is beyond the scope of this paper.
In addition, there are issues remaining which are not captured
by our conjecture:
for a complete understanding of the effective low energy theory the combined
presence of discrete levels (\ref{conjecture_full}) and a possible continuous
component in the conformal spectrum leading to the existence of towers of
levels starting at scaling dimensions (\ref{conj_tower}) in the lattice model
needs to be explained.  Another open question is why the states with lowest
energy of the lattice model, Eq.~(\ref{conj10-tower}), are found in a sector
with non-zero charge quantum numbers, i.e.\ $(n_1,n_2)=(1,0)$.

\begin{figure}[t]
  \begin{tikzpicture}[scale=2.,>=stealth]

    \tikzstyle{every loop}=[]
    \tikzstyle{every node}=[circle,draw,thin,fill=blue!40,minimum
    size=5pt,inner sep=0pt] 
    
    \node [label=below left:${(0,0)}$] at (0,0) {};
    \node [label=below right:${(0,\pi)}$] at (2.5,0) {};
    \node [label=above right:${(\pi,\pi)}$] at (2.5,1.5) {};
    \node [label=above left:${(\pi,0)}$] at (0,1.5) {};
    
    \draw[->, semithick]  (0.1,0) -- (2.4,0);
    \coordinate [label=below:$\phi_2$] () at (1.25,0);
    \draw[->, semithick]  (0,0.1) -- (0,1.4);
    \coordinate [label=left:$\phi_1$] () at (0,0.75);
    \draw[dashed]  (2.5,0.1) -- (2.5,1.4);
    \coordinate [label=right:{$S=1$ XXZ chain}] () at (2.6,0.85);
    \coordinate [label=right:{with twist $\varphi=\phi_1-\pi$}] ()
    at (2.6,0.65);

    \draw[->, semithick]  (0.3,0.3) -- (0.1,0.1);
    \coordinate [label=above right:{$\mathcal{R}$}] () at (0.3,0.3);
    
    \draw[->, semithick]  (0.3,1.2) -- (0.1,1.4);
    \coordinate [label=below right:{$\mathcal{NS}$}] () at (0.3,1.2);
    
  \end{tikzpicture}
  \caption{Models connected by variation of the twists (\ref{boundMA}):
    the symbol $\mathcal{R}$ ($\mathcal{NS}$) denote periodic (anti-periodic) boundary conditions for the fermionic degrees of freedom of the $U_q[OSp(3|2)]$ superspin chain in analogy to the Ramond (Neveu-Schwarz)
    sector of conformal field theories.  On the dashed line for $\phi_2=\pi$ the spectrum of the superspin
    chain contains the eigenenergies of the integrable spin $S=1$ Heisenberg
    chain. }
  \label{fig:twist-phase-diagram}
\end{figure}
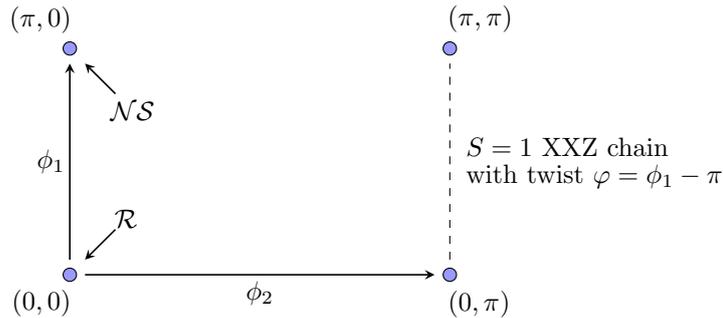

In other models showing such peculiar features studies of the spectral flow under the change of toroidal boundary conditions have provided further insights \cite{VeJS14,FrHo17}.
For the
$U_q[OSp(3|2)]$ superspin chain this amounts to an extension of the
finite-size analysis presented in this paper to its integrable modification
obtained by including the generic toroidal twists (\ref{boundMA}).  Following
the evolution of the low energy levels under varying twist angles
$(\phi_1,\phi_2)$ one can connect the superspin chain to several, closely
related models and thereby obtain additional evidence supporting the
conjecture for the conformal spectrum, see
Figure~\ref{fig:twist-phase-diagram}.
An exhaustive discussion of these relations is beyond the scope of this work.
Here we limit ourselves to list a few observations, mostly concerning the
lowest state in the zero charge sector:

We have pointed out already in our discussion of the Bethe ansatz solution in
$fbbbf$ grading, Eqs.~(\ref{betheFBBBF}), that the spectrum of the
$U_q[OSp(3|2)]$ chain contains the eigenenergies of the integrable $S=1$ XXZ
Heisenberg model on the line $(\phi_1,\phi_2)=(\pi+\varphi,\pi)$.  The
Virasoro central charge of the latter model
varies between
$c=3/2$ for periodic ($\varphi=\pi-\phi_1=0$) and $c=0$ for anti-periodic
($\varphi=\pi-\phi_1=\pi$) boundary conditions for the spin-$1$ chain, see
Appendix~\ref{appS1}.
The energy of the latter state has no corrections to scaling and does not
change under the spectral flow $(\phi_1,\phi_2)=(0,\pi)\to(0,0)$. It therefore
connects to that of the lowest state in the zero charge sector of the periodic
superspin chain, Eq.~(\ref{e00exact}), that we have used as reference state for our finite-size
analysis.

\begin{figure}[t]
  \includegraphics[width=0.7\textwidth]
  {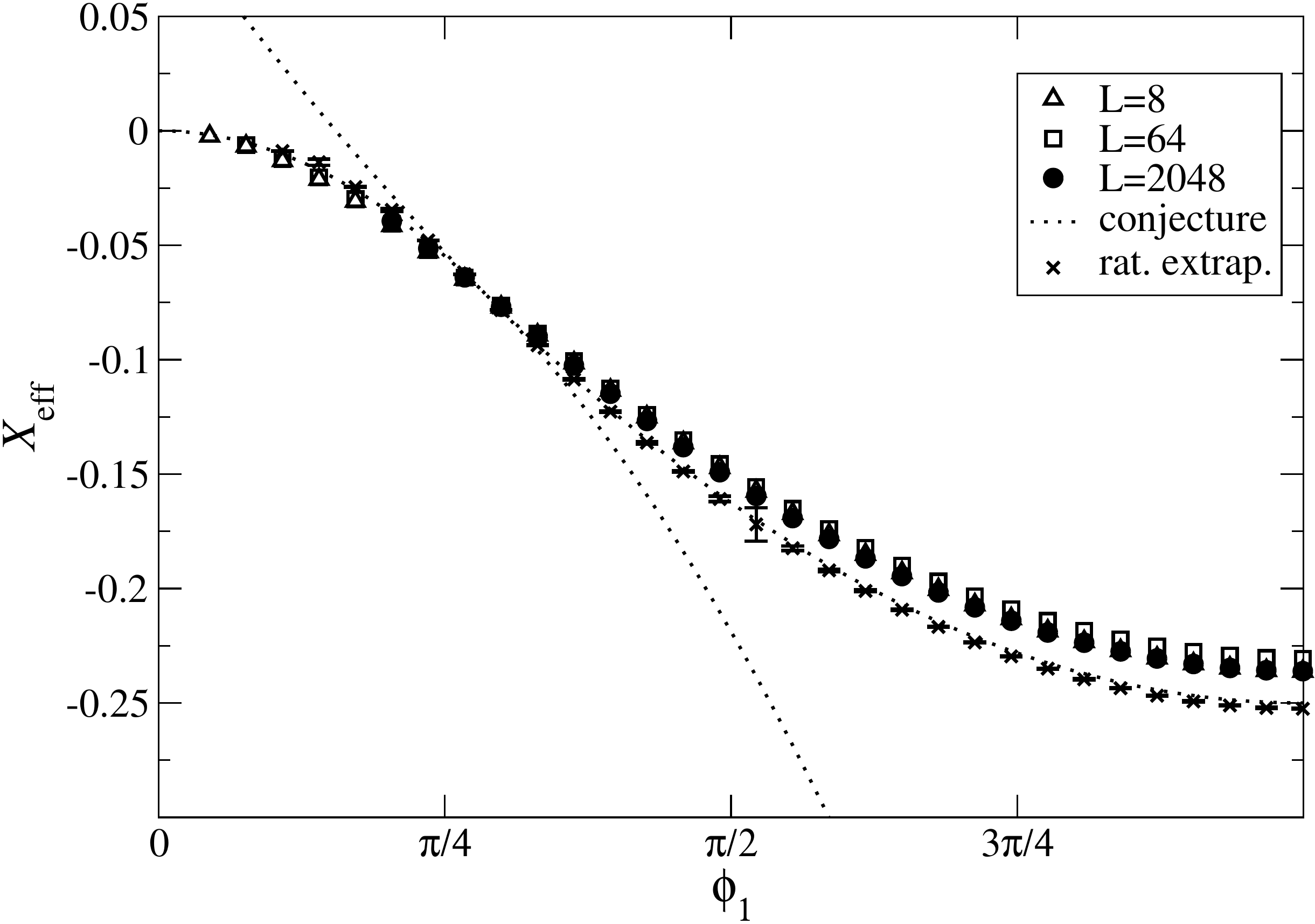}
  \caption{Spectral flow of the lowest level in the charge sector $(0,0)$ as
    function of the twist angle $\phi_1$ for $\gamma = 2\pi/7$. The effective
    scaling dimension for $\phi_1=0$ ($\pi$) appears in the Ramond
    (Neveu-Schwarz) of the low energy effective theory, respectively.  Symbols are finite-size data from the solution of the Bethe equations (\ref{betheFBBBF}), lines show the conjectured analytical behaviour (\ref{conj00-twist}).}
  \label{fig:sec00-twist}
\end{figure}

The variation of the twist along the line $(\phi_1,\phi_2)=(0,0)\to(\pi,0)$
corresponds to an adiabatic change of the boundary conditions for the
fermionic degrees of freedom from periodic to anti-periodic ones. In the field
theory describing the continuum limit of the superspin chain this corresponds
to the Ramond ($\mathcal{R}$) and Neveu-Schwarz ($\mathcal{NS}$) sector,
respectively.  The spectral flow connects the reference state with
energy (\ref{e00exact}) and  the zero charge ground state of the lattice model for twists $(\phi_1,\phi_2)=(\pi,0)$.  In the $fbbbf$ grading the latter is parameterized by roots arranged in $L/2$ complexes (\ref{stringFBBBF}). As shown in Figure~\ref{fig:sec00-twist} for $\gamma=2\pi/7$ its effective scaling dimension extrapolates to $X_{0,\text{eff}}^{\mathcal{NS}}=-1/4$ which coincides with the observation of a central charge $c_{\mathcal{NS}}=3$ in the isotropic model 
\cite{MaNR98,JaRS03,FrMa15}. 
Curiously, following the scaling as a function
of the twist $\phi_1$ we find different analytical expressions near
$\phi_1=0$ and $\pi$, respectively
\begin{equation}
  \label{conj00-twist}
  X(\phi_1) = \begin{cases}
    X_{\text{eff}}^{\mathcal{R}}(\phi_1)=
    -\frac{\pi}{4\gamma}\,\left(\frac{\phi_1}{\pi}\right)^2 \,, &
    0\le\phi_1\le \gamma\,,\\
    X_{\text{eff}}^{\mathcal{NS}}(\phi_1)=
    -\frac14 +
    \frac{\pi}{4(\pi-\gamma)}\,\left(\frac{\pi-\phi_1}{\pi}\right)^2\,, &
    \gamma\le\phi_1\le \pi\,.
  \end{cases}
\end{equation}
Note that the dependence on the twist near $\phi_1=\pi$ can be related to the
vortex contribution $m_1$ in our proposal for the scaling dimensions
(\ref{conjecture_summary}).  Near $\phi_1=0$, however, the flow under the
twist resembles the $\gamma$-dependence or the $m_2$ vortices albeit with the
'wrong' sign. Now suppose 
that it is possible to
extend the amplitude $X_{\text{eff}}^{\mathcal{R}}(\phi_1)$ to the 
twist angle domain 
of the Neveu-Schwarz sector by means of a well defined analytical 
continuation procedure. Under this hypothesis we observe that
$X_{\text{eff}}^{\mathcal{R}}(\pi)  
=-\pi/4\gamma$ is in fact smaller than the lowest observed scaling dimension
in the Neveu-Schwarz sector, i.e $X_{\text{eff}}^{\mathcal{NS}}(\pi)=-1/4$. 
Following the arguments of Refs.~\cite{EsFS05,VeJS14,FrHo17} we may speculate that
this can be taken as an indication for the presence of operators in the non-unitary effective field theory for the Neveu-Schwarz sector that correspond to non-normalizable states and therefore are absent in the spectrum of the lattice model.  To put this on firm ground, however, further studies are
required.

\begin{acknowledgments}
  Funding for this work has been provided by the research unit
  \emph{Correlations in Integrable Quantum Many-Body Systems} (FOR2316).
  Funding by the Deutsche Forschungsgemeinschaft under grant No.\ Fr~737/9-1
  is gratefully acknowledged.  M.\,J.\,Martins thanks the Brazilian research
  agency CNPq under the projects 304798-2017/401694-2016 for partial financial
  support.
\end{acknowledgments}
\newpage
\appendix

\section{The integrable spin $S=1$ XXZ model}
\label{appS1}
We start by recalling the Bethe ansatz solution of the integrable $S=1$ XXZ model with generic toroidal boundary conditions \cite{ZaFa80}. The Hilbert space of this model can be separated in the disjoint sectors labeled by the total magnetization $S^z=n$.  The spectrum of the lattice model with $L$ sites in this sector is parameterized by $L-n$ roots $\mu_j$ of the Bethe equations
\begin{equation}
\label{S1bethe}
    \left[\frac{\sinh(\mu_{j}+i\gamma/2)}
    {\sinh(\mu_{j}-i\gamma/2)}\right]^{{L}}=
    \mathrm{e}^{i\varphi} \prod_{\stackrel{k=1}{k \neq j}}^{L-n}
    \frac{\sinh(\mu_{j}-\mu_{k}+i\gamma/2)}
    {\sinh(\mu_{j}-\mu_{k}-i\gamma/2)},\quad j=1,\cdots,{L-n}\,,
\end{equation}
where $0\le\varphi\le\pi$ corresponds to the angle of rotation around the $z$-axis.  Given a solution $\{\mu_j\}$ to Eqs.~(\ref{S1bethe}) the corresponding eigenenergy is given by
\begin{equation}
\label{S1energy}
  E^{\mathrm{XXZ}}_{n}(L,\varphi)= \sum^{L-n}_{j=1} \frac{2 \sin\gamma}{\cos\gamma-\cosh(2\mu_j)}\,.
\end{equation}
The analysis of the system in the thermodynamic limit, $L\to\infty$,  is facilitated by the fact that the roots of the Bethe equations can be grouped into sets forming so-called strings \cite{TaSu72,KiRe87}.  For $0\le\gamma<\pi$ the root configurations for the ground state and low lying excitations are dominated by pairs of complex conjugate rapidities
\begin{equation}
\label{S1string}
    \mu_{j\pm} = \xi_j\pm i\frac{\gamma}{4}\,
\end{equation}
continuously distributed along the real $\xi$-axis.  Based on this observation one can use the root density formalism \cite{YaYa69} to compute the density of these two-strings \cite{Sogo84}
\begin{equation}
\label{S1dens}
    \sigma(\xi) = \frac{1}{\gamma}\, \frac{1}{\cosh\left(2\pi\xi/\gamma\right)}\,
\end{equation}
and the ground state energy per site 
\begin{equation}
\label{S1einf}
    \varepsilon_\infty = \lim_{L\to\infty} E_0(L)/L = -2\cot\frac{\gamma}{2}\,.
\end{equation}
In this approach the elementary low-lying excitations over this ground state are found to have dressed energy and momentum
\begin{equation}
\label{S1disp}
    \varepsilon(\xi) = 2\pi\sigma(\xi)\,,\quad
    p(\xi) =\int_\xi^\infty \mathrm{d}x\,\varepsilon(x) \,,
\end{equation}
giving a dispersion $\varepsilon(p)\sim v_F\, |\sin p|$ with Fermi velocity $v_F=2\pi/\gamma$.  The complete finite-size spectrum of this model has been studied using a combination of analytical and numerical methods giving \cite{AlMa89,AlMa90,FrYF90}
\begin{equation}
\label{S1fse}
    E_{n,m}^{\mathrm{XXZ}}(L,\varphi) - L\varepsilon_\infty = \frac{2\pi v_F}{L} \left[ n^2\, \frac{\pi-\gamma}{4\pi} +\left(m+\frac{\varphi}{\pi}\right)^2 \frac{\pi}{4(\pi-\gamma)} + X_I(r,j)
    -\frac{c}{12}\right]\,.
\end{equation}
The conformal field theory describing the low energy behaviour of the integrable spin $S=1$ XXZ chain has central charge $c=3/2$ \cite{BaTs86}.  The operators of this theory are given in terms of composites of a $U(1)$ Kac-Moody field with charge $n$ and vorticity $m$ and an Ising (or $Z(2)$) operator with scaling dimension $X_I(r,j)$. The latter take values
\begin{equation}
  X_I(0,0)\in\{0,1\}\,,\quad X_I(1,0) = X_I(0,1) = \frac18\,,\quad X_I(1,1)=\frac12\,,
\end{equation}
depending on $n$, $m$ and the parity of the system size through the selection rules $r=n+L \bmod 2$, $j=m+L \bmod 2$.  In particular, $X_I=1/8$ for $n+m$ odd.
Note that within the root density approach based on the string hypothesis (\ref{S1string}) only the contributions from the Kac-Moody field to the finite-size energies (\ref{S1fse}) are obtained \cite{VeWo85,Hame86}.  The differences between the true root configurations solving the Bethe equations (\ref{S1bethe}) and the string hypothesis add up to give the Ising part $X_I$ \cite{AlMa89}.

\section{Degeneration of root configurations}
\label{app:roots}
Usually, the finite-size analysis of a particular level in a Bethe ansatz
solvable lattice system relies on the fact that the roots to the Bethe
equations form characteristic patterns which allow to characterize this level
uniquely for any finite system and in the thermodynamic limit $L\to\infty$.
For most of the low energy states studied in this paper this is also true for
the $OSp(3|2)$ superspin chain.  However, we have encountered a number of
situations in which at a finite lattice size $L_*$ some of the Bethe roots
either diverge or degenerate leading to a qualitative change in the
corresponding root pattern.
It turns out that it was not always possible to identify the new
pattern of roots in order to follow the state for larger systems
sizes.  Moreover, such degeneration occurs for sizes which cannot be
reached by Hamiltonian exact diagonalization preventing us to make the
identification of the new pattern of roots.
In this appendix we present such degenerations which have been observed in our finite-size studies in more detail.

\begin{figure}[ht]
   \includegraphics[width=0.9\textwidth]{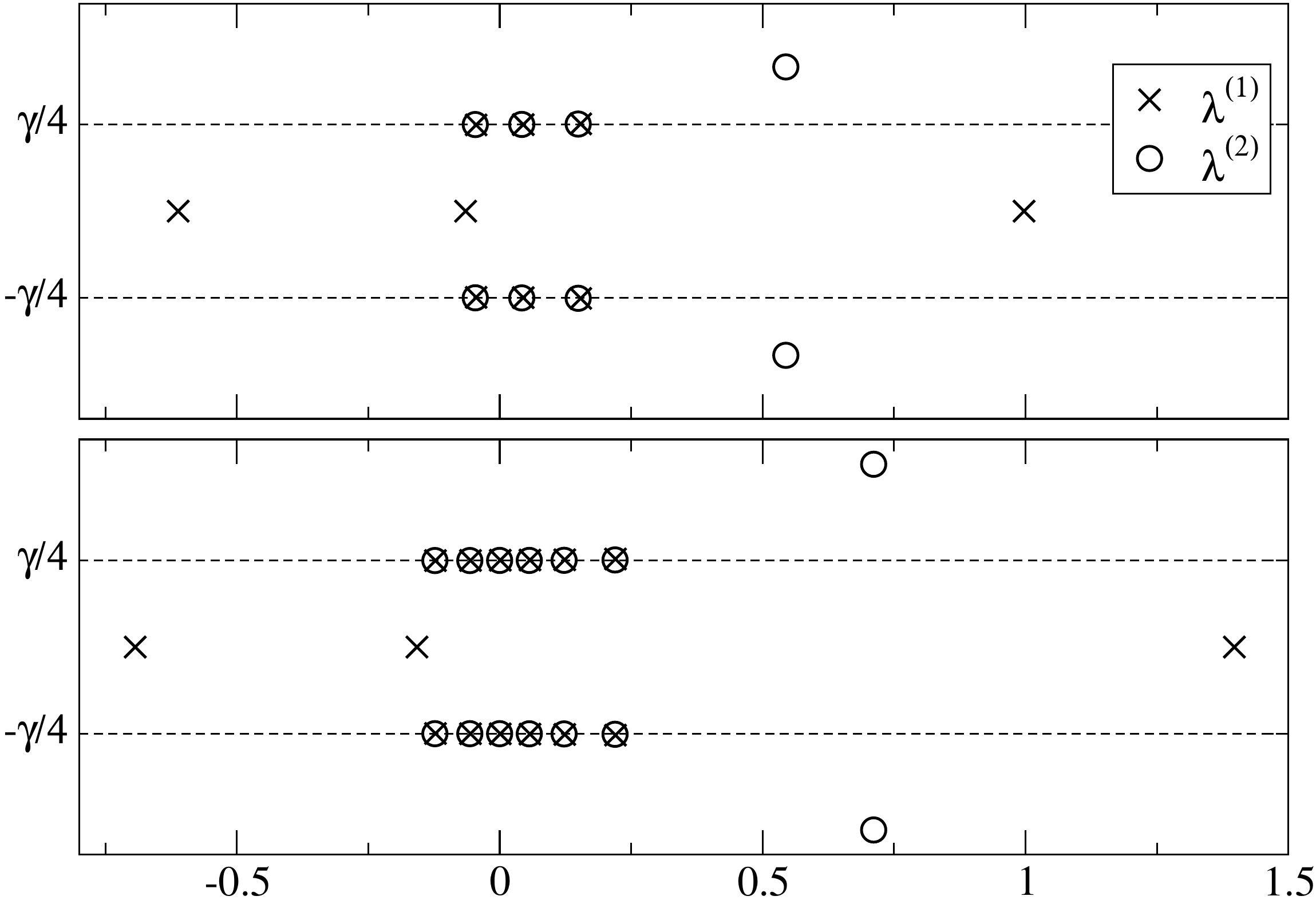}
   \caption{Finite part of the $fbbbf$ root configurations for the spin $s=1$
     state in sector $(0,0)$ for $L = 10$ (top) and $16$ (bottom) and
     $\gamma = 2\pi/7$.
     As $L$ is increased we note the growth of the real part of one of the
     roots at first level and of the real part of a complex pair at the second
     level.}
  \label{fig:roots_00-s6}
\end{figure}

At the end of Section~\ref{sec:sub00} we have discussed a level in charge sector $(0,0)$ where we have found that the corresponding Bethe root configuration changes as the system size is increased.  For small $L$ the $fbbbf$ roots for this state are arranged as $f:[(1_1^+)^3,z_2] \oplus[1_1,2_2]_{-\infty}$, the finite ones are shown in Figure~\ref{fig:roots_00-s6} for $L=10$ and $16$ and anisotropy $\gamma=2\pi/7$.  
As can be seen from this Figure three of the roots which are not part of the complexes (\ref{stringFBBBF}), namely one of the real first level roots $[1_1^+]$ and the pair of complex conjugate level-$2$ roots, $[z_2]$, increase considerably as the system size grows.  We can follow this behaviour based on our numerical solution of the Bethe equations (\ref{betheFBBBF}) up to some finite system size $L_*$ which depends on the anisotropy, e.g.\ $L_*=26$ for the parameters used in Figure~\ref{fig:roots_00-s6}. Beyond $L_*$ the root configuration degenerates and it is likely that it changes forming a different pattern.  In principle it might be possible to identify such a new pattern by solving the Bethe equations for $L\gtrsim L_*$. However, whether this describes an eigenstate of the superspin chain cannot be checked since an exact diagonalization of the Hamiltonian for system of that size is not feasible. 
In some cases it may be possible to avoid the degenerations described above by working in the other grading.  Here, however, the $bfbfb$ root configuration degenerates at the same system size $L_*$.  As a consequence of this scenario we do not have sufficient data for a reliable finite-size analysis.

A second example for a state where the patterns formed by the Bethe roots
changes with the system size has been observed in the charge sector $(1,0)$.
\begin{figure}[ht]
   \includegraphics[width=0.9\textwidth]{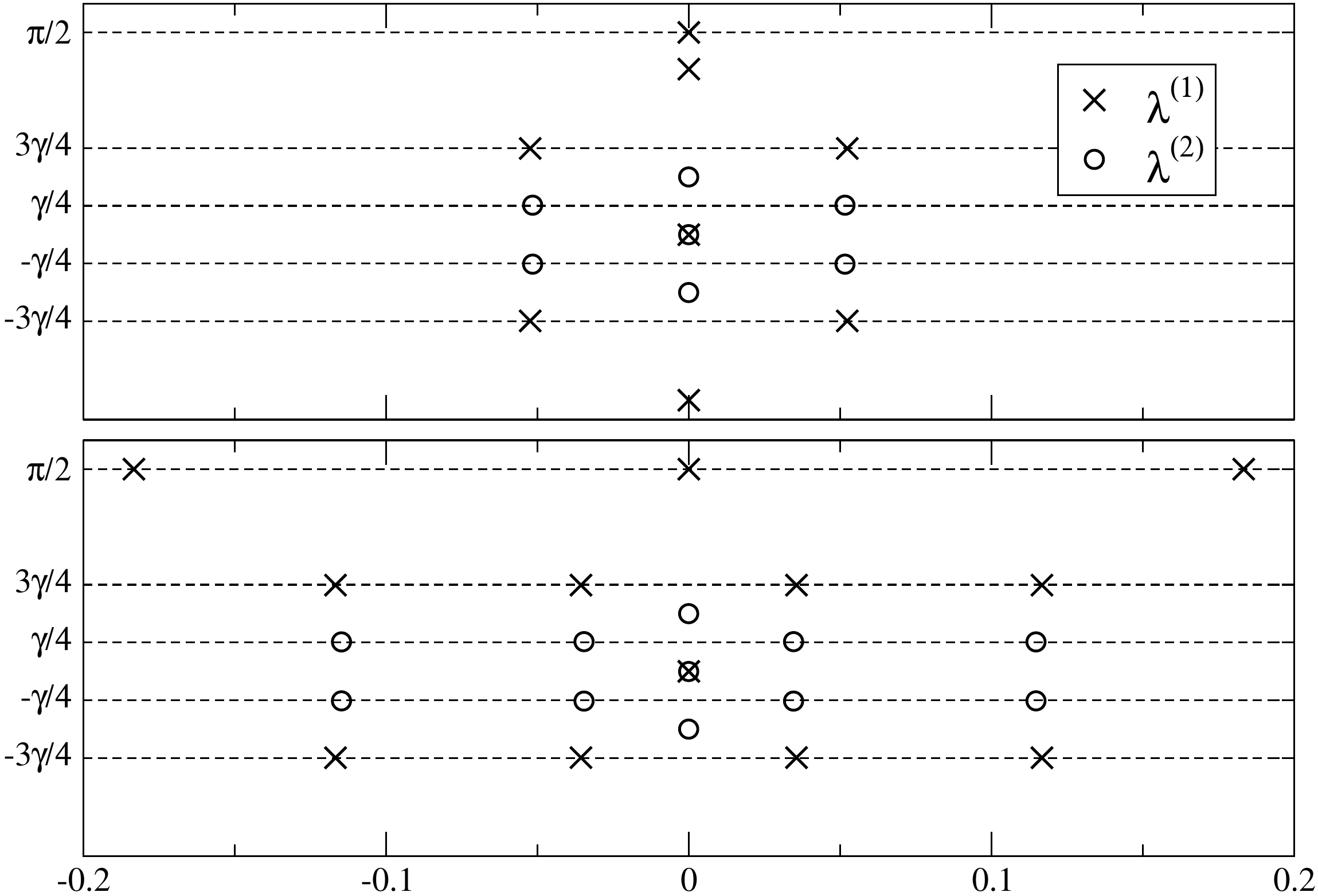}
   \caption{Degeneration of Bethe roots for a state in charge sector $(1,0)$:
     for anisotropy $\gamma=2\pi/7$ the root configuration changes from
     $b:[1_1^-,1_2^+,3_{21}^+,z_1]$ for system size $L = 10$ (top) to
     $b:[(1_1^-)^3, 1_2^+, 3_{21}^+]$ for $L=12$ (bottom).  }
  \label{fig:roots_10-s3}
\end{figure}
In Fig.~\ref{fig:roots_10-s3} we show how the $bfbfb$ root configuration of
this state changes when the system size is increased.  We first note that a
complex pair $[z_1]$ at the first level degenerates at $L_{*}=12$ into the
root pattern $[(1_1^-)^2]$ with a rather large real part.  This configuration
remains unchanged until we reach another finite system size $\bar{L}_*$ which
again depends on the anisotropy, e.g.\ for $\gamma=2\pi/7$ we have
$\bar{L}_{*}=42$.  Now for $L \gtrsim \bar{L}_* $ we find that the second
level roots in two of the $bfbfb$ complexes degenerate giving rise to yet
another pattern of roots configurations making it difficult to follow this
state for large $L$. We remark that this kind of degeneracies is also present
when we use the $fbbbf$ grading and therefore we are in a situation similar to
the one described previously.

\begin{figure}[ht]
  \includegraphics[width=\textwidth]{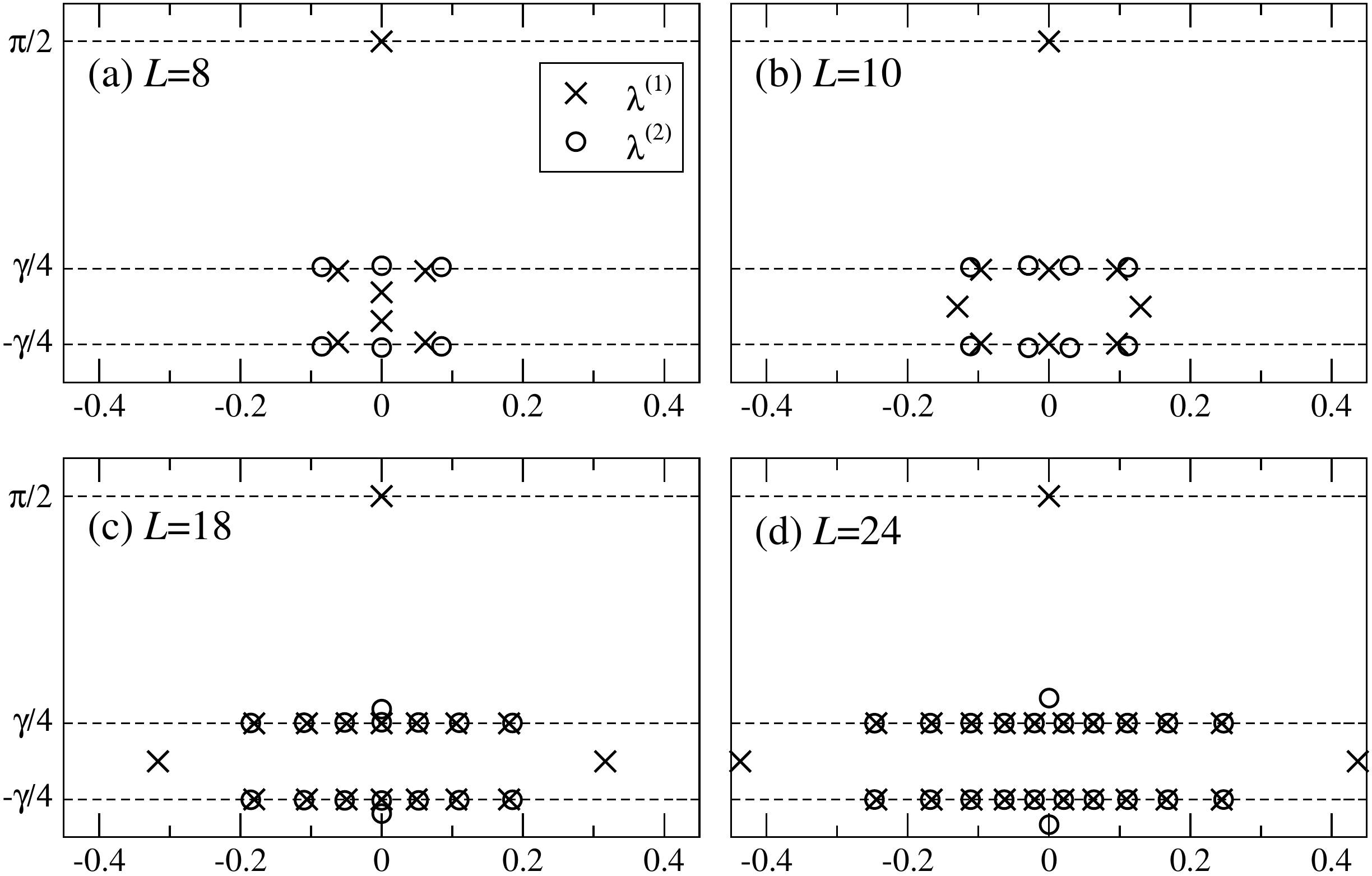}
  \caption{
    Sequence of degenerations of the Bethe root patterns for a state in
    charge sector $(1,1)$ at anisotropy $\gamma = 2\pi/7$: the root
    configuration changes from $f:[1_1^-]$ for $L=8$ (a) to $f:[(1_1^+)^2,
    1_1^-,2_2^+]$ for $L=10$ (b), and further to $f:[(1_1^+)^2,1_1^-,z_2]$ for
    $L=18$ (c) and $L=24$ (d). 
  }
  \label{fig:roots_11-s2}
\end{figure}

Another example of a state whose root configuration changes twice, at distinct
lattice sizes, has been observed in the $(1,1)$ sector. The degenerations are
exhibited in Fig.~\ref{fig:roots_11-s2} for $\gamma=2\pi/7$.
In (a) we show the root pattern for $L=8$ which is built out of a
configuration of type $f:[1_1^-]$.  By increasing the system size to $L=10$ we
see that one of the 2-strings at level one splits into two real roots giving
rise to new configuration $f:[(1_1^+)^2, 1_1^-,2_2^+]$, see
Fig.~\ref{fig:roots_11-s2}(b).  In addition to that, for $L> 12$ we
note that one of the two-strings at level two starts to be deformed into a
$[z_2]$ root configuration.  This latter behaviour has been displayed in
Fig.~\ref{fig:roots_11-s2}(c) and (d).  This root pattern remains
stable for system sizes up to $L=36$.  Beyond this size the solution of the
Bethe equations failed due to numerical instabilities.
%

\section{Missing and unclassified states}
\label{app:missing_states}
In addition to states changing their root configuration which have been
discussed in Appendix~\ref{app:roots} we also found states for which we could
find the root configuration only for specific values of the anisotropy
$\gamma$ and system size $L$. In this appendix we list these states and their
root configurations if known.

The first of these states belongs to the charge sector $(1,0)$ and
is described in terms of a $f: [1_1^-, 1_2^+, z_1, z_2]$ or
$b: [ 3_{12}^+, z_1, z_2]$ root configuration. The Bethe roots for both
gradings and anisotropy $\gamma = 2\pi/7$ can be found in
table~\ref{tab:roots-10-7}.
\begin{table}[ht]
\begin{tabular}{c | c || c }
    $\lambda^{(1)}/\gamma$ ($bfbfb$)& $\lambda^{(1)}/ \gamma$ ($fbbbf$) & $\lambda^{(2)}/\gamma$\\ \hline
  $-0.0614037 + 0.7497763i$ & $-0.0497337 + 0.2308725i$ & $-0.0594249 + 0.2575524i$ \\
  $-0.0614037 - 0.7497763i$ & $-0.0497337 - 0.2308725i$ & $-0.0594249 - 0.2575524i$ \\
  $~0.2775207 + 0.7511673i$ & $-0.1104048 + 0.1852444i$ & $~0.2586970 + 0.2364768i$ \\
  $~0.2775207 - 0.7511673i$ & $-0.1104048 - 0.1852444i$ & $~0.2586970 - 0.2364768i$ \\
  $-0.1083635 + 0.4999904i$ & $~0.2434987 + 0.2732079i$ & $-0.1052276 + 0.3794159i$ \\
  $-0.1083635 - 0.4999904i$ & $~0.2434987 - 0.2732079i$ & $-0.1052276 - 0.3794159i$ \\
  $-0.1077535 + 1.0052139i$ & $-0.0869949 + \pi/(2\gamma)i$ & $-0.1083634 $ \\
  $-0.1077535 - 1.0052139i$ & & \\ \hline
\end{tabular}
\caption{Root configuration of the $b: [3_{12}^+, z_1, z_2]$ or equivalently $f: [1_1^-, 1_2^+, z_1, z_2]$ state in the $(1,0)$ sector for $L=8$ and $\gamma = 2\pi/7$ \cite{FrHM19.data}.}
\label{tab:roots-10-7}
\end{table}

In the $(1,1)$ charge sector there are two states for which we know the root
configurations only for certain parameters. The first of these states has a $b: [1_1^-, z_1, z_2]$ root configuration at $L=6$ and
$\gamma < \pi/4$, see tab.~\ref{tab:roots-11-5-bfbfb}. When $\gamma \to \pi/4$
the imaginary part of the $[z_1]$ configuration shrinks such that at
$\gamma = \pi/4$ it passes into $[(1_1^+)^2]$ to avoid degenerations. For
$\gamma > \pi/4$ we weren't able to find a root configuration.
\begin{table}[ht]
\begin{tabular}{c | c}
    $\lambda^{(1)}/\gamma$ & $\lambda^{(2)}/\gamma$  \\ \hline
    $~0.0802213 + 0.7496060i$ & $0.0802934 + 0.2503739i$ \\
    $~0.0802213 - 0.7496060i$ & $0.0802934 - 0.2503739i$ \\
    $~0.7922719 + 0.6098025i$ & $0.9907590 + 0.3621826i$ \\
    $~0.7922719 - 0.6098025i$ & $0.9907590 - 0.3621826i$ \\ 
    $-0.6739339 + \pi/(2\gamma)i$ &  \\ \hline
\end{tabular}
\caption{Root configuration of the $b: [1_1^+, z_1, z_2]$ state in the $(1,1)$ sector for $L=6$ and $\gamma = 2\pi/9$ \cite{FrHM19.data}.}
\label{tab:roots-11-5-bfbfb}
\end{table}

For the charges $(n_1,n_2) = (2,1)$ we found one additional state with a very peculiar root configuration which does not fit to our notation introduced in Sec.~\ref{secNUMERICS}. Using the $fbbbf$ grading it consist of two (one) purely imaginary roots on the first (second) level. In contrast to the string complexes occurring at the other states here the remaining roots on both levels are not complex conjugates. Instead their imaginary parts are slightly shifted which can be seen explicitly in tab.~\ref{tab:roots-21-4-fbbbf} where we list the corresponding Bethe roots for $L=8$ and $\gamma = 2\pi/7$. We were able to calculate the Bethe roots for this state for $L \leq 16$ and $0 < \gamma \leq \pi/3$.
\begin{table}[ht]
\begin{tabular}{c | c}
    $\lambda^{(1)}/\gamma$ & $\lambda^{(2)}/\gamma$  \\ \hline
    $-0.0574347 + 0.2561136i$ & $-0.0576961 + 0.2563855i$ \\
    $~0.0574347 + 0.2561136i$ & $~0.0576961 + 0.2563855i$ \\
    $-0.0635931 - 0.2340169i$ & $-0.0573882 - 0.2442878i$ \\
    $~0.0635931 - 0.2340169i$ & $~0.0573882 - 0.2442878i$ \\
    $-0.1620614i$ & $0.5803344i$ \\
    $~1.3092483i$ & \\ \hline
\end{tabular}
\caption{Root configuration of the discussed state in the $(2,1)$ sector using the $fbbbf$ grading for $L=8$ and $\gamma = 2\pi/7$ \cite{FrHM19.data}.}
\label{tab:roots-21-4-fbbbf}
\end{table}


%

\end{document}